\documentclass{article}
\usepackage[top=1in, bottom=1in, left=1.25in, right=1.25in]{geometry}
\usepackage{graphicx}
\usepackage{amsmath} 
\usepackage{amsthm}
\usepackage{amsfonts}
\usepackage{amssymb}
\usepackage{comment}
\usepackage{braket}
\usepackage{enumitem}
\usepackage{xcolor}
\usepackage{soul}

\usepackage[
    backend=biber, 
    style=phys,
    doi=false, 
    url=false,
    eprint=false,
    biblabel=brackets,
    sorting=none,
]{biblatex}

\AtEveryBibitem{%
  \ifentrytype{book}{%
  }{%
    \clearlist{publisher}%
    \clearfield{note}%
    \clearfield{edition}%
    \clearlist{language}%
  }%
}

\DeclareFieldFormat[article]{pages}{\mkfirstpage{#1}}
\DeclareFieldFormat{titlecase}{#1}
\DeclareFieldFormat[article]{title}{\mkbibemph{#1}}
\DeclareFieldFormat[article]{journaltitle}{%
  \iffieldundef{doi}{%
    #1%
  }{%
    \href{https://doi.org/\thefield{doi}}{#1}%
  }%
}

\DeclareSourcemap{
  \maps[datatype=bibtex]{
    \map{
      \step[fieldsource=journal, match=\regexp{(?i)Journal\s+of\s+Statistical\s+Mechanics.*}, final]
      \step[fieldsource=journal, match=\regexp{(?i)Journal\s+of\s+Statistical\s+Mechanics.*}, replace={J. Stat. Mech.}]
      \step[fieldset=volume, null]
    }
    \map{
      \step[fieldsource=journal, match=\regexp{(?i)Physical\s+Review\s+Letters}, replace={Phys. Rev. Lett.}]
      \step[fieldsource=journal, match=\regexp{(?i)Physical\s+Review\s+A}, replace={Phys. Rev. A}]
      \step[fieldsource=journal, match=\regexp{(?i)Physical\s+Review\s+B}, replace={Phys. Rev. B}]
      \step[fieldsource=journal, match=\regexp{(?i)Physical\s+Review\s+X}, replace={Phys. Rev. X}]
      \step[fieldsource=journal, match=\regexp{(?i)Physical\s+Review\s+E}, replace={Phys. Rev. E}]
      \step[fieldsource=journal, match=\regexp{(?i)Physical\s+Review\s+Research}, replace={Phys. Rev. Research}]
      \step[fieldsource=journal, match=\regexp{(?i)Reviews\s+of\s+Modern\s+Physics}, replace={Rev. Mod. Phys.}]
      \step[fieldsource=journal, match=\regexp{(?i)Nature\s+Communications}, replace={Nat. Commun.}]
    }
  }
}

\usepackage[colorlinks=true,
            linkcolor=blue,
            citecolor=blue,
            urlcolor=blue]{hyperref}
\usepackage[capitalise]{cleveref}
\usepackage[font=small]{caption}
\usepackage{tikz} 

\newcommand{\bs}[1]{\boldsymbol{#1}} 
\newcommand{\rhoGE}{\rho_{\text{\tiny GE}}}

\newcommand{\rhoGGE}{\rho_{\text{\tiny GGE}}}

\title{\bf Athermality of generalized Gibbs ensembles}
\author{Riccardo Senese$^1$, Bruno Bertini$^2$, Katja Klobas$^2$, Pasquale Calabrese$^1$}

\begin{document}

\maketitle

{\small
\vspace{-5mm}  \ \\
{$^{1}$}  SISSA and INFN, via Bonomea 265, 34136 Trieste, Italy\\
\medskip
{$^{2}$} School of Physics and Astronomy, University of Birmingham, Edgbaston, Birmingham, B15 2TT, UK
\medskip
}

\begin{abstract}
Integrable quantum systems evolving from non-equilibrium initial states do not thermalize to conventional Gibbs ensembles (GE). Instead, at long times they relax to generalized Gibbs ensembles (GGEs), which incorporate the full set of local and quasi-local conserved quantities. While GGEs have been extensively studied in the literature, a quantitative analytic characterization of how different they are from ordinary GEs is still lacking. 
In this work, we address this question by employing the concept of \emph{athermality}, which we define within quantum resource theory as the relative entropy between a given state and the closest thermal state.
By means of integrability techniques we compute the athermality for several quantum quenches in paradigmatic integrable models, including the free XY spin chain, the interacting Lieb-Liniger model, the XXZ spin chain, and the harmonic chain. We find that often the athermality becomes anomalously small when the post-quench Hamiltonian is critical in its ground state, despite probing physics at a finite energy density. We also prove that it systematically develops a singularity at criticality, which is inherited from the entropy of the GGE. 
\end{abstract}

\tableofcontents

\section{Introduction}
\label{sec:intro}

Isolated quantum systems evolving unitarily from non-equilibrium initial states generically thermalize~\cite{Deutsch_91,Srednicki_1994,polkovnikov_colloquium_2011, eisert_quantum_2015, dalessio_quantum_2016, calabrese_introduction_2016, bastianello_introduction_2022}. 
This means that, at long times, the reduced density matrix of a finite subsystem embedded in a thermodynamically large system approaches the reduced density matrix of a thermal ensemble, with the temperature uniquely fixed by the energy density of the initial state. As a consequence, although the full system remains at all times in a pure state characterized by zero von Neumann entropy, finite subsystems behave locally as if they were in thermal equilibrium. In this sense, entanglement with the rest of the system effectively acts as a thermal bath for local observables~\cite{Kaufman_2016, essler_quench_2016, alba_entanglement_2017,calabrese_ln}. This mechanism provides the modern understanding of thermalization in isolated quantum many-body systems.

From a dynamical perspective, one of the most intriguing aspects of quantum integrable models~\cite{korepin_quantum_1993} is their failure to thermalize under quantum unitary dynamics. In these systems, the existence of an \emph{extensive} number of mutually compatible conservation laws strongly suppresses the scrambling of local quantum information, providing a natural mechanism for local memory retention. Importantly, despite the fragile nature of integrability, this behavior persists over long prethermal timescales~\cite{moeckel_interaction_2008, kollar_generalized_2011,  marcuzzi_prethermalization_2013, essler_quench_2014, bertini_prethermalization_2015,  brandino_glimmers_2015, bertini_prerelaxation_2015, babadi_far_2015, bertini_thermalization_2016, fagotti_universal_2015, alba_prethermalization_2017, mallayya_prethermalization_2019, Bertini_2020_PT, durnin_2021_nonequilibrium, lopezpiqueres_hydrodynamics_2021} even in experimentally realistic settings~\cite{kinoshita_quantum_2006, gring_relaxation_2012, Langen_2015}.
For this reason, at long times an integrable system does not thermalize to the standard Gibbs ensemble (GE). Instead, it relaxes to a special statistical ensemble that incorporates the constraints imposed by the full set of local and quasi-local conserved quantities~\cite{essler_quench_2016, vidmar_generalized_2016, ilievski_quasilocal_2016}. This ensemble is known as the \emph{generalized Gibbs ensemble} (GGE)~\cite{rigol_relaxation_2007}. 
The GGE extends the conventional Gibbs description by assigning a Lagrange multiplier to each conserved quantity, thereby encoding the strong memory that integrable systems retain of their initial conditions. It provides the correct stationary description for local observables after quantum quenches in a broad class of integrable models and has become one of the central concepts in the modern understanding of non-equilibrium quantum dynamics.

GGEs have been extensively studied in the literature~\cite{essler_quench_2016, vidmar_generalized_2016}, in particular with regard to their thermodynamic properties~\cite{caux_constructing_2012, mossel_generalized_2012, doyon_thermalization_2017, dymarsky_generalized_2019, ilievski_equilibrium_2019}, entropies~\cite{santos_weak_2012, collura_stationary_2014, kormos_stationary_2014, gurarie_global_2013, fagotti_finite_2013, dora_escort_2014, piroli_correlations_2017, alba_quench_2017, bertini_entanglement_2018}, expectation values of local observables~\cite{kormos_exact_2011, pozsgay_local_2011, pozsgay_mean_2011, negro_on_2013, mestyan_short_2014,  bertini_quantum_2016, pozsgay_excited_2017, bastianello_exact_2018,Bastianello_2018}, and correlation functions~\cite{calabrese_quantum_2012a, kozlowski_on_2018, granet_finite_2020, granet_low_2021, granet_a_2020, denardis_correlation_2022, doyon_emergence_2023}. 

A very natural question, which has nevertheless received little attention and has been addressed mostly by means of numerical calculations in finite systems \cite{rigol_initialstate_2011, he_initialstate_2012, he_initialstate_2013} or rather indirectly by integrability techniques (typically through comparisons of expectation values and correlation functions)~\cite{Rossini_2010,Rossini_2009,fagotti_reduced_2013,fagotti_relaxation_2014,pozsgay_generalized_2013}, concerns the precise relation between GGEs and ordinary thermal ensembles. In particular, how different is a GGE from a standard GE? Is there a simple and physically transparent way to quantify this difference? Addressing these questions is important not only for understanding the structure of integrable stationary states, but also for clarifying in which sense integrable systems fail to thermalize in the conventional way.

The goal of the present work is to address these questions. We will show that the concept of {\it athermality}~\cite{brandao_resource_2013, lostaglio_description_2015, brandao_second_2015, gour_resource_2015}, rooted in quantum resource theory~\cite{coecke_a_2016, chitambar2019quantum}, provides a particularly natural and effective framework for this purpose. Here we define the athermality as the relative entropy between a given state (or ensemble) and the {\it closest} GE, where closeness is measured by the minimization of the same relative entropy. Such closest thermal ensemble is simply the GE with the same average energy of the chosen state, and as a consequence the athermality is simplified to equal the difference of their von Neumann entropies. In the context of integrable quenches we consider the GGE as the given state (cf.~Ref.~\cite{sels_stationary_2015} for the use of a minimized relative entropy in a similar context), and hence the closest GE is precisely the thermal ensemble to which the system would have locally relaxed in the absence of integrability. For this reason, the athermality coincides with the difference between the GE and GGE entropies employed in the numerical studies of Refs.~\cite{rigol_initialstate_2011, he_initialstate_2012, he_initialstate_2013, rigol_fundamental_2016}. We stress, however, that by means of integrability techniques we obtain exact thermodynamic-limit results for the athermality, which enable us to shed new light on the fundamental questions above.

We compute the GGE athermality for several different quench protocols in integrable models, including the free XY spin chain in a transverse field, the interacting XXZ spin chain, the interacting  Lieb–Liniger gas, and the harmonic chain. The derivations, technical details, and general formalism are presented in the main body of the paper. Here, however, we would like to emphasize the most surprising outcomes of our analysis.\\

\ (i)  \ The athermality, as a function of the post-quench parameters in the Hamiltonian, is always singular at quantum critical points \cite{sachdev_quantum_2011}. This appears at first rather remarkable because the athermality after a quench probes physics at a finite energy density. 
This sharp signature of quantum criticality in the middle of the many-body spectrum is due to the singular behaviour of the GGE entropy, while the GE entropy remains analytic everywhere as expected at any nonzero temperature~\cite{araki_gibbs_1969}. Singularities in the GGE entropy at quantum criticality were observed in Ref.~\cite{porta_effective_2018} (cf.~also Refs.~\cite{Calabrese_ev_2005, fagotti_evolution_2008, russomanno_entanglement_2016, paul_hidden_2024, mitra_quantum_2025}). Here we analytically prove their emergence in the XY chain in a transverse field, and demonstrate that they arise also in the XXZ chain. For the XXZ chain we also uncover the presence of a fractal behaviour of the GGE entropy in the gapless phase, similarly to what found for other physical quantities~\cite{ilievski_microscopic_2017, Collura_2020,ilievski_popcorn_2022}. 
Singular behaviours at criticality in the context of GGEs have also been uncovered by probing local observables~\cite{sengupta_quench_2004, bhattacharyya_signature_2015, porta_effective_2018, rossini_dynamics_2020}, while influence of quantum criticality has been observed in the context of integrable transport~\cite{ilievski_quasilocal_2016, bertini_finite_2021, bulchandani_superdiffision_2021}. More generally, signatures of quantum criticality emerging at intermediate and late times following quantum quenches have been observed in both integrable~\cite{pappalardi_multipartite_2017, heyl_detecting_2018,titum_probing_2019,surace_operator_2020,vernier_integrable_2023,csepanyi_dynamical_2024,csepanyi2026observing} and non-integrable settings~\cite{haldar_signatures_2021,robertson_simple_2023,dag_detecting_2023}.  
Despite the growing body of evidence for such phenomena, a unified picture and theoretical explanation for their microscopic origin and significance is still missing. In the present integrable context, however, sharp singularities can be characterized exactly thanks to the analytical tractability of the models. \\

\ (ii) \ We find that the athermality is generically an extensive quantity in the subsystem size, with a prefactor of order unity. However, often the athermality becomes surprisingly small at quantum criticality, where it takes the form of a sharp local-minimum cusp. 
Importantly, the dip at criticality emerges from the interplay between the GE and GGE entropies, and hence it is in general a distinctive property only of the athermality. We uncover that this enhanced proximity between the GGE and a thermal ensemble is generally associated with a {\it pinning mechanism}:  the GE and GGE root densities (or mode occupation functions in free models) get pinned to a common maximal value at a critical momentum; moreover, the two distributions exhibit very similar behaviour in a broad neighbourhood of this point. As a consequence, the differences between the two ensembles are strongly suppressed over the region of momentum space that contributes most significantly to thermodynamic quantities and athermality.\\

We conclude this introduction by highlighting a connection with known results in conformal field theory (CFT) that provide us with a heuristic explanation for why the athermality often becomes surprisingly small at quantum criticality.
Our starting point is the well-known and rather remarkable observation that, for quantum quenches in CFTs from a special class of initial states of the form
$|\psi_0\rangle = e^{-\tau_0 H}|B\rangle,
$
the stationary state is exactly thermal, with an effective inverse temperature $\beta = 4\tau_0$~\cite{Calabrese_2006,calabrese_quantum_2007, calabrese_quantum_2016}. This result is particularly striking because the underlying dynamics remains fully integrable. The origin of this emergent thermal behavior can be traced back to the structure of the initial state, which excites only the Hamiltonian while leaving all higher conserved charges effectively unactivated~\cite{Cardy_2016,calabrese_quantum_2016}. As a consequence, the corresponding GGE reduces to an ordinary GE.
We expect that many of the quenches analyzed in this work admit, in the scaling limit and close to a quantum critical point, an effective description that is very similar to the one above. In this regime, the initial states become approximately conformal boundary states and the higher conserved quantities are only weakly activated. The resulting stationary state is therefore expected to be very close to the corresponding thermal ensemble, naturally leading to a small value of the athermality.
While this argument provides a qualitative understanding of the suppression of athermality near criticality, the actual values that we find are often extraordinarily small. This remains somewhat surprising.

The remainder of the work is organized as follows. In \cref{sec:quantumathermdef} we define the athermality and highlight the simplifying features emerging from our minimization prescription. In \cref{sec:athermandTBA} we discuss how the athermality can be calculated exactly in the framework of the thermodynamic Bethe ansatz (TBA)~\cite{yang_thermodynamics_1969, takahashi_thermodynamics_1999} and its relevance in the context of quantum quenches. In \cref{sec:XY} we study the athermality for integrable quantum quenches in the XY chain in a transverse field, and analyze its behaviour across both quantum critical lines. In \cref{sec:XXZsec} we turn to the XXZ chain and demonstrate that the main conclusions found for the free XY chain survive in the presence of interactions. In \cref{sec:LL} we consider integrable quenches in the Lieb-Liniger gas, and analyze the athermality for repulsive interactions and the GGE entropy for both repulsive and attractive couplings. The absence of a quantum phase transition in the repulsive gas is reflected in the absence of striking features in the athermality. In \cref{sec:harmonicchain} we consider the harmonic chain, and study the differences in athermality due to the bosonic statistics of the elementary modes. Finally, in \cref{sec:timevathermality}, we provide some remarks about the time evolution of the athermality after quantum quenches.
Several appendices follow which review basic concepts, give further details of the derivations and present additional results.

\section{Quantum athermality}
\label{sec:quantumathermdef}

Consider a system described by a Hamiltonian $H$, and a generic (reduced) density matrix $\rho$.
We are interested in quantifying how far $\rho$ is from a GE at inverse temperature $\beta$
\begin{equation}
\label{eq:GibbsRDM}
    \rhoGE = \frac{e^{-\beta H}}{Z_\text{\tiny GE}} \qquad \qquad Z_\text{\tiny GE} = {\rm Tr}[e^{-\beta H}] \ .
\end{equation}
A possible choice would be to employ a proper distance between $\rho$ and $\rhoGE$, such as the trace distance, fidelity or Hilbert-Schmidt distance~\cite{nielsen_quantum_2010}. The former two possess a clear operational meaning in experimental settings but are usually hard to calculate, while the latter lacks a simple physically relevant operational interpretation. For these reasons we turn our attention to the \emph{quantum relative entropy} (QRE)~\cite{vedral_role_2002, nielsen_quantum_2010}, defined as
\begin{equation}
    S(\rho||\sigma)={\rm Tr}[\rho \ln \rho] - {\rm Tr}[\rho \ln \sigma] \ ,
\end{equation}
where $\rho$ and $\sigma$ are two arbitrary density matrices (see also, e.g., Refs.~\cite{yungerhalpern_noncommuting_2020, kranzl_experimental_2023} for the use of the QRE in similar contexts). The QRE is not a metric, as it is not symmetric under $\rho \leftrightarrow \sigma$ and it violates the triangle inequality. However, it respects positivity
\begin{equation}
\label{eq:ineqrelentr}
    S(\rho||\sigma) \ge 0 \ \ \ \ \forall \ \sigma, \rho \qquad \qquad S(\rho||\sigma)=0 \ \ {\rm iff} \ \ \sigma = \rho \ ,
\end{equation}
and has a clear operational interpretation via the quantum Stein's lemma~\cite{hiai_proper_1991, nagaoka_strong_2000, vedral_role_2002}. 

It is well known that the use of the QRE with respect to GEs can be drastically simplified~\cite{brandao_resource_2013}. This is a consequence of the equality
\begin{equation}
\label{eq:SrelrhoGE}
\begin{aligned}
    S(\rho||\rhoGE) &= -S_{\rm vN}(\rho) + \beta \, {\rm Tr}[\rho \, H] + \ln Z_\text{\tiny GE} \\
    &= \beta\big({\rm Tr}[\rho \, H]-{\rm Tr}[\rhoGE \, H]\big)- \big(S_{\rm vN}(\rho)-S_{\rm vN}(\rhoGE)\big) \ ,
\end{aligned}
\end{equation}
where $S_{\rm vN}$ denotes the von Neumann entanglement entropy and we have used $\ln Z_\text{\tiny GE} = -\beta\, {\rm Tr}[\rhoGE \, H]+S_{\rm vN}(\rhoGE)$. 
We define the athermality $\mathcal{A}(\rho)$ of a generic density matrix $\rho$ in terms of the QRE as
\begin{equation}
\label{eq:athermality}
    \mathcal{A}(\rho) = \min_{\beta} S(\rho||\rhoGE) \ ,
\end{equation}
where as in \cref{eq:GibbsRDM} the dependence of $\rhoGE$ on $\beta$ is implicit. From the perspective of quantum resource theories~\cite{chitambar2019quantum}, the athermality $\mathcal{A}(\rho)$ quantifies how resourceful a state $\rho$ that differs from a GE is within an experimental setup in which preparing a GE at \emph{any} inverse temperature $\beta$ has negligible cost. We stress that, due to the presence of the minimization over $\beta$, the definition \eqref{eq:athermality} differs from the one relevant to situations in which the only resource-free states that can be prepared at negligible cost are GEs at a \emph{fixed} inverse temperature $\beta_{\rm fixed}$~\cite{brandao_resource_2013, brandao_second_2015, gour_resource_2015, goold_role_2016}.
To find the global minimum of $S\big(\rho||\rhoGE\big)$ as a function of $\beta$ we consider its derivatives
\begin{equation}
\label{eq:firstderofSrho}
    \frac{d S\big(\rho||\rhoGE\big)}{d \beta} = {\rm Tr}[\rho \, H] + \frac{d \ln Z_\text{\tiny GE}}{d \beta} = {\rm Tr}[\rho \, H] - {\rm Tr}[\rhoGE \, H] \ ,
\end{equation}
\begin{equation}
\label{eq:secondderofSrho}
    \frac{d^2 S\big(\rho||\rhoGE\big)}{d \beta^2} = - \frac{d {\rm Tr}[\rhoGE \, H]}{d \beta} = {\rm Tr}[\rhoGE \, H^2] - {\rm Tr}[\rhoGE \, H]^2  ,
\end{equation}
and note
\begin{equation}
\label{eq:secondderofSrho2}
    \frac{d^2 S\big(\rho||\rhoGE\big)}{d \beta^2} > 0  \quad  \text{for}
    \quad  |\beta |< \infty \ .
\end{equation}
Setting \cref{eq:firstderofSrho} to zero, and given \cref{eq:secondderofSrho2}, we find that the minimum of the QRE is obtained at the inverse temperature $\beta^*$ for which the average energy of $\rho$ and $\rhoGE(\beta^*)$ is the same. From this it follows that the quantum athermality \eqref{eq:athermality} is simply given by a difference of two von Neumann entropies
\begin{equation} \label{eq:athermalitysimple}
    \mathcal{A}(\rho) = S\big(\rho||\rhoGE(\beta^*)\big) 
    = S_{\rm vN}\big(\rhoGE(\beta^*)\big)-S_{\rm vN}(\rho) \ , 
    \quad \qquad {\rm Tr}[\rhoGE(\beta^*) \, H] = {\rm Tr}[\rho \, H] \equiv E_\rho\ .
\end{equation}
From \cref{eq:secondderofSrho,eq:secondderofSrho2} we see that the GE energy is a monotonically decreasing function of $\beta$,  therefore $\beta^*$ is always unique (it can be negative if the spectrum is bounded from above). The semipositive definiteness of $\mathcal{A}(\rho)$,  guaranteed by the inequality \eqref{eq:ineqrelentr}, is re-interpreted in \cref{eq:athermalitysimple} as a consequence of the GE being the highest-entropy ensemble at a fixed average energy.

In many cases of interest the Hamiltonian $H$ might possess a $\mathrm{U}(1)$ symmetry generated by an observable $N$ ($[N,H]=0$) that is of operational importance. For example, $N$ might indicate the number of particles within the system subject to full experimental control. In such cases it is also meaningful to consider a GE with the addition of an extra Lagrange multiplier (grand canonical ensemble)
\begin{equation} \label{eq:grandGibbsRDM}
    \rhoGE = \frac{e^{-\beta (H-\mu N)}}{Z_\text{\tiny GE}}, \qquad
    Z_\text{\tiny GE} = {\rm Tr}[e^{-\beta (H - \mu N)}] \ .
\end{equation}
The QRE with respect to such GE retains the same form of \cref{eq:SrelrhoGE} but with the substitution $H \to H - \mu N$. Following similar logic and simple minimization steps as above we define the athermality $\mathcal{A}(\rho)$ as
\begin{equation} \label{eq:athermality2}
  \begin{gathered}
    \mathcal{A}(\rho) = \min_{\beta,\, \mu} S(\rho||\rhoGE) 
    = S_{\rm vN}\big(\rhoGE(\beta^*,\mu^*)\big)-S_{\rm vN}(\rho)\ , \\
    {\rm Tr}[\rhoGE(\beta^*,\mu^*) \, H] = {\rm Tr}[\rho \, H] \equiv E_\rho\ , \qquad 
    {\rm Tr}[\rhoGE(\beta^*,\mu^*) \, N] = {\rm Tr}[\rho \, N] \equiv N_\rho \ .
  \end{gathered}
\end{equation}
It is straightforward to generalize these statements in order to include an arbitrary finite number of operationally relevant $\mathrm{U}(1)$ symmetries.

\section{Athermality of Generalized Gibbs Ensembles}
\label{sec:athermandTBA}

The distinguishing feature of quantum integrable Hamiltonians~\cite{korepin_quantum_1993} is that they possess an \emph{extensive} number of algebraically independent charges $Q^{(n)}$ ($[H,Q^{(n)}]=0\,, \ \forall \ n=1, 2, \ldots$), where each $Q^{(n)}$ can be written as a sum or integral of (quasi)local~\cite{ilievski_quasilocal_2016, prosen_families_2013, prosen_quasilocal_2014, pereira_exactly_2014, ilievski_quasilocal_2015, ilievski_complete_2015, essler_generalized_2015, vernier_quasilocal_2017} densities $q^{(n)}(x)$~\cite{korepin_quantum_1993, slavnov_algebraic_2022}.
For this reason, the Generalized Gibbs Ensemble (GGE)
\begin{equation}
\label{eq:defGGE}
    \rhoGGE = \frac{e^{-\sum_n \beta_n Q^{(n)}}}{Z_\text{\tiny GGE}}  
\end{equation}
plays a special role in quantum integrability (note that we implicitly included $H$ in the set $Q^{(n)}$). In the framework of quantum quenches in integrable systems, the GGE correctly captures local properties of the system at late times~\cite{essler_quench_2016, vidmar_generalized_2016}. In particular, given an initial lowly entangled state $\ket{\psi(0)}$ (often the ground state of a pre-quench Hamiltonian $H'$) that for times $t>0$ evolves unitarily under a post-quench integrable Hamiltonian $H$, in the thermodynamic limit and at late times one expects in any finite (or at most subextensive) subsystem $A$ (with complement $B$)~\cite{essler_quench_2016}
\begin{equation}
\label{eq:relaxIntModels}
    \lim_{t \to \infty} \lim_{L \to \infty} {\rm Tr}_{ B}\big[\ket{\psi(t)}\!\bra{\psi(t)}\big] = \lim_{L \to \infty}{\rm Tr}_{ B}[\rhoGGE]=\rho_{A} \ ,
\end{equation}
where $L$ denotes the system size and the order of limits on the left-hand side is important. The Lagrange multipliers $\beta_n$ in \cref{eq:defGGE} enforce that the average densities $\braket{Q^{(n)}}/L$ at infinite times are the same as in the initial state $\ket{\psi(0)}$. We stress that in the quench context, where the partial trace over $B$ plays a crucial role, the GGE is \emph{only one} of many possible choices of ensembles capturing the correct local properties at late times~\cite{essler_quench_2016}.

In the following we focus on the GGE density of athermality 
\begin{equation}
\label{eq:athermdens}
    \alpha = \lim_{L \to \infty}\frac{\mathcal{A}(\rhoGGE)}{L} = \lim_{L \to \infty}\frac{S_{\rm vN}\big(\rhoGE(\beta^*)\big)-S_{\rm vN}(\rhoGGE)}{L} \ .
\end{equation}
This choice is convenient for several reasons: (i) it captures the leading order (in system size) value of the athermality between arbitrary GEs and GGEs, which is an interesting quantity irrespective of the many (isolated, open, weakly nonintegrable or driven) setups where GGEs emerge~\cite{essler_quench_2016,vidmar_generalized_2016,kollar_generalized_2011,lazarides_periodic_2014,bertini_transport_2016,castro-alvaredo_emergent_2016,Lange2018,lenarcic_perturbative_2018,durnin_2021_nonequilibrium,starchl_relaxation_2022,buca_unified_2023,ulcakar_iterative_2024,ulcakar_generalized_2025}; (ii) the intensive value $\alpha$ describes the leading order (in subsystem size $|A|$) of the athermality in the reduced density matrix $\rho_A$ from \cref{eq:relaxIntModels}, relevant in the context of quantum quenches in isolated systems; (iii) $\alpha$ can be computed solely within the framework of the TBA~\cite{yang_thermodynamics_1969, korepin_quantum_1993, takahashi_thermodynamics_1999}, as we explain in the next section.
Generalizations of \cref{eq:athermdens} to accommodate more Lagrange multipliers in the GE, as in \cref{eq:grandGibbsRDM,eq:athermality2}, are straightforward.

\subsection{Entropy density of GGEs and TBA}
Here we just recall the fundamental ingredients we will need in the rest of the discussion, while a brief review of some of the basic concepts can be found in Appendix~\ref{appendix:RDMabelian}. 

Eigenstates of Bethe-integrable models can be parametrized in terms of sets of rapidities $\bs \lambda = (\lambda_1, \lambda_2, \ldots)$. In the limit $L \to \infty$, the essence of TBA is to switch attention from individual eigenstates to classes of eigenstates, or \emph{macrostates}, with the same local properties. For the simplest models hosting only one type of quasiparticles, in the limit of large $L$ two eigenstates are said to belong to the same macrostate if they possess an identical distribution $\varrho(\lambda)$ of rapidities, where $\varrho(\lambda)$ is known as root density. In models hosting multiple species of quasiparticles, e.g.~bound states of different length, a macrostate is specified by a set $\bs \varrho(\lambda)=\{\varrho_\ell(\lambda)\}_{\ell = 1}^{\ell_{\rm max}}$ of root densities corresponding to different quasiparticle species labelled by $\ell=1,2,\ldots$. The insight of TBA is that many physical quantities depend only on macrostate information. For example, the density of thermal entropy from \cref{eq:athermdens} is given by
\begin{equation} \label{eq:GEvNTBA}
    \lim_{L \to \infty}\frac{S_{\rm vN}\big(\rhoGE(\beta^*)\big)}{L} 
    = s_\text{\tiny YY}[\bs \varrho_{\beta^*}(\lambda)] \ ,
\end{equation}
where $s_\text{\tiny YY}[\bs \varrho_{\bs \beta^*}(\lambda)]$ denotes the Yang-Yang entropy density of the thermal macrostate $\bs \varrho_{\beta^*}(\lambda)$ at inverse temperature $\beta^*$~\cite{korepin_quantum_1993} (see Appendix~\ref{appendix:RDMabelian}). 

Crucially, also the GGE entropy density in \cref{eq:athermdens} depends solely on macrostate information~\cite{mossel_generalized_2012, fagotti_conservation_2014, ares_lack_2023}. 
This is particularly simple to see in cases where the charges $Q^{(n)}$ with \emph{nonzero} Lagrange multipliers $\beta_n$ in \cref{eq:defGGE} form an Abelian algebra
\begin{equation}
\label{eq:AbAssumption}
    [H, Q^{(n)}] = 0 \ \ \ \ \forall \ n \ , \qquad \qquad [Q^{(m)},Q^{(n)}] = 0 \ \ \ \ \forall \ m, n  \ \ .
\end{equation}
In such cases ones speaks of an Abelian GGE. Averages with respect to such GGEs are dominated by eigenstates belonging to a specific non-thermal saddle-point macrostate $\bs \varrho_{\text{\tiny GGE}}(\lambda)$. Similarly to \cref{eq:GEvNTBA}, the GGE entropy density depends solely on the macrostate information encoded in $\bs \varrho_{\text{\tiny GGE}}(\lambda)$
\begin{equation}
\label{eq:SvNrhoATBA}
\lim_{L \to \infty}\frac{S_{\rm vN}\big(\rhoGGE\big)}{L} = s_\text{\tiny YY}[\bs \varrho_{\text{\tiny GGE}}(\lambda)] \ .
\end{equation}
When the GGE emerges from quantum quenches as in \cref{eq:relaxIntModels}, the macrostate $\bs \varrho_{\text{\tiny GGE}}(\lambda)$ can be obtained from the quench action method~\cite{caux_time_2013, caux_quench_2016} if expressions for overlaps $\braket{\psi(0)|\bs \lambda}$ are known. In the absence of the overlaps, it is possible to reconstruct $\bs \varrho_{\text{\tiny GGE}}(\lambda)$ if the expectation values $\braket{\psi(0)|Q^{(n)}|\psi(0)}$ of all the charges entering the GGE \eqref{eq:defGGE} are available~\cite{ilievski_complete_2015, ilievski_stringcharge_2016}. Moreover, for integrable initial states~\cite{piroli_what_2017, ghoshal_boundary_1994, fioretto_quantum_2010, sotiriadis_zamolodchikov_2012, bertini_quantum_2014}, one can apply the quantum transfer matrix approach to find analytical expressions for the rapidity distributions~\cite{piroli_from_2017, piroli_non_2018, piroli_integrable_2019a, piroli_integrable_2019b}. In the following sections we will make use of results obtained from all these approaches (see Appendix~\ref{appendix:RDMabelian} for further details). 

We stress that it has been shown explicitly in many contexts~\cite{fagotti_evolution_2008, alba_entanglement_2009, calabrese_quantum_2012a, fagotti_reduced_2013, collura_stationary_2014, ares_excited_2014, alba_entanglement_2017, bertini_entanglement_2018} that the entropy density \eqref{eq:SvNrhoATBA} does indeed describe the leading order (in subsystem size $A$) saturation value of the entanglement entropy at late times after quantum quenches in integrable models. 

In summary, we see from \cref{eq:athermdens,eq:GEvNTBA,eq:SvNrhoATBA} that the density of GGE athermality $\alpha$ is simply given by
\begin{equation}
\label{eq:defAthDensity}
    \alpha= s_\text{\tiny YY}[\bs \varrho_{\beta^*}] - s_\text{\tiny YY}[\bs \varrho_{\text{\tiny GGE}}] \ .
\end{equation}
As before, these results generalize easily to cases in which we include additional charges in the GE description, cf.~\cref{eq:grandGibbsRDM,eq:athermality2}.\\

We remark that integrable models in general possess discrete and continuous non-Abelian symmetries. In some models, even the full extensive algebra of higher (quasi)local conserved charges $Q^{(n)}$ might be non-Abelian. This is for example the case in the XY model (although in the absence of a transverse field)~\cite{essler_quench_2016, fagotti_conservation_2014, ares_lack_2023} and in the XXZ chain (only at specific roots of unity values of the anisotropy $\Delta$)~\cite{deguchi_sl2_2001, korff_auxiliary_2003, korff_twisted_2004, zadnik_quasilocal_2016, ilievski_quasilocal_2016}, considered respectively in Sections \ref{sec:XY} and \ref{sec:XXZsec}. As pointed out in Ref.~\cite{fagotti_conservation_2014} (see also Refs.~\cite{essler_quench_2016, ares_lack_2023}), however, for any given quench the initial state selects an Abelian subset of charges that can be used to construct an Abelian GGE. Therefore, a description in terms of rapidity distributions should always be possible~\cite{ilievski_interacting_2017, ilievski_equilibrium_2019}. 

For sake of completeness we mention that in certain cases the GGE functional form in \cref{eq:defGGE} cannot be directly applied for interacting models (at least not without modifications)~\cite{de_nardis_solution_2014, ilievski_interacting_2017, ilievski_equilibrium_2019, cecile_squeezed_2024}. This is for example the case in the BEC quench~\cite{de_nardis_solution_2014} in the Lieb-Liniger model analyzed in \cref{sec:LL} and in the Néel quench in the gapped XXZ chain of \cref{sec:XXZsec}~\cite{cecile_squeezed_2024}. In such cases, however, the correct ensemble $\rho_{\text{\tiny GGE}}$ behind the reduced density matrix $\rho_A$ of \cref{eq:relaxIntModels}, even if not of the form \eqref{eq:defGGE}, is exactly captured within the TBA framework we employ.

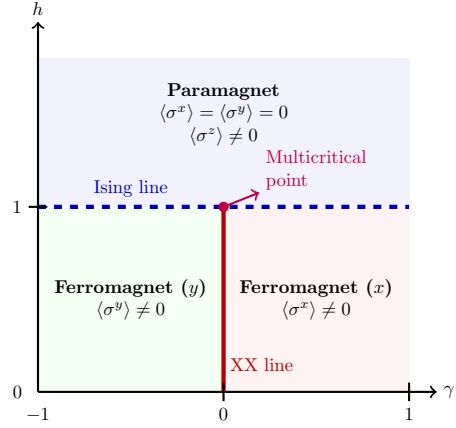
\begin{figure}[t!]
    \centering
    \begin{tikzpicture}[scale=0.7, transform shape, x=3.5cm, y=3.5cm]
        \fill[blue!5] (-1, 1) rectangle (1, 1.8);  
        \fill[green!5] (-1, 0) rectangle (0, 1);   
        \fill[red!5] (0, 0) rectangle (1, 1);      

        \draw[->, thick] (-1, 0) -- (1.15, 0) node[right] {$\large \gamma$};
        
        \draw[->, thick] (-1, 0) -- (-1, 2.0) node[above] {$\large h$};

        \draw[ultra thick, dashed, blue!70!black] (-1, 1) -- (1, 1); 
        \draw[ultra thick, red!70!black] (0, 0) -- (0, 1);           

        \draw[thick] (1, 0.05) -- (1, -0.05) node[below] {$1$};
        \draw[thick] (0, 0.05) -- (0, -0.05) node[below] {$0$};
        \node[below] at (-1, -0.05) {$-1$};

        \draw[thick] (-0.95, 1) -- (-1.05, 1) node[left] {$1$};
        \node[left] at (-1.05, 0) {$0$};

        \node[align=center] at (0, 1.5) {\textbf{Paramagnet} \\ $\langle \sigma^x \rangle = \langle \sigma^y \rangle = 0$ \\ $\langle \sigma^z \rangle \neq 0$};
        \node[align=center] at (0.5, 0.5) {\textbf{Ferromagnet ($x$)} \\ $\langle \sigma^x \rangle \neq 0$};
        \node[align=center] at (-0.5, 0.5) {\textbf{Ferromagnet ($y$)} \\ $\langle \sigma^y \rangle \neq 0$};

        \node[blue!70!black, align=center, above=2pt] at (-0.5, 1) {Ising line};
        \node[red!70!black, align=center, right=2pt] at (-0.02, 0.15) {XX line};
        
        \filldraw[purple] (0,1) circle (0.025) coordinate (mcp);
        \node[purple, align=left] (mcp_label) at (0.5, 1.2) {Multicritical \\ point};
        \draw[->, thick, purple] (mcp) -- (mcp_label);
    \end{tikzpicture}
    \caption{ $T=0$ quantum phase diagram of the XY chain in a transverse field with Hamiltonian $H_{\rm XY}(\gamma,h)$ from \cref{eq:HXY}. See text for description of (i) Ising critical line (blue dashed) and (ii) XX critical line (red).}
    \label{fig:xy_phase_diagram}
\end{figure}
\section{XY chain in a transverse field}
\label{sec:XY}

We consider the XY spin chain in a transverse magnetic field~\cite{lieb_two_1961, katsura_statistical_1962, pfeuty_onedimensional_1970} with periodic boundary conditions (PBC), given by the spin-$1/2$ Hamiltonian
\begin{equation}
\label{eq:HXY}
    H_{\rm XY}(\gamma,h) = - \frac{J}{2} \sum_{j=1}^L \bigg[ \frac{1+\gamma}{2}\sigma_j^x \sigma_{j+1}^x +\frac{1-\gamma}{2}\sigma_j^y \sigma_{j+1}^y + h \sigma_j^z   \bigg] \ .
\end{equation}
Here $\sigma^{\delta}_j$ denotes a Pauli matrix at site $j$, $\gamma \in [-1,1]$ the anisotropy parameter and we restrict our analysis to the regime $J > 0$, $h>0$. $H_{\rm XY}$ features a $\mathbb Z_2$ symmetry of rotations by $\pi$ around the $\hat z$ axis, i.e.~$\sigma_j^\delta \to - \sigma_j^\delta$ for $\delta = x,y$. For the special choices $\gamma = 0$ and $\gamma = 1$ one obtains, respectively, the XX chain in a transverse field (where $\mathbb Z_2$ is promoted to $\mathrm{U}(1)$) and the transverse field Ising chain (TFIC). 
The $T = 0$ phase diagram of the model, represented in \cref{fig:xy_phase_diagram}, hosts 2 different classes of quantum critical lines~\cite{dutta_quantum_2015}:
\begin{enumerate}[label=(\roman*)]
    \item Each point on the line $(\gamma \neq 0, h =1)$ is a quantum critical point in the Ising universality class, separating the ferromagnetic ordered phase of $h<1$ (order along $\hat x$ for $\gamma > 0$ and along $\hat y$ for $\gamma <0$) from the paramagnetic phase of $h>1$.
    \item The XX line $(\gamma = 0, h < 1)$ represents a gapless Luttinger liquid~\cite{mattis_exact_1965, haldane_luttinger_1981} phase with quasi-long-range order. Each point on the line is a quantum critical point  in the compact free boson universality class, separating the two ($\gamma > 0$ and $\gamma<0$) gapped ferromagnetic phases in (i).
\end{enumerate}

The spin-$1/2$ operators in \cref{eq:HXY} can be mapped to spinless fermions $c_j, c_j^\dag$ by the nonlocal Jordan-Wigner transformation~\cite{jordan_ber_1928}, which leaves $H_{\rm XY}$ as a sum of local terms in the fermions~\cite{lieb_two_1961}. Given the $\mathbb Z_2$ symmetry $[H_{\rm XY},e^{i \pi N}]=0$, where $N =\sum_{j=1}^L c_j^\dag c_j$, $H_{\rm XY}$ is block diagonal in the subspaces of the fermionic Fock space with even (e) and odd (o) numbers of fermions, i.e.~$H_{\rm XY} = H_{\rm e} \oplus H_{\rm o}$. The diagonalization of $H_{\rm e}$ and $H_{\rm o}$ by Fourier and Bogoliubov transformations is standard and leads to~\cite{lieb_two_1961, katsura_statistical_1962}
\begin{align} \label{eq:XYHdiag}
  H_{S}&=\sum_{k \in S} \varepsilon_{\gamma,h}(k) \, \alpha_k^\dag \alpha_k + E_{S}\, ,
  \qquad \qquad S = {\rm e, o} \, ,\\
  \label{eq:varepsXY}
  \varepsilon_{\gamma,h}(k) &= J \sqrt{\big[h-\cos(k)\big]^2+\gamma^2 \sin^2(k)} \ .
\end{align}
Here $\alpha_k$ and $\alpha_k^\dag$ (for which the dependence on $\gamma$ and $h$ is implicit) indicate fermionic Bogoliubov operators associated with the free momenta ($L$ even)
\begin{equation}
    k_n = \frac{2\pi(n + 1/2 \, \delta_{S,{\rm e}})}{L} \qquad n 
    = -\frac{L}{2}, \ldots, \frac{L}{2}-1  \, , \quad   S = {\rm e,o}  \, .
\end{equation}
The eigenstates of $H_{\rm XY}$ are generated by repeated actions of $\alpha^\dag_{k_j}$ on the Bogoliubov vacua $\ket{0;\gamma,h}_{S}$, defined by $\alpha_k\ket{0;\gamma,h}_S = 0 \ \  \forall \ k \in S$. For $h>1$ the ground state of $H_{\rm XY}$ coincides with $\ket{0;\gamma,h}_{\rm e}$, while for $h < 1$ the two $S=$ e, o vacua $\ket{0;\gamma,h}_{S}$ become degenerate ground states for $L \to \infty$ and for $\gamma\neq 0$ hybridize by spontaneous symmetry breaking to yield ferromagnetic order.

\subsection{Quench protocol and GGE root density}

We focus on the standard quench protocol in which the initial state is chosen to be the ground state $\ket{\psi(0)}=\ket{{\rm GS};\gamma_0,h_0}$ of the pre-quench Hamiltonian $H_{\rm XY}(\gamma_0,h_0)$, and we consider the stationary state locally reached at late times after evolution with $H_{\rm XY}(\gamma,h)$, where $(\gamma_0,h_0)\neq(\gamma,h)$. These initial states $\ket{\psi(0)}$ are a paradigmatic example of integrable initial states~\cite{piroli_what_2017}, for which the overlaps with the eigenstates of $H_{\rm XY}(\gamma,h)$ can be easily obtained~\cite{calabrese_quantum_2012}. For $h_0>1$, where $\ket{\psi(0)} = \ket{0;\gamma_0,h_0}_{\rm e}$, one finds
\begin{align}
\label{eq:GSexpansionXY}
    \ket{\psi(0)}&=\ket{0;\gamma_0,h_0}_{\rm e} = \prod_{k\in {\rm e}, \, k >0} \frac{1}{\sqrt{1+|\Lambda_k|^2}} (1+\Lambda_k \alpha_k^\dag \alpha_{-k}^\dag) \ket{0;\gamma,h}_{\rm e} \ , \\
    \Lambda_k &= i \frac{\big(h_0 \gamma - h \gamma_0+(\gamma_0-\gamma)\cos(k)\big)\sin(k)}{\varepsilon_{\gamma,h}(k)\varepsilon_{\gamma_0,h_0}(k)/J^2 + \cos^2(k) + \gamma \gamma_0 \sin^2(k)+h h_0 - (h+h_0)\cos(k)} \ ,
\end{align}
where $\alpha_k$ refer to the post-quench Hamiltonian $H_{\rm XY}(\gamma,h)$. 
From \cref{eq:GSexpansionXY} one readily obtains overlaps with the even eigenstates $ \ket{k_1,\ldots,k_{2m};\gamma,h}_{\rm e} = \alpha_{k_1}^\dag \ldots \alpha_{k_{2m}}^\dag \ket{0;\gamma,h}_{\rm e}$ of $H_{\rm XY}(\gamma,h)$, from which we see that only parity-invariant eigenstates in the $S = {\rm e}$ sector constructed by creation of pairs $(k,-k)$ on top of the vacuum $\ket{0;\gamma,h}_{\rm e}$ have nonzero overlap with $\ket{\psi(0)}$ (this being the characteristic property of integrable initial states~\cite{piroli_what_2017} in free models). 
In free models $\varrho_{\text{\tiny GGE}}(k)$ is easily determined by exploiting the fact that
\begin{align}
\label{eq:rhospXY}
    \varrho_{\text{\tiny GGE}}(k) &= \frac{1}{2\pi} \braket{\psi(0)|\alpha_k^\dag \alpha_k|\psi(0)} = \frac{1-\Omega_k}{4\pi} \ , \\
    \label{eq:rhospXY2}
    \Omega_k &= \frac{\cos^2(k) + \gamma \gamma_0 \sin^2(k)+h h_0 - (h+h_0)\cos(k)}{\varepsilon_{\gamma,h}(k)\varepsilon_{\gamma_0,h_0}(k)/J^2} \ .
\end{align}

For $h_0<1$, the physical ground state at $L \to \infty$ spontaneously breaks the $\mathbb Z_2$ symmetry and assumes the form
\begin{align}
\label{eq:psih0m1}
    \ket{\psi(0)}=\ket{{\rm GS};\gamma_0,h_0} = \frac{\ket{0;\gamma_0,h_0}_{\rm e}+ \ket{0;\gamma_0,h_0}_{\rm o}}{\sqrt{2}} \ .
\end{align}
Taking into account the impossibility of thermodynamic phase transitions at finite energy densities in 1D, we deduce that all local operators connecting the even and odd sectors have vanishing expectation value at infinite times after the quench from \eqref{eq:psih0m1}. 
Therefore, the macrostate $\varrho_{\text{\tiny GGE}}(k)$ associated with the stationary reduced density matrix reached at late times is given again by \cref{eq:rhospXY,eq:rhospXY2}, because both the even and odd sectors give identical diagonal contributions for $L \to \infty$.
Having access to $\varrho_{\text{\tiny GGE}}(k)$ from \cref{eq:rhospXY}, and to the thermal root density
\begin{equation}
\label{eq:thermRDxy}
    \varrho_{\beta^*}(k) = \frac{1}{2\pi}\frac{1}{e^{\varepsilon_{\gamma,h}(k)\beta^*}+1},
\end{equation}
at the inverse temperature $\beta^*$ fixed by \cref{eq:athermalitysimple}, one obtains the entropy densities $s_\text{\tiny YY}[\varrho_{\text{\tiny GGE}}]$ and $s_\text{\tiny YY}[\varrho_{\beta^*}]$ needed in \cref{eq:defAthDensity} for the density of athermality $\alpha$ as
\begin{equation}
\label{eq:YYxy}
    s_\text{\tiny YY}[\varrho] = -\int_{-\pi}^\pi \frac{dk}{2\pi}\big[n(k)\ln n(k)+\big(1-n(k)\big)\ln\big(1 -n(k)\big)\big]\, , 
    \qquad \ n(k) \equiv 2 \pi \varrho(k) \ .
\end{equation}

\begin{figure}[t!]
  \centering
  \begin{minipage}{0.328\textwidth}
    \centering
    \includegraphics[width=\linewidth]{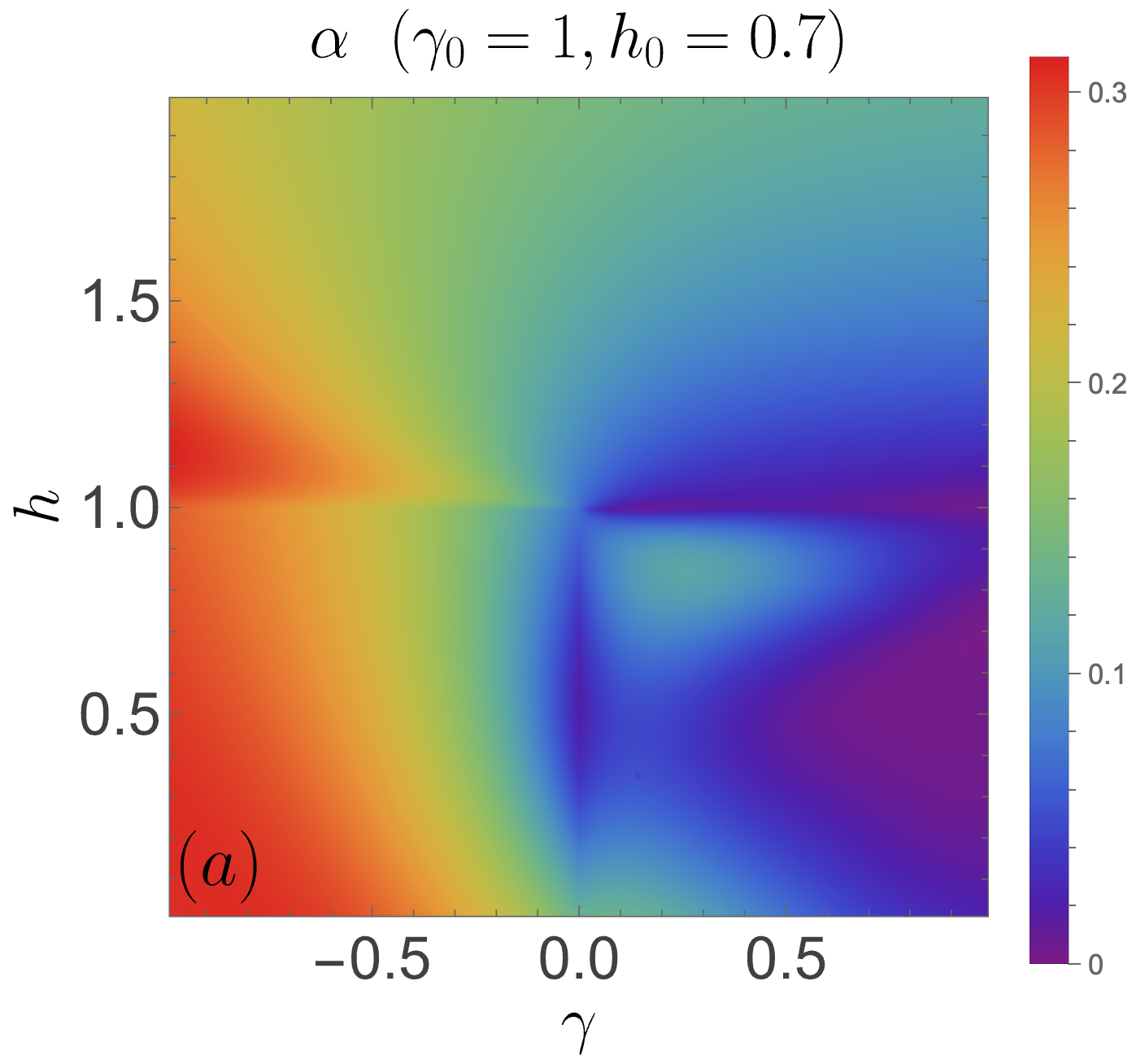}
  \end{minipage}
  \hfill 
  \begin{minipage}{0.328\textwidth}
    \centering
    \includegraphics[width=\linewidth]{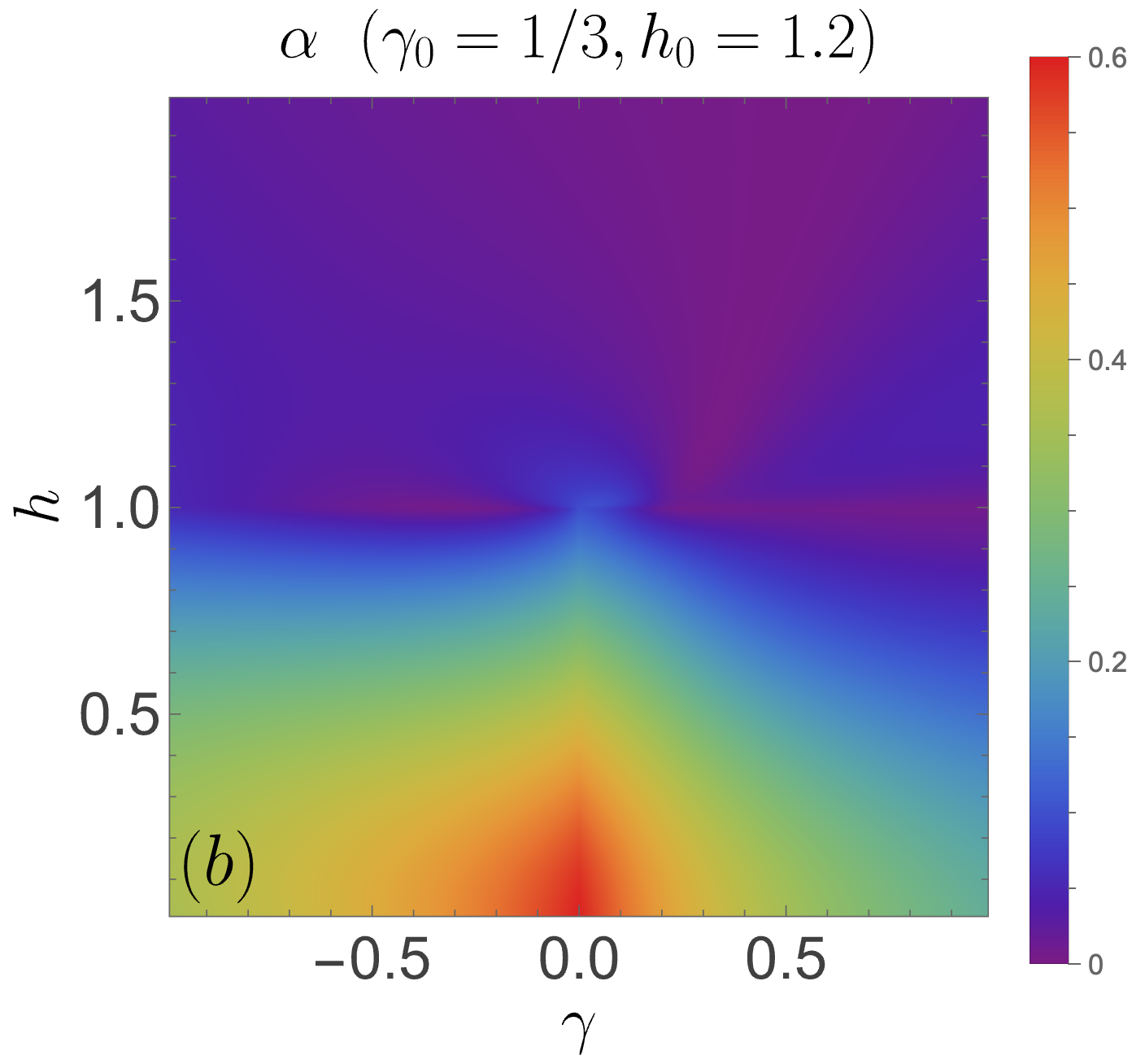}
  \end{minipage}
  \hfill 
  \begin{minipage}{0.328\textwidth}
    \centering
    \includegraphics[width=\linewidth]{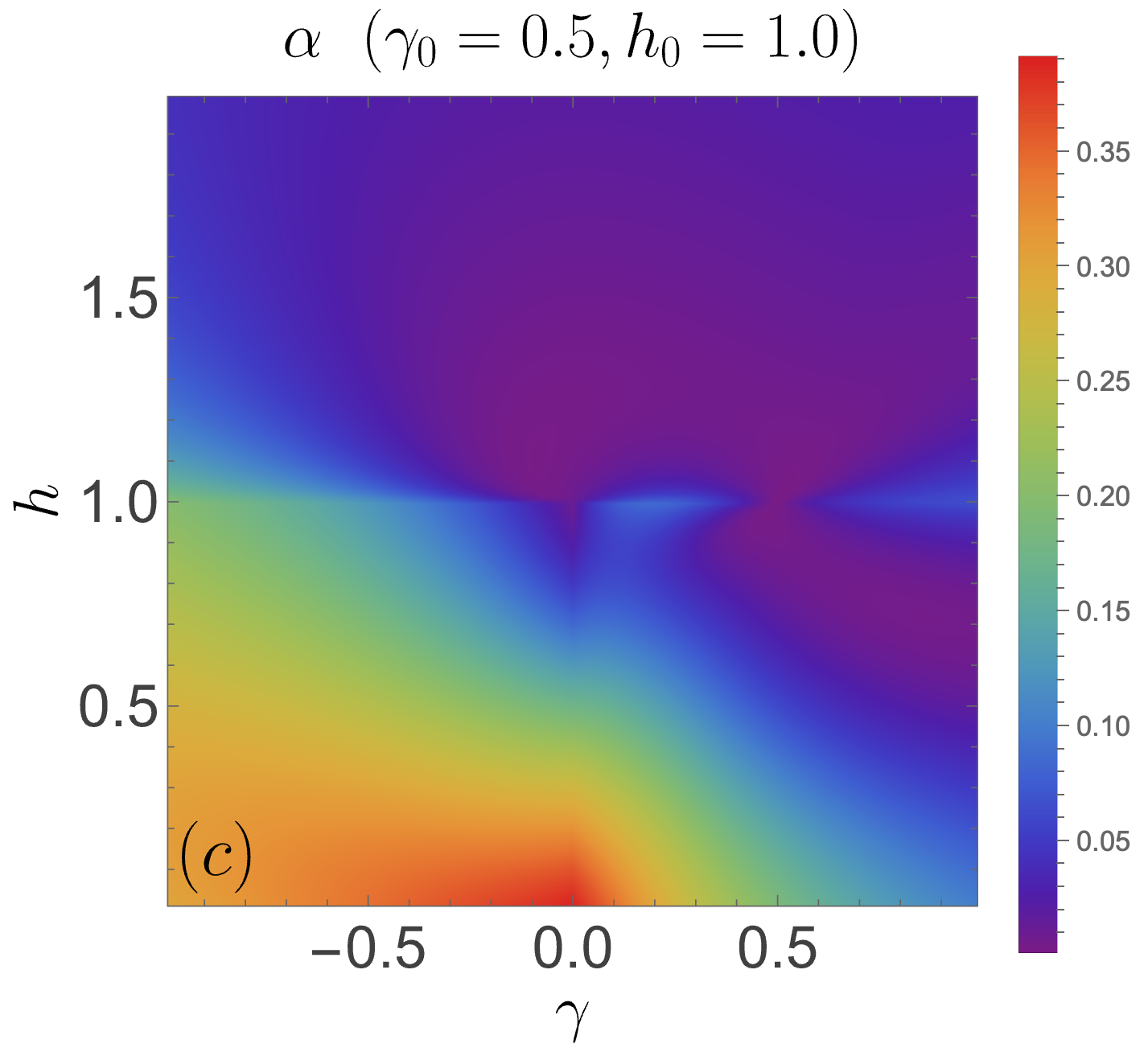}
  \end{minipage}
  \caption{Density plots for the density of athermality $\alpha$ in the plane $(\gamma,h)$, for fixed prequench parameters $(\gamma_0,h_0)$. (a) and (b) correspond to noncritical initial states, while (c) shows results for a critical initial state on the Ising line $h_0 =1$. See \cref{fig:XY_e_sGE_sGGE} in Appendix~\ref{appendix:XYchain} for corresponding plots for the energy density $e$ and the entropy densities $s_{\text{\tiny YY}}[\varrho_{\text{\tiny GGE}}]$ and $s_{\text{\tiny YY}}[\varrho_{\beta^*}]$.}
  \label{fig:main_XY_DP}
\end{figure}

\subsection{Varying the postquench Hamiltonian}
\label{sec:XYvarypostH}
Given the large number of available parameters ($\gamma,h,\gamma_0,h_0$), we start by considering the athermality as a function of the postquench $(\gamma,h)$ when the prequench ones $(\gamma_0,h_0)$ are fixed. 

\begin{figure}[b!]
  \centering
  \begin{minipage}{0.328\textwidth}
    \centering
    \includegraphics[width=\linewidth]{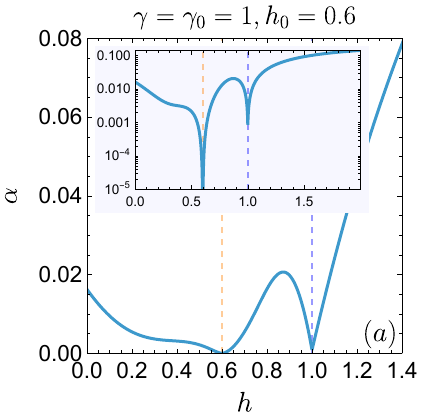}
  \end{minipage}
  \hfill 
  \begin{minipage}{0.328\textwidth}
    \centering
    \includegraphics[width=\linewidth]{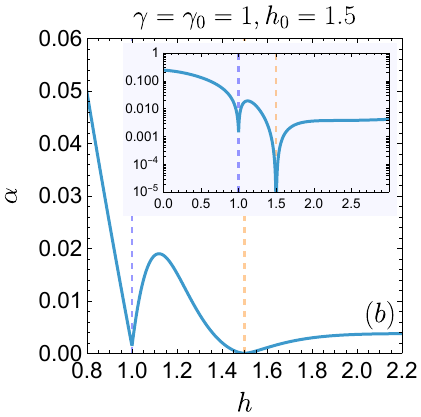}
  \end{minipage}
  \hfill 
  \begin{minipage}{0.328\textwidth}
    \centering
    \includegraphics[width=\linewidth]{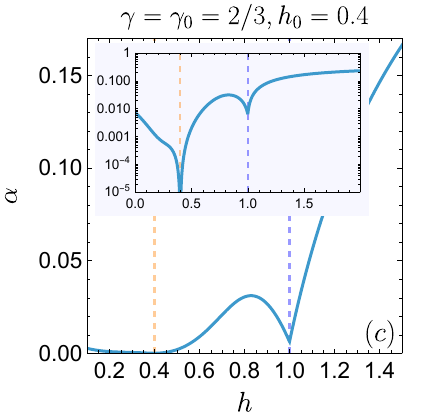}
  \end{minipage}
  \caption{Density of athermality $\alpha$ as a function of the postquench transverse field $h$, for fixed $\gamma=\gamma_0$ ($h$-driven quench) and $h_0$. The orange and blue dashed lines indicate, respectively, the trivial zero at $h =h_0$ and the critical Ising point $h = 1$. (a) and (b) correspond to quenches in the TFIC ($\gamma = \gamma_0 = 1$), while in (c) the anisotropy is present. Insets: same data in log-scale.}
  \label{fig:XY_cuts_h}
\end{figure}

\begin{figure}[t!]
  \centering
  \begin{minipage}{0.328\textwidth}
    \centering
    \includegraphics[width=\linewidth]{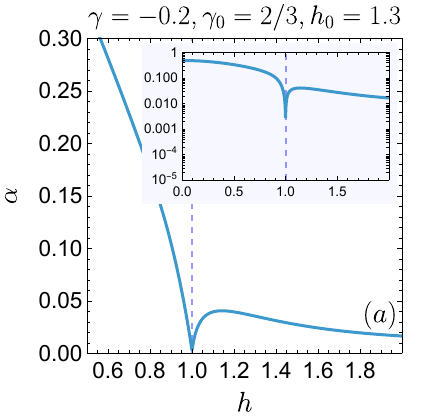}
  \end{minipage}
  \hfill 
  \begin{minipage}{0.328\textwidth}
    \centering
    \includegraphics[width=\linewidth]{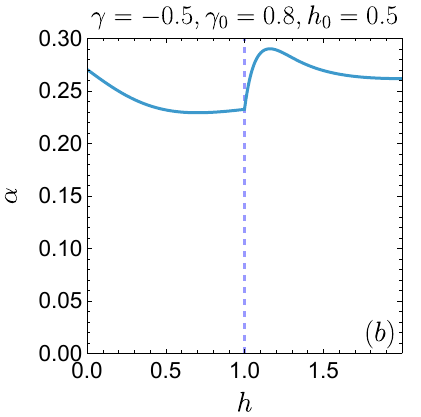}
  \end{minipage}
  \hfill 
  \begin{minipage}{0.328\textwidth}
    \centering
    \includegraphics[width=\linewidth]{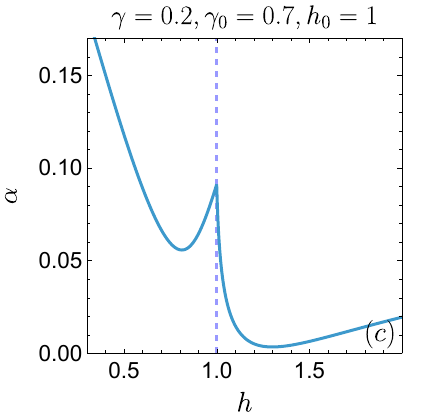}
  \end{minipage}
  \caption{
  $\alpha$ as a function of $h$, for fixed $\gamma\neq\gamma_0$ ($h$- and $\gamma$-driven quench) and $h_0$. The blue dashed lines indicates the critical Ising point $h = 1$. The inset in (a) shows the same data in log-scale.}
  \label{fig:XY_cuts_h_2}
\end{figure}

In \cref{fig:main_XY_DP} we show a few density plots for the density of athermality $\alpha$ in the $(\gamma,h)$ plane. The most important feature, already anticipated in the Introduction, is that from these plots one is able to clearly identify the critical lines of the $T = 0$ quantum phase diagram (see \cref{fig:xy_phase_diagram}), despite the fact that $\alpha$ probes only physics at a finite energy density above the ground state. They emerge as \emph{non-analytic} features of $\alpha$ at the $T=0$ critical lines, as evident from \cref{fig:XY_cuts_h,fig:XY_cuts_h_2,fig:XY_cuts_gamma} where we plot some cuts in the $(\gamma,h)$ plane. In addition to these singular points, all subplots in \cref{fig:main_XY_DP} feature trivial zeros due to the absence of the quench for $(\gamma,h)=(\gamma_0,h_0)$. 

The singular behaviour of $\alpha$ comes from the GGE entropy density $s_{\text{\tiny YY}}[\varrho_{\text{\tiny GGE}}]$, see \cref{sec:nonanbh} below for further details and Appendix~\ref{appendix:XYchainProof} for a proof. The thermal entropy density $s_{\text{\tiny YY}}[\varrho_{\beta^*}]$ is instead a smooth function of $(\gamma,h)$ as a consequence of the absence of phase transitions in 1D system with short-range interactions at finite temperatures (see, e.g., Ref.~\cite{araki_gibbs_1969}). In the context of the XY chain, the smoothness of $s_{\text{\tiny YY}}[\varrho_{\beta^*}]$ with respect to $\gamma$ and $h$ is easily proved by means of Leibniz's integral rule by noticing that the integrand in \cref{eq:YYxy} is an even function of $\varepsilon_{\gamma,h}(k)$ (hence its Taylor expansion involves only the analytic function $\varepsilon_{\gamma,h}^2(k)$) and that $\beta^*=\beta^*(\gamma,h)$ is also a smooth function (this comes from the fact that the energy density $e$ injected by the quench is trivially smooth).
Apart from the singular points, a natural question is whether the qualitative behaviour of $\alpha$ in the $(\gamma,h)$ plane merely reflects analogous behaviour of the energy density $e$ injected by the quench, which we show in \cref{fig:XY_e_sGE_sGGE} of Appendix~\ref{appendix:XYchain}. We find that $\alpha$ possesses much more structure than $e$, and that there is no strong correlation between the behaviour of $e$ and $\alpha$.

\subsubsection{$h$ dependence}
\cref{fig:XY_cuts_h} shows cuts where $\gamma = \gamma_0$ and hence the quench is entirely driven by $h_0 \to h$. \cref{fig:XY_cuts_h}(a) and (b) correspond to $h$-quenches in the TFIC ($\gamma = \gamma_0=1$). We see that, in addition to the trivial zeros at $h_0 = h$, the plots for $\alpha$ present a very pronounced local minimum at criticality $h = 1$, where a cusp marks non-analytic behaviour and $\alpha$ is anomalously close (but not equal) to zero. \cref{fig:XY_cuts_h}(c) shows that analogous behaviour emerges in $h$-quenches even away from the TFIC limit ($\gamma_0=\gamma<1$). In the density plots of \cref{fig:main_XY_DP}(a) and (b) these dips at criticality coincide with the narrow purple regions around the critical line $(\gamma>0,h=1)$. 

\cref{fig:XY_cuts_h_2} considers again the dependence of $\alpha$ on $h$, but now in quenches where $\gamma_0 \neq \gamma$ in addition to $h_0 \neq h$ (2-parameter quenches). While for $\gamma_0,\gamma>0$ one again generally finds a dip at criticality $h = 1$ (cf.~\cref{fig:main_XY_DP}(a) and (b)), the behaviour is richer when one allows large jumps $0<\gamma_0 \to 0>\gamma$. In such cases one can find again a dip at criticality $h = 1$, as shown in \cref{fig:XY_cuts_h_2}(a) (cf.~\cref{fig:main_XY_DP}(b)), but it is also possible to find situations where $h = 1$ features a non-analytic point that does not represent a pronounced local minimum, as shown in \cref{fig:XY_cuts_h_2}(b) (cf.~\cref{fig:main_XY_DP}(a)). 

In \cref{fig:main_XY_DP}(c) and \cref{fig:XY_cuts_h_2}(c) we show, respectively, a density plot and a cut for quenches from the critical point $h_0=1$. In this case the situation is yet again different, with the critical point $h = 1$ exhibiting a local maximum (instead of a minimum) emerging as a non-analytic cusp.

We stress that while the singular nature of $\alpha$ at the critical points is due solely to $s_{\text{\tiny YY}}[\varrho_{\text{\tiny GGE}}]$, the behaviour as a pronounced local minimum or maximum depends in general on the interplay between $s_{\text{\tiny YY}}[\varrho_{\beta^*}]$ and $s_{\text{\tiny YY}}[\varrho_{\text{\tiny GGE}}]$ (cf.~density plots in Appendix~\ref{appendix:XYchain}), and it is therefore in general only a property of the athermality.

\subsubsection{$\gamma$ dependence}
\cref{fig:XY_cuts_gamma} shows cuts for the dependence of $\alpha$ on $\gamma$, with $h_0$, $h$ and $\gamma_0$ fixed. Based on the features of  \cref{fig:main_XY_DP}(a) around the XX critical line we expect situations where the critical point $\gamma = 0$ corresponds to a local-minimum cusp. This is shown in the cut of \cref{fig:XY_cuts_gamma}(a), for choices of $(\gamma_0,h_0)$ similar to those of \cref{fig:main_XY_DP}(a). On the other hand, \cref{fig:XY_cuts_gamma}(b) shows a cut for parameters choices similar to those of \cref{fig:main_XY_DP}(b), and indeed we find a local-maximum cusp, as expected from the density plot. The density plot of \cref{fig:main_XY_DP}(c), corresponding to a critical initial state $(\gamma_0,h_0=1)$ features both regions where the critical point $\gamma = 0$ corresponds to a minimum cusp ($h$ close to 1), and regions where $\gamma = 0$ corresponds to a maximum cusp ($h$ close to $0$), which we have also verified by appropriate cuts. As for the case of the Ising critical line, also here the non-analytic behaviours are entirely due to $s_{\text{\tiny YY}}[\varrho_{\text{\tiny GGE}}]$, while the local-minimum (local-maximum) nature of the critical points arises in general from the nontrivial interplay between $s_{\text{\tiny YY}}[\varrho_{\text{\tiny GGE}}]$ and $s_{\text{\tiny YY}}[\varrho_{\beta^*}]$ (cf.~density plots in Appendix~\ref{appendix:XYchain}). 

\subsubsection{Non-analytic behaviour from root densities and pinning mechanism}
\label{sec:nonanbh}
The non-analytic behaviour of $s_{\text{\tiny YY}}[\varrho_{\text{\tiny GGE}}]$ at the critical lines (cf.~density plots in Appendix~\ref{appendix:XYchain}) is inherited from the singular points in the $\gamma$ and $h$ dependence of $\varrho_{\text{\tiny GGE}}(k)$  (\cref{eq:rhospXY,eq:rhospXY2}), which in turn can be traced back to the dispersion $\varepsilon_{\gamma,h}(k)$ becoming gapless at criticality. For example, focusing directly on the density of occupation numbers $n_{\text{\tiny GGE}}(k)= 2 \pi \varrho_{\text{\tiny GGE}}(k)$, it is easy to see that
\begin{equation}
\lim_{k \to 0}n_{\text{\tiny GGE}}(k) = \frac{1 - {\rm sgn}\big[(h-1)(h_0-1)\big]}{2} \qquad \quad \text{for } h, h_0 \neq 1 \ ,
\end{equation}
where ${\rm sgn}(x)$ is the sign function. Therefore $n_{\text{\tiny GGE}}(k)$ is subject to a discontinuous jump from $0$ to $1$ whenever $h$ or $h_0$ cross the critical points $h=1$ or $h_0=1$. Exactly on the critical Ising line we find 
\begin{equation}
\label{eq:pinnsp}
    \lim_{k \to 0}\Big[\lim_{h \to 1}n_{\text{\tiny GGE}}(k)\Big] = \frac{1}{2} \qquad \quad \text{for } \ h_0 \neq 1, \gamma \neq 0 \ ,
\end{equation}
and analogously for $h_0 = 1$, $h\neq 1, \gamma_0 \neq 0$ by the symmetry $(\gamma,h)\leftrightarrow(\gamma_0,h_0)$ of $n_{\text{\tiny GGE}}(k)$. 
Similar discontinuities are found along the XX line when $\gamma$ or $\gamma_0$ cross zero, at the values of $k^*$ for which $\varepsilon_{0,h}(k^*)=0$. Examples of the discontinuous jumps across the Ising line are given in \cref{fig:root_densities_1} (see \cref{fig:root_densities_2} for additional plots). In Appendix~\ref{appendix:XYchainProof} we prove analytically that the existence of singular points for $n_{\text{\tiny GGE}}(k)$ translates into non-analytic features at criticality of $s_{\text{\tiny YY}}[\varrho_{\text{\tiny GGE}}]$. 

\begin{figure}[t!]
  \centering
  \begin{minipage}{0.328\textwidth}
    \centering
    \includegraphics[width=\linewidth]{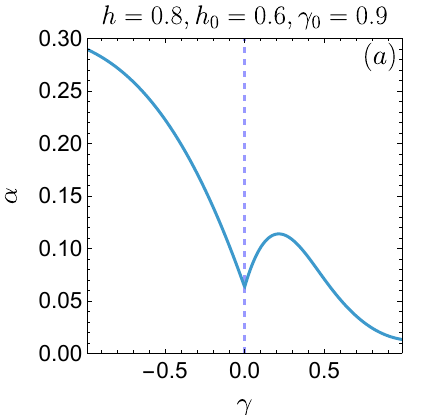}
  \end{minipage}
  \begin{minipage}{0.328\textwidth}
    \centering
    \includegraphics[width=\linewidth]{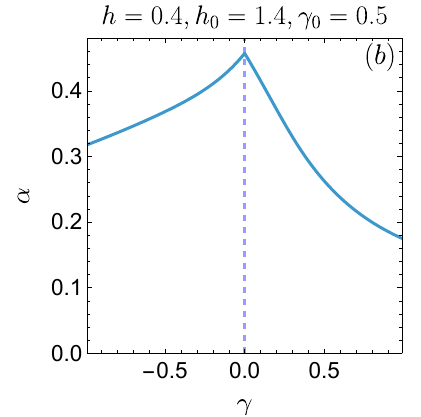}
  \end{minipage}
  \caption{ $\alpha$ as a function of $\gamma$, for fixed $h\neq h_0$ ($h$- and $\gamma$-driven quench) and $\gamma_0$. The blue dashed lines indicates the critical XX point $\gamma = 0$.}
  \label{fig:XY_cuts_gamma}
\end{figure}

\begin{figure}[!b]
    \centering
    \includegraphics[width=1.0\linewidth]{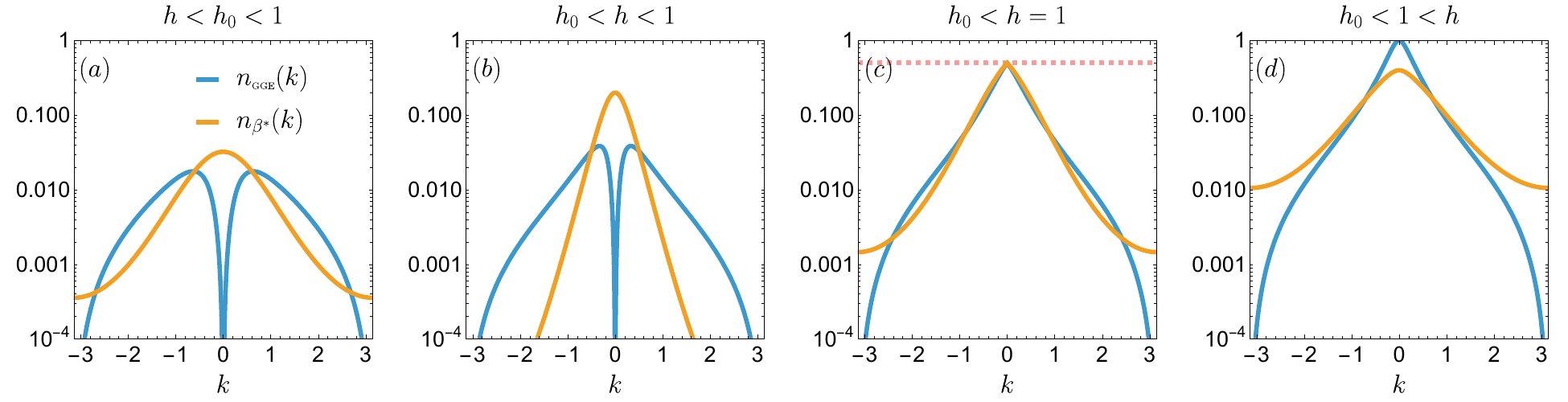}
    \caption{Log-scale comparison between the occupation functions $n_{\text{\tiny GGE}}(k)$ and $n_{\beta^*}(k)$, for different choices of $h$ in the TFIC ($\gamma = \gamma_0=1$) with $h_0=0.6$: (a) $h = 0.4$; (b) $h = 0.8$; (c) $h = 1$ (critical), with red dashed line indicating the pinning value of $1/2$; (d) $h = 1.2$.}
    \label{fig:root_densities_1}
\end{figure}

Some of the most striking features arising from the previous athermality plots are the dips at criticality for which $\alpha \approx 0$, see \cref{fig:XY_cuts_h} and \cref{fig:XY_cuts_h_2}(a). In the case of the Ising line, they emerge due to a pinning mechanism for which both $n_{\text{\tiny GGE}}(k)$ (see \cref{eq:pinnsp}) and $n_{\beta^*}(k)=2 \pi \varrho_{\beta^*}(k)$ (see \cref{eq:thermRDxy} and remember $\varepsilon_{\gamma,1}(0)=0$) are pinned to the value of $1/2$ for $h \to 1$ and $k \to 0$, while also featuring very similar decays away from $k = 0$. This forces the curves $n_{\text{\tiny GGE}}(k)$ and $n_{\beta^*}(k)$ to be anomalously close to each other, drastically suppressing $\alpha$. An example of the Ising-line pinning is shown in log-scale in \cref{fig:root_densities_1}(c). Similar (quasi)pinning mechanisms are also responsible for the dips along the XX line, like the one in \cref{fig:XY_cuts_gamma}(a).

\subsection{Varying the prequench Hamiltonian}
\label{sec:XYvaryingH}

\begin{figure}[b!]
  \centering
  \begin{minipage}{0.328\textwidth}
    \centering
    \includegraphics[width=\linewidth]{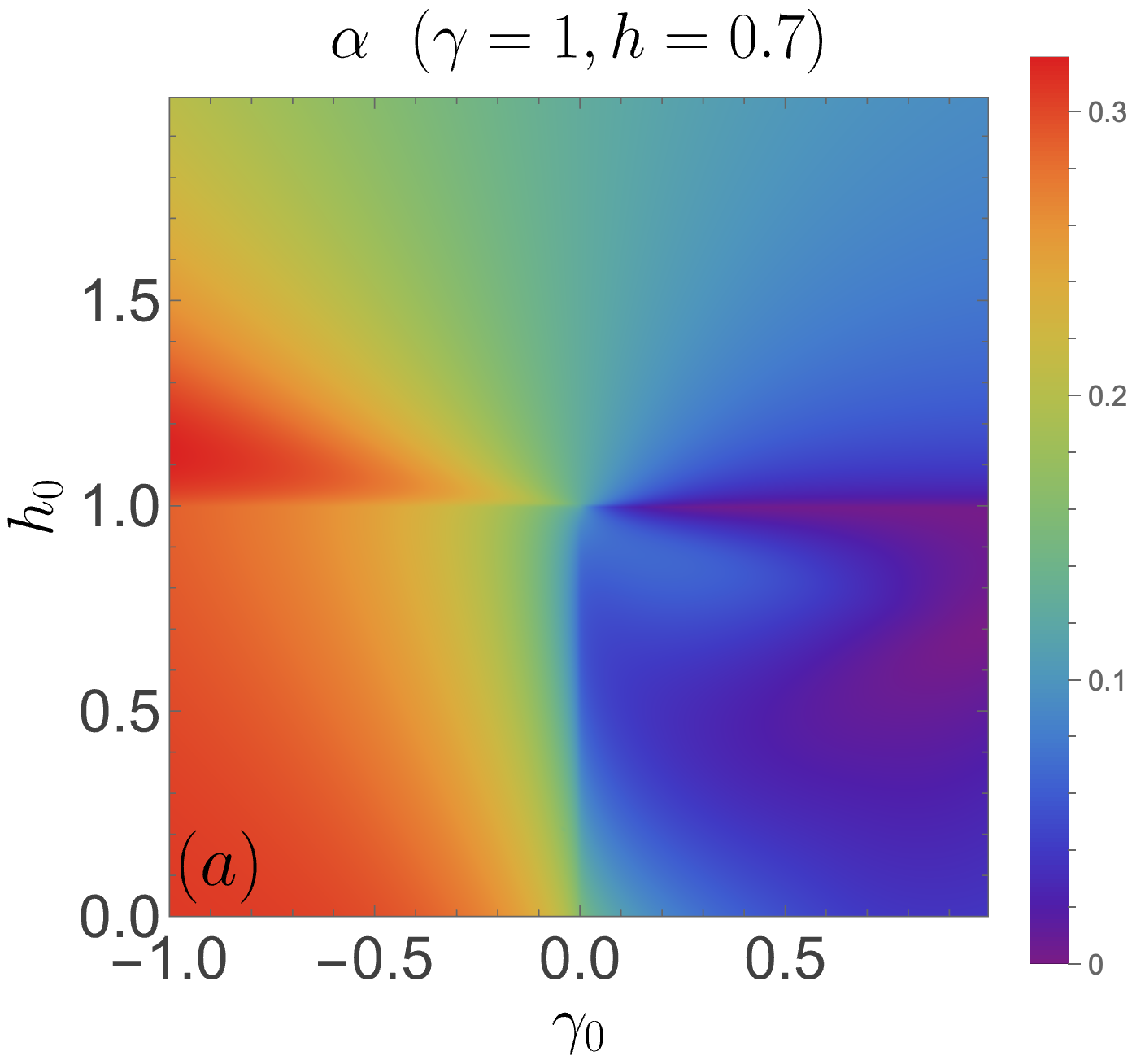}
  \end{minipage}
  \hfill 
  \begin{minipage}{0.328\textwidth}
    \centering
    \includegraphics[width=\linewidth]{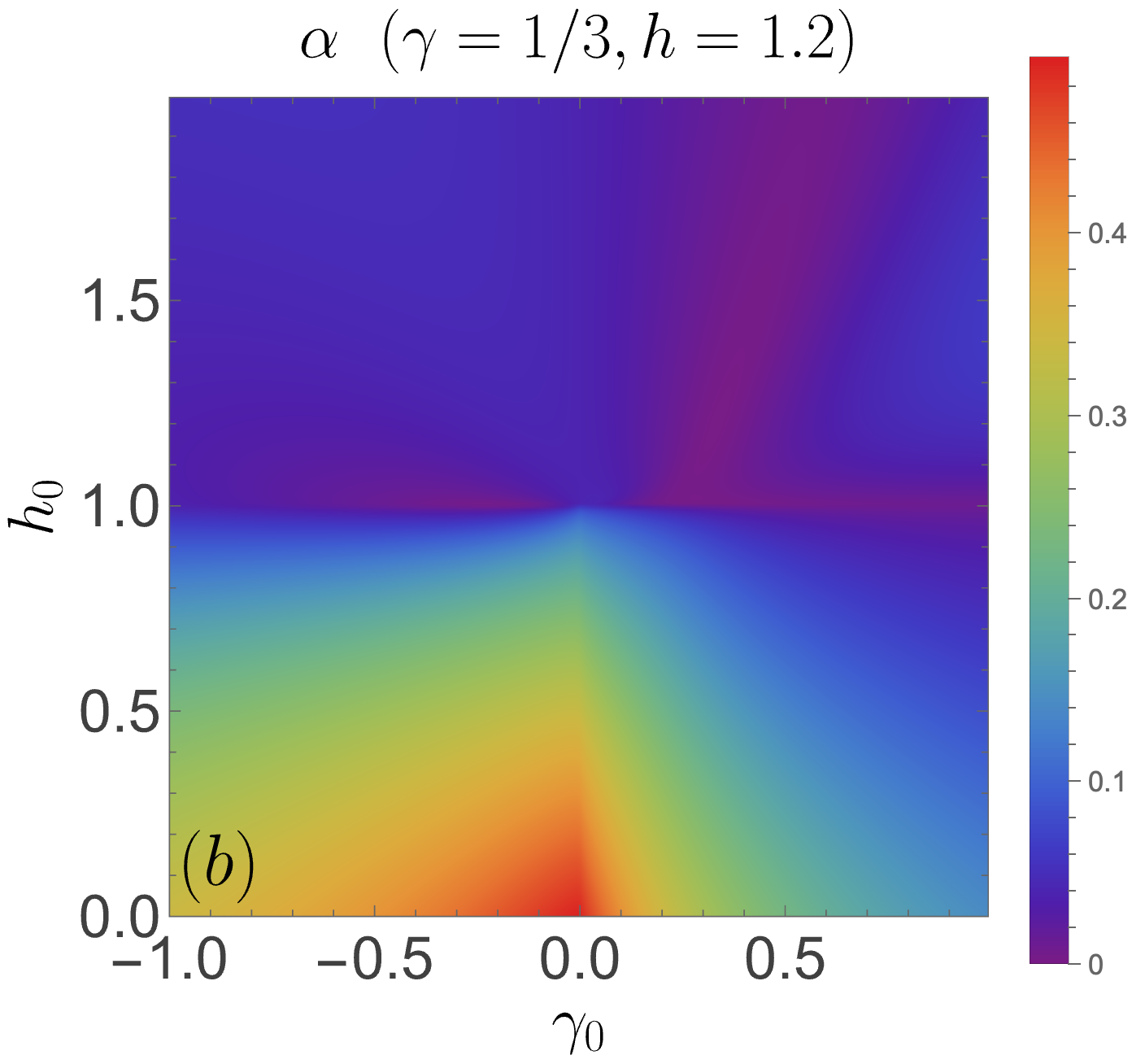}
  \end{minipage}
  \hfill 
  \begin{minipage}{0.328\textwidth}
    \centering
    \includegraphics[width=\linewidth]{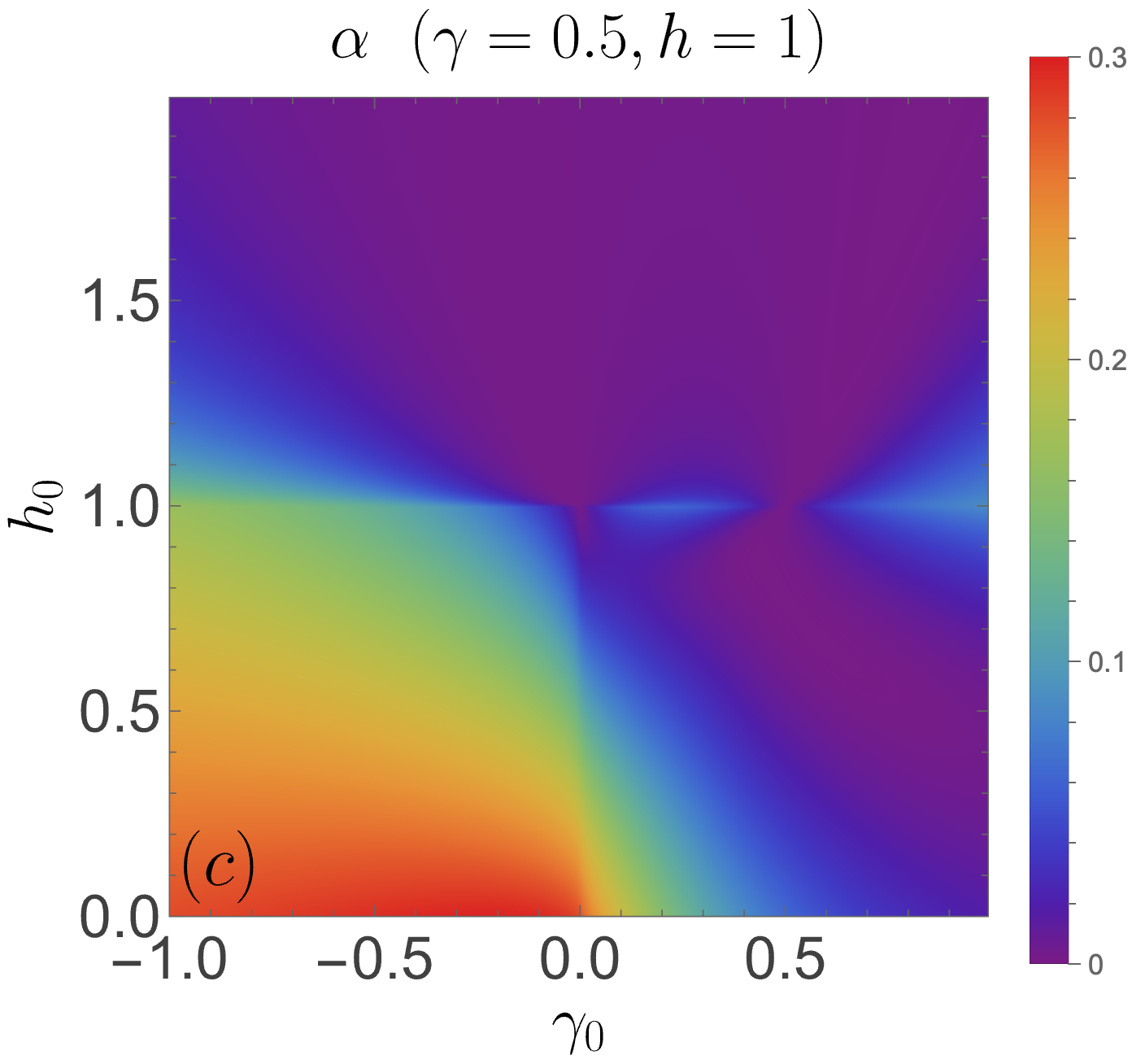}
  \end{minipage}
  \caption{Density plots for $\alpha$ in the plane $(\gamma_0,h_0)$, for fixed postquench parameters $(\gamma,h)$. The subplots (a), (b) and (c) correspond to the (a), (b) and (c) density subplots of \cref{fig:main_XY_DP} under the swap $(\gamma,h)\leftrightarrow(\gamma_0,h_0)$. In particular, current subplot (c) corresponds to a quench to a critical Hamiltonian on the Ising line.}
  \label{fig:main_XY_DP_2}
\end{figure}

We now investigate the structure of $\alpha$ as a function of the \emph{prequench} parameters $(\gamma_0,h_0)$, for fixed postquench $(\gamma,h)$. In this setup it is perhaps more natural to expect signatures of quantum criticality in the plots for $\alpha$, given that we vary the parameters $\gamma_0$ and $h_0$ that define the $T = 0$ ground state. 

Due to the invariance of $\varrho_{\text{\tiny GGE}}(k)$ under the swap $(\gamma_0,h_0) \leftrightarrow (\gamma,h)$, for any density plot or cut of $\alpha$ presented in \cref{sec:XYvarypostH} as a function of $(\gamma,h)$ (for fixed $\gamma_0$ and $h_0$), the contribution $-s_\text{\tiny YY}[\varrho_{\text{\tiny GGE}}]$ to $\alpha$ remains identical if we consider now the athermality as a function of $(\gamma_0,h_0)$ (for fixed $(\gamma,h)$ chosen to be identical to the $(\gamma_0,h_0)$ used in the figures of~\cref{sec:XYvarypostH}). However, the overall $\alpha$ changes with respect to the plots of \cref{sec:XYvarypostH} because of the now different contribution $s_\text{\tiny YY}[\varrho_{\beta^*}]$, where unlike before $\beta^*$ is determined by the average energy of the \emph{fixed} Hamiltonian $H_{\rm XY}(\gamma,h)$ in the \emph{varying} initial state $\ket{{\rm GS};\gamma_0,h_0}$. Indeed, while in \cref{sec:XYvarypostH} the postquench energy density $e$, and by consequence $s_\text{\tiny YY}[\varrho_{\beta^*}]$, did not feature singular points as we varied $\gamma$ and $h$, here $e$ is singular at the critical lines as we vary $\gamma_0$ and $h_0$. For fixed postquench $\gamma$ and $h$ we have 
\begin{align}
    e(\gamma_0,h_0) &= \braket{{\rm GS}; \gamma_0,h_0|H_{\rm XY}(\gamma,h)|{\rm GS}; \gamma_0,h_0} \ , \\
    H_{\rm XY}(\gamma,h)& = H_{\rm XY}(\gamma_0,h_0) + (\gamma-\gamma_0)\frac{\partial H_{\rm XY}(\gamma_0,h_0)}{\partial \gamma_0} + (h-h_0)\frac{\partial H_{\rm XY}(\gamma_0,h_0)}{\partial h_0} \ ,
\end{align}
and by application of the Hellmann-Feynman theorem we obtain
\begin{equation}
\label{eq:eGSHF}
    e(\gamma_0,h_0) = e_{\text{\tiny GS}}(\gamma_0,h_0) + (\gamma-\gamma_0)\frac{\partial e_{\text{\tiny GS}}(\gamma_0,h_0)}{\partial \gamma_0} + (h-h_0)\frac{\partial e_{\text{\tiny GS}}(\gamma_0,h_0)}{\partial h_0} \ ,
\end{equation}
where $e_{\text{\tiny GS}}(\gamma_0,h_0)$ is the ground state energy of $H_{\rm XY}(\gamma_0,h_0)$. By definition of quantum phase transition $e_{\text{\tiny GS}}(\gamma_0,h_0)$ is singular at the critical points (in the XY chain logarithmic divergences appear in the second-order derivatives of $e_{\text{\tiny GS}}$), therefore \cref{eq:eGSHF} proves that the postquench energy density $e(\gamma_0,h_0)$, and by consequence $s_\text{\tiny YY}[\varrho_{\beta^*}]$ via $\beta^*$, is singular too (see Appendix~\ref{appendix:XYchain} for some plots). 

In \cref{fig:main_XY_DP_2} we show density plots analogous to those of \cref{fig:main_XY_DP} but swapping $(\gamma_0,h_0)\leftrightarrow(\gamma,h)$, i.e.~we now vary $(\gamma_0,h_0)$ and fix $(\gamma,h)$ equal to the values $(\gamma_0,h_0)$ from \cref{fig:main_XY_DP}. By visual comparison we see that the structure of $\alpha$ is very similar to that of \cref{fig:main_XY_DP}, in particular with the critical lines emerging as non-analytic features. However, as discussed, the results are not expected to be equal and indeed one can easily identify differences, e.g.~the absence of a local-minimum cusp (which we verified also by means of cuts) on the XX line of \cref{fig:main_XY_DP_2}(a) as compared to \cref{fig:main_XY_DP}(a) (cf.~also \cref{fig:XY_cuts_gamma}(a)). Density plots and cuts for $e$, $s_{\text{\tiny YY}}[\varrho_{\beta^*}]$ and $s_{\text{\tiny YY}}[\varrho_{\text{\tiny GGE}}]$ corresponding to the parameter choice of \cref{fig:main_XY_DP_2}(a) are shown in \cref{fig:XY_e_sGE_sGGE_2} of Appendix~\ref{appendix:XYchain}. Similarly to before, we find that in $\alpha$ some interesting structure (see e.g. \cref{fig:main_XY_DP_2}(a) for $\gamma_0>0$ and $h_0<1$) emerges from the interplay (i.e.~the difference) between $s_{\text{\tiny YY}}[\varrho_{\beta^*}]$ and $s_{\text{\tiny YY}}[\varrho_{\text{\tiny GGE}}]$ (cf.~same parameter region in \cref{fig:XY_e_sGE_sGGE_2}, where the behaviour appears qualitatively different).

\section{XXZ chain}
\label{sec:XXZsec}
The spin-1/2 XXZ chain is a paradigmatic interacting integrable model solvable by Bethe ansatz~\cite{bethe_zur_1931, orbach_linear_1958, yang_onedimensional_1966, yang_onedimensional_1966a, yang_onedimensional_1966b, takahashi_thermodynamics_1999, korepin_quantum_1993}. In the absence of external magnetic fields its Hamiltonian reads
\begin{equation}
\label{eq:HXXZ}
    H_{\rm XXZ}(\Delta) =  \frac{J}{4} \sum_{j=1}^L \bigg[\sigma_j^x \sigma_{j+1}^x +\sigma_j^y \sigma_{j+1}^y + \Delta \Big( \sigma_j^z \sigma_{j+1}^z-1\Big)   \bigg] \ ,
\end{equation}
where we assume PBC and focus on antiferromagnetic couplings $J=1$, $\Delta > 0$. The quantum phase diagram for anisotropies $\Delta > 0$ features two phases separated by a quantum critical point at $\Delta =1$: 
\begin{enumerate}
    \item $\Delta > 1$ is a gapped phase where the GS spontaneously breaks the $\mathbb Z_2$ symmetry $\sigma^z \to -\sigma^z$, giving rise to antiferromagnetic order along the $\hat z$ axis.
    \item $0<\Delta < 1$ is a gapless Luttinger liquid with quasi long-range order.
\end{enumerate}
The Hamiltonian \eqref{eq:HXXZ} possesses a $\mathrm{U}(1)$ symmetry associated with conservation of the magnetization $\sigma_{\rm tot}^z = \sum_{j=1}^L \sigma_j^z$. The eigenvalues of $\sigma_{\rm tot}^z$ are $L - 2 N$, where $N$ denotes the number of spins flips relative to the fully polarized state with all spins up. Below we focus only on quenches from initial states $\ket{\psi(0)}$ that have zero magnetization $(N = L/2)$ and therefore \emph{no} additional chemical potential $\mu$ (i.e.~external magnetic field) must be included in the GE ($\mu=0$ automatically in \cref{eq:grandGibbsRDM}).

The spectrum of the XXZ chain features bound states~\cite{takahashi_thermodynamics_1999}, i.e.~solutions $\bs \lambda$ of the Bethe equations forming regular string patterns in the complex plane (see Appendix~\ref{appendix:XXZ}). Therefore each macrostate is identified by a set of root densities $\bs \varrho(\lambda)=\{\varrho_\ell(\lambda)\}_{\ell=1}^{\ell_{\rm max}}$, which specify the distribution of the real string centers for bound states composed of $n_\ell$ particles (i.e.~$n_\ell$ spin flips relative to the fully polarized state). In Appendix~\ref{appendix:XXZ} we report the TBA equations determining the thermal macrostates $\bs\varrho_{\beta^*}(\lambda)$ at inverse temperature $\beta^*$~\cite{takahashi_thermodynamics_1999}. For our purposes, the most important facts about the TBA description of the XXZ chain are the following:
\begin{itemize}
\item In the regime $\Delta > 1$ there are an infinite number ($\ell_{\rm max}=\infty$) of distinct string types  and $n_\ell=\ell$. The string centers are restricted to the real interval $\lambda \in (-\pi/2,\pi/2]$. 
\item In the regime $0<\Delta<1$ we restrict our attention to values of the anisotropy parametrized by 
\begin{equation}
\label{eq:Deltam1}
    \Delta = \cos\left(\frac{\pi}{p + 1}\right) \quad \text{with} \quad p \in \mathbb{Z}^+ \  .
\end{equation}
This subclass of the roots of unity points is characterized by a relatively simple TBA description~\cite{takahashi_thermodynamics_1999} (due to a significantly enriched symmetry algebra~\cite{deguchi_sl2_2001, korff_auxiliary_2003, korff_twisted_2004}), in which only $\ell_{\rm max}=p+1$ distinct string types are allowed. The number of particles in each string fulfils $n_\ell = \ell (1-\delta_{\ell,p+1}) + \delta_{\ell,p+1}$. The cases of $\ell = 1, p+1$ (simple unbounded magnons) are distinguished by their parities $v_1 = 1$ and $v_{p+1} = -1$ (see Appendix~\ref{appendix:XXZ}), which are associated respectively with a real and complex rapidity. For for $\ell \neq 1, p+1$ we have instead $v_\ell=1$. The string centers live on the full real line $\lambda \in (-\infty, \infty)$. 
\end{itemize}

Given a macrostate specified by the root densities $\varrho_\ell(\lambda)$ for $\ell = 1, 2, \ldots, \ell_{\rm max}$, the Yang-Yang entropy density is given by
\begin{equation}
\label{eq:YYentdensXXZ}
    s_{\text{\tiny YY}}[\bs \varrho]=\sum_{\ell = 1}^{\ell_{\rm max}}\int_{-A}^{A}d\lambda \, \bigg[\varrho_\ell(\lambda)\ln \bigg(1+ \frac{\varrho_{\ell}^{(h)}(\lambda)} {\varrho_\ell(\lambda)} \bigg)+\varrho_\ell^{(h)}(\lambda)\ln \bigg(1+ \frac{\varrho_\ell(\lambda)}{\varrho_\ell^{(h)}(\lambda)} \bigg)\bigg] \ .
\end{equation}
Here $\varrho^{(h)}_\ell(\lambda)$ denotes the density of hole centers associated with $\varrho_\ell(\lambda)$ (see Appendix~\ref{appendix:XXZ}), and $A = \pi/2$ ($A=\infty$) for $\Delta>1$ ($0<\Delta<1$ fulfilling \cref{eq:Deltam1}).

\subsection{Initial states and GGE root densities}

We focus on quenches from the class of tilted-Néel product states
\begin{equation}
\label{eq:tiltedNeel}
    \ket{\psi(0)}=\ket{\theta; \nearrow \swarrow}= \left[ \left( \cos \frac{\theta}{2} \ket{\uparrow} + i \sin \frac{\theta}{2} \ket{\downarrow}\right) \otimes \left( \sin \frac{\theta}{2} \ket{\uparrow} - i \cos \frac{\theta}{2} \ket{\downarrow}\right)\right]^{\otimes L/2} \ .
\end{equation}
These have zero magnetization $\sigma^z_{\rm tot}$ irrespective of the tilting angle $\theta$, and their average energy density $e$ is given for arbitrary anisotropies $\Delta$ and tilting angles $\theta$ by
\begin{equation}
    e(\theta,\Delta) = -\frac{1}{4} \left[ \sin^2(\theta) + \Delta\big(\cos^2(\theta) +1\big)\right] \ .
\end{equation}
Exact formulas for the overlaps between the Néel state ($\theta = 0$) and the XXZ eigenstates have been derived in Refs.~\cite{brockmann_gaudinlike_2014, brockmann_neelxxz_2014}, while a conjecture for the overlaps in the full class \eqref{eq:tiltedNeel} has been proposed in Ref.~\cite{pozsgay_overlaps_2018}. 

Actually, in the regime $\Delta > 1$ Ref.~\cite{piroli_exact_2016} (see also Ref.~\cite{piroli_from_2017}) obtained explicit results for the GGE root densities $\bs{\varrho}_{\text{\tiny GGE}}(\lambda)$ for the full class \eqref{eq:tiltedNeel}, by reconstructing the rapidity distributions from the initial state~\cite{ilievski_stringcharge_2016, ilievski_complete_2015} (see also Refs.~\cite{fagotti_stationary_2013, fagotti_relaxation_2014}). Simple and compact recursive formulas to obtain the GGE $\varrho_{\ell}(\lambda)$ and $\varrho_{\ell}^{(h)}(\lambda)$ up to arbitrary $\ell$ can be found directly in Ref.~\cite{piroli_exact_2016}. For arbitrary tilting angles $\theta$ the majority of particles are unbounded magnons ($\ell = 1$), while bound states ($\ell\ge2$) contribute a finite but lower amount to the total density $d = \sum_\ell d_\ell= \lim_{L \to \infty} N/L = 1/2$.

In the regime $0<\Delta < 1$ of \cref{eq:Deltam1} we restrict our attention to the Néel state ($\theta = 0$) for which the quench action TBA equations have been obtained for the anisotropies of \cref{eq:Deltam1} in Ref.~\cite{piroli_non_2018} (see also Ref.~\cite{bertini_nonequilibrium_2023}). The equations are reported in Appendix~\ref{appendix:XXZ}. They can be numerically solved to determine the GGE root and hole densities $\varrho_{\ell}(\lambda)$ and $\varrho_{\ell}^{(h)}(\lambda)$. Also in this regime unbounded magnons are generally the dominant contribution. 

Having access to $\varrho_{\ell}(\lambda)$ and $\varrho_{\ell}^{(h)}(\lambda)$ both for the GGE and the thermal macrostate, we compute the athermality $\alpha$ by applying \cref{eq:YYentdensXXZ,eq:defAthDensity}.

\subsection{Results}
\label{sec:resultsXXZ}
\begin{figure}[!b]
  \centering
  \begin{minipage}{0.328\textwidth}
    \centering
    \includegraphics[width=\linewidth]{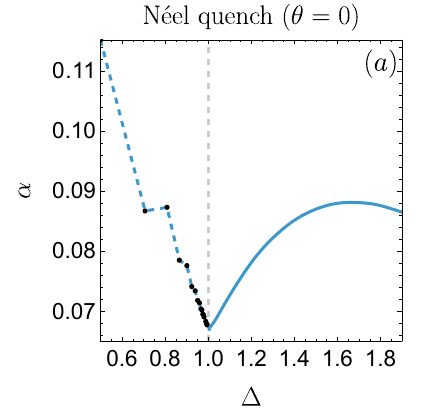}
  \end{minipage}
  \hfill 
  \begin{minipage}{0.328\textwidth}
    \centering
    \includegraphics[width=\linewidth]{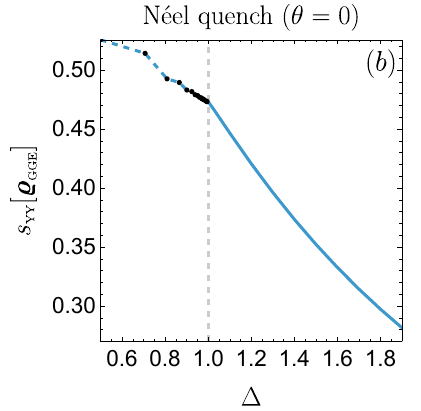}
  \end{minipage}
  \hfill 
  \begin{minipage}{0.328\textwidth}
    \centering
    \includegraphics[width=\linewidth]{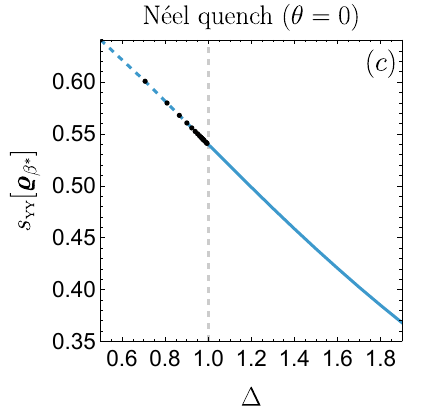}
  \end{minipage}
  \caption{Densities of athermality $\alpha$ (a), GGE entropy $s_{\text{\tiny YY}}[\bs\varrho_{\text{\tiny GGE}}]$ (b) and GE entropy $s_{\text{\tiny YY}}[\bs\varrho_{\beta^*}]$ (c) as a function of $\Delta$ in the quench from the Néel state (\cref{eq:tiltedNeel} for $\theta = 0$). The dashed blue lines in the interval $0<\Delta<1$ are just guides to the eye that connect results for the anisotropies in \cref{eq:Deltam1}. The vertical dashed line indicates the critical point $\Delta = 1$.}
  \label{fig:XXZ_Neel}
\end{figure}
In \cref{fig:XXZ_Neel}(a) we report the density of athermality $\alpha$ as a function of $\Delta$ in a quench from the Néel state ($\theta=0$). Remarkably, we observe the same phenomenology frequently found in the XY chain, namely the presence of a dip at criticality ($\Delta = 1$) in the form of a pronounced local-minimum cusp. The singular nature at $\Delta = 1$ is again due to the GGE entropy density $s_{\text{\tiny YY}}[\bs\varrho_{\text{\tiny GGE}}]$, as visible in \cref{fig:XXZ_Neel}(b), while the GE entropy in \cref{fig:XXZ_Neel}(c) appears smooth, as expected. These results show that a reduction in the degree of athermality at quantum critical points is not an artifact of free models like XY, but can occur also in the presence of interactions. \cref{fig:XXZ_Neel}(a) and (b) also uncover in the gapless regime $0<\Delta<1$ what appears as a strongly non-smooth behaviour of $s_{\text{\tiny YY}}[\bs \varrho_{\text{\tiny GGE}}]$ and $\alpha$ as functions of $\Delta$. This is analogous to the fractal behaviour found in other physical quantities like Drude weights~\cite{ilievski_microscopic_2017, ilievski_popcorn_2022}, which emerges due to the discontinuous quasiparticle content at different roots of unity points for $\Delta$ (\cref{eq:Deltam1} being the simplest class of such points). On the other hand, $s_{\text{\tiny YY}}[\bs \varrho_{\beta^*}]$ shown in \cref{fig:XXZ_Neel}(c) appears smooth also for $0<\Delta < 1$, again due to the impossibility of finite-temperature phase transitions in 1D. Indeed, it is analytically well-understood that discontinuities in the quasiparticle content at the roots of unity points play no role in GE thermodynamics~\cite{destri_new_1992}.

\begin{figure}[t!]
    \centering
    \begin{minipage}[b]{0.235\textwidth}
        \centering
        \includegraphics[width=\textwidth]{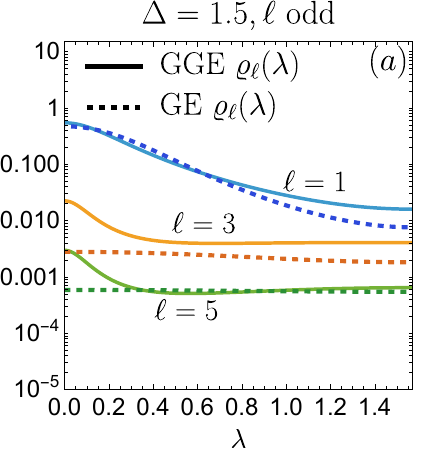}
    \end{minipage}
    \begin{minipage}[b]{0.235\textwidth}
        \centering
        \includegraphics[width=\textwidth]{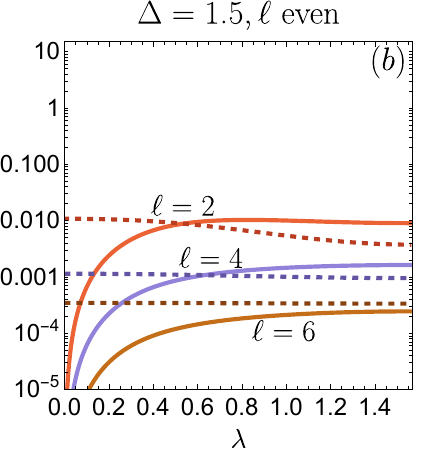}
    \end{minipage}
    \hfill
    \begin{minipage}[b]{0.235\textwidth}
        \centering
        \includegraphics[width=\textwidth]{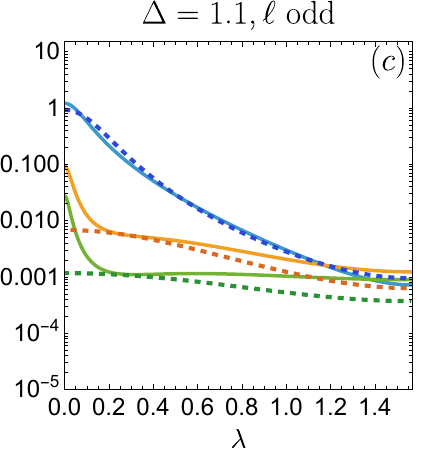}
    \end{minipage}
    \begin{minipage}[b]{0.235\textwidth}
        \centering
        \includegraphics[width=\textwidth]{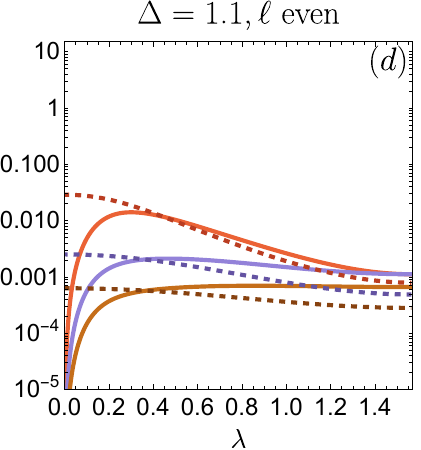}
    \end{minipage}
    
    \vspace{0.2cm} 

    \begin{minipage}[b]{0.235\textwidth}
        \centering
        \includegraphics[width=\textwidth]{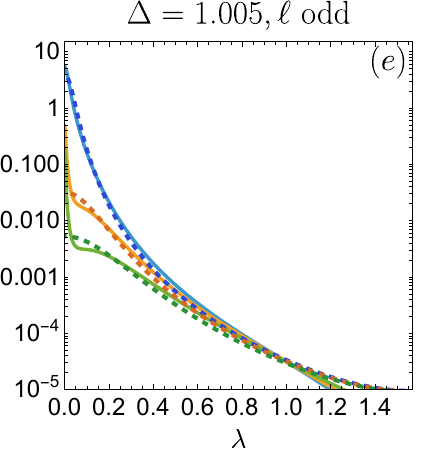}
    \end{minipage}
    \begin{minipage}[b]{0.235\textwidth}
        \centering
        \includegraphics[width=\textwidth]{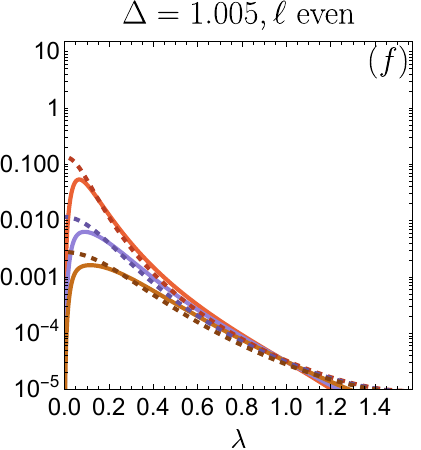}
    \end{minipage}
    \hfill
    \begin{minipage}[b]{0.235\textwidth}
        \centering
        \includegraphics[width=\textwidth]{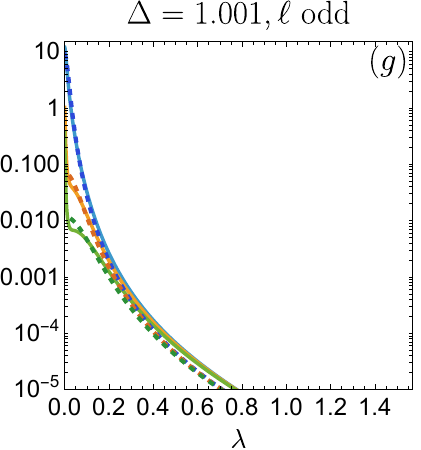}
    \end{minipage}
    \begin{minipage}[b]{0.235\textwidth}
        \centering
        \includegraphics[width=\textwidth]{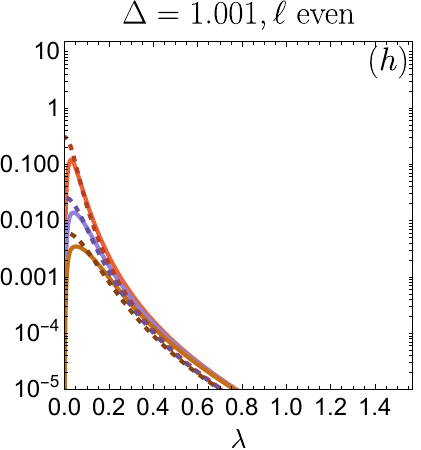}
    \end{minipage}
    
    \caption{Log-scale plots for the GE (dashed lines) and GGE (continuous lines) root densities $\varrho_{\ell}(\lambda)$ for $1\le\ell\le 6$ and $\Delta>1$ in the Néel quench ($\theta = 0$). The curves are shown only for $\lambda \ge0$ as they are symmetric around $\lambda = 0$.}
    \label{fig:XXZ_root_densities_Dg1}
\end{figure}

Given the simple pinning mechanism behind the dips at criticality discussed for XY, it is natural to wonder if something similar happens for the XXZ chain. In \cref{fig:XXZ_root_densities_Dg1} we plot in log-scale the GE and GGE root densities $\varrho_{\ell}(\lambda)$ for a few values of $\ell\ge 1$ and $\Delta>1$ in the Néel quench ($\theta = 0$). We observe a behaviour that is qualitatively similar to the pinning mechanism of \cref{fig:root_densities_1} for the XY chain. In particular, as we approach the quantum critical point  $\Delta = 1$, both the GE and GGE root densities for each $\ell$ shrink around $\lambda = 0$ and develop similar pronounced peaks. An important difference compared to XY is that the pinning value is unbounded as the limit $\Delta \to 1$ is approached. This is a consequence of the $\lambda$-dependent density of total vacancies $\varrho_{\ell}^{\rm (tot)}(\lambda) = \varrho_\ell(\lambda)+\varrho_\ell^{(h)}(\lambda)$, which is instead a constant in free theories like XY. In particular, $\varrho_\ell^{(\rm tot)}(\lambda)$ develops a divergence at $\lambda = 0$ in the limit $\Delta\to1$. This is due to the fact that, in the limit considered, the driving term $a_\ell(\lambda)$ in TBA (see \cref{eq:rhorhohXXZDg1} for $\Delta > 1$ and \cref{eq:rhorhohXXZDl1} for $\Delta < 1$ in Appendix~\ref{appendix:XXZ}) approaches a Dirac delta centred at zero. Given that the entropies \eqref{eq:YYentdensXXZ} depend also on the hole densities, in Appendix~\ref{appendix:XXZ} we report plots identical to those in \cref{fig:XXZ_root_densities_Dg1} but for $\varrho_\ell^{(h)}(\lambda)$, and find that a similar pinning mechanism emerges. It is natural to wonder if a pinning mechanism to finite values, hence more similar to the one seen in XY, appears if one focuses on the bounded filling functions $\vartheta_\ell(\lambda)=\varrho_\ell(\lambda)/\varrho_{\ell}^{(\rm tot)}(\lambda)$. We find this to not be the case, given that the similarity between the GE and GGE filling functions $\vartheta_\ell(\lambda)$ emerges on a large range of $\lambda$ values as an almost constant plateau, see plots in Appendix~\ref{appendix:XXZ}. 
In Appendix~\ref{appendix:XXZ} we show that a pinning mechanism for $\varrho_{\ell}(\lambda)$ arises also in the gapless regime $0<\Delta<1$ as $\Delta \to 1$ is approached (i.e. $p\gg1$ in the parametrization \eqref{eq:Deltam1}), giving rise to the dip at criticality for $\alpha$ on the left side of \cref{fig:XXZ_Neel}(a). 

\begin{figure}[t!]
  \centering
  \begin{minipage}{0.328\textwidth}
    \centering
    \includegraphics[width=\linewidth]{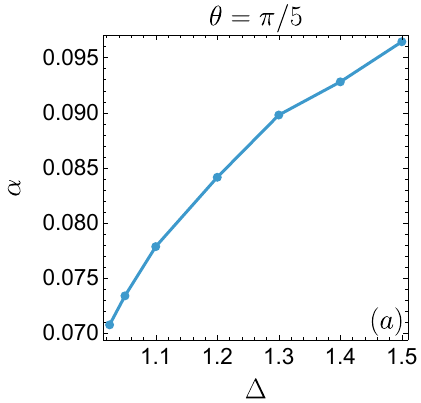}
  \end{minipage}
  \hfill 
  \begin{minipage}{0.328\textwidth}
    \centering
    \includegraphics[width=\linewidth]{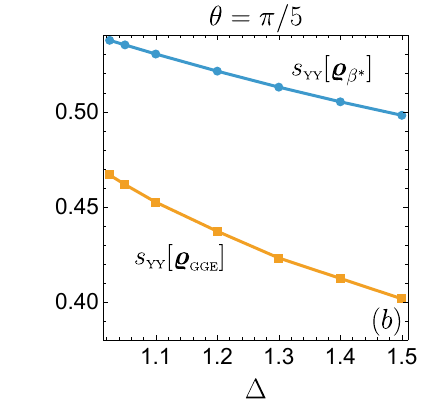}
  \end{minipage}
  \hfill 
  \begin{minipage}{0.328\textwidth}
    \centering
    \includegraphics[width=\linewidth]{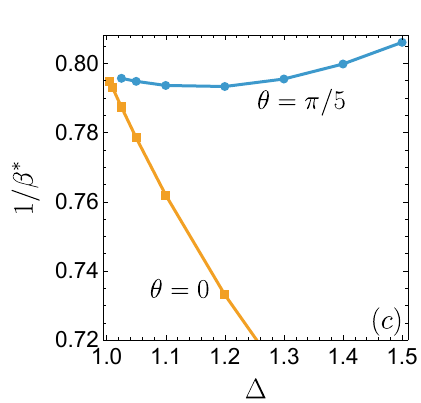}
  \end{minipage}
  \caption{Density of athermality $\alpha$ (a), GGE and GE entropy densities (b) and temperatures $1/\beta^*$ (c) as a function of $\Delta$ in the quench from the tilted-Néel state \eqref{eq:tiltedNeel} with $\theta = \pi/5$. In (c) we also plot for comparison the temperature of the Néel quench ($\theta = 0$). The continuous lines are just guides to the eye.}
  \label{fig:XXZ_tiltedNeel}
\end{figure}

Thanks to the results of Ref.~\cite{piroli_exact_2016}, in the gapped regime $\Delta > 1$ we can also consider nonzero tilting angles $\theta \neq 0$ for the initial states \eqref{eq:tiltedNeel}. From \cref{fig:XXZ_tiltedNeel}(a) we observe that a decrease in the athermality $\alpha$ close to criticality is present also for $\theta = \pi/5$. In \cref{fig:XXZ_tiltedNeel}(b) we plot the GE and GGE entropy densities giving rise to the dip. \cref{fig:XXZ_tiltedNeel}(c) compares the temperatures $1/\beta^*$ as a function of $\Delta$ in both the $\theta=0$ and $\theta = \pi/5$ quenches, demonstrating that in general there is no simple direct correlation between the behaviour of the temperature and that of $\alpha$ (cf.~the minimum of $1/\beta^*$ at intermediate values of $\Delta>1$ in the $\theta = \pi/5$ case).

Overall, the results of this section demonstrate that sharp signatures of quantum criticality in the athermality and GGE entropy survive also in the presence of interactions.

\section{Lieb-Liniger model}
\label{sec:LL}

The Lieb-Liniger (LL) model~\cite{lieb_exact_1963, lieb_exact_1963a} describes a one-dimensional gas of bosons interacting with Dirac delta potentials (we set $\hbar = 2m = 1$)
\begin{equation}
H_{\rm LL}(c) = \int_{0}^L dx \Big[ -\phi^\dag(x) \partial_x^2 \phi(x) + c \, \phi(x)^\dag\phi(x)^\dag \phi(x)\phi(x)\Big] \ .
\end{equation}
Here $\phi(x)$ denotes the annihilation operator for a canonical Bose field, $c$ is the interaction strength and we assume PBC. The model cannot be mapped to a free theory aside for the limits $c \to 0$ (free bosons) and $c \to \infty$ (free fermions via a continuum Jordan-Wigner transformation~\cite{girardeau_relationship_1960, creamer_quantum_1980}). In the following we consider both the repulsive ($c>0$) and attractive ($c<0$) regimes~\cite{takahashi_thermodynamics_1999, mcguire_study_1964}. 

For $c>0$ the model admits no bound states in the spectrum, and the eigenstates are parametrized by \emph{real} rapidities $\bs \lambda$ only. Therefore, in the TBA description the macrostates are specified by a single root density $\varrho(\lambda)$ (see Appendix~\ref{appendix:LL} for details). The ground state sector is gapless irrespective of the interaction strength $c>0$, and the low-energy physics is described by a Luttinger liquid~\cite{cazalilla_one_2011, giamarchi_quantum_2003}. 

Given the absence of a quantum critical point, based on the results from the previous sections we can already anticipate that the athermality will feature less structure than in the XY and XXZ cases. We stress that in repulsive LL the thermal macrostate $\varrho_{\beta^*,\mu^*}(\lambda)$ features an additional experimentally relevant Lagrange multiplier $\mu$, associated with the conservation of the number of particles (see \cref{eq:athermality2}). The  TBA nonlinear integral equations defining $\varrho_{\beta^*,\mu^*}(\lambda)$ are reported in Appendix~\ref{appendix:LL}.

In the attractive regime $c<0$ bound states of arbitrary number of particles exist (corresponding to strings of arbitrary length)~\cite{takahashi_thermodynamics_1999, mcguire_study_1964, piroli_multiparticle_2016, piroli_quantum_2016} and macrostates are specified by an infinite sets of root densities $\varrho_\ell(\lambda)$, $\ell = 1, 2, \ldots, \infty$ corresponding to the densities of $n_\ell = \ell$ bound particles with string center $\lambda$. Crucially, the attractive LL does not admit a stable thermodynamic limit in equilibrium. For $c < 0$, the system undergoes a collapse into a macroscopic bound state whose ground-state energy is superextensive ($\propto -L^3$ at fixed nonzero density)~\cite{takahashi_thermodynamics_1999}. Because the entropy scales only extensively ($\propto L$), temperature cannot stabilize the free energy, precluding the existence of a well-defined thermal ensemble. Therefore at $c<0$ the athermality cannot be defined. However, quantum quenches to the attractive LL are perfectly meaningful~\cite{piroli_multiparticle_2016, piroli_quantum_2016} and can be experimentally realized~\cite{khaykovich_formation_2002, strecker_formation_2002, haller_realization_2009, horvath2025observing, zeng2026realization}. Indeed, in the quench setup the extensive charges (including the energy) associated with the initial state $\ket{\psi(0)}$ are conserved and therefore at late times the system is locally described by a well-defined macrostate of $H_{\rm LL}(c<0)$. This enables us to consider the behaviour of both $s_{\text{\tiny YY}}[\varrho_{\text{\tiny GGE}}]$ ($c>0$) and $s_{\text{\tiny YY}}[\bs\varrho_{\text{\tiny GGE}}]$ ($c<0$). These, together with the thermal Yang-Yang density $(c>0)$, are obtained from \cref{eq:YYentdensXXZ} by setting $A = \infty$ and $\ell_{\rm max}=1$ ($\ell_{\rm max}=\infty$) for $c>0$ ($c<0$). 

\subsection{Initial state and GGE root density}

We focus on a quench with Hamiltonian $H_{\rm LL}(c)$ starting from the ground state of the $c = 0$ free bosonic theory
\begin{equation}
    H_{\rm free} = - \int_0^L dx \, \phi^\dag(x) \partial_x^2 \phi(x) = \sum_k k^2 \tilde\phi_k^\dag \tilde \phi_k \ ,
\end{equation}
where $\tilde \phi_k$ are the standard Fourier modes $\tilde \phi_k = \int_0^L dx\, e^{i x k} \phi(x)/\sqrt{L}$. The Bose-Einstein condensate (BEC) ground state of $H_{\rm free}$ in the $N$-particle sector is
\begin{equation}
\label{eq:GSNLL}
    \ket{\psi(0)}=\ket{{\rm GS};N} = \frac{1}{\sqrt{N!}}(\tilde \phi_0^\dag)^N \ket 0 \ .
\end{equation}
It possesses off-diagonal long-range order~\cite{yang_concept_1962} and it is a famous example of an integrable initial state~\cite{piroli_what_2017}. Exact expressions for the overlaps between $\ket{\psi(0)}$ and the eigenstates of $H_{\rm LL}(c)$ (for arbitrary $c$) have been obtained in Refs.~\cite{de_nardis_solution_2014,brockmann_overlaps_2014,brockmann_neelxxz_2014}. 

\begin{figure}[b!]
  \centering
  \begin{minipage}{0.4\textwidth}
    \centering
    \includegraphics[width=\linewidth]{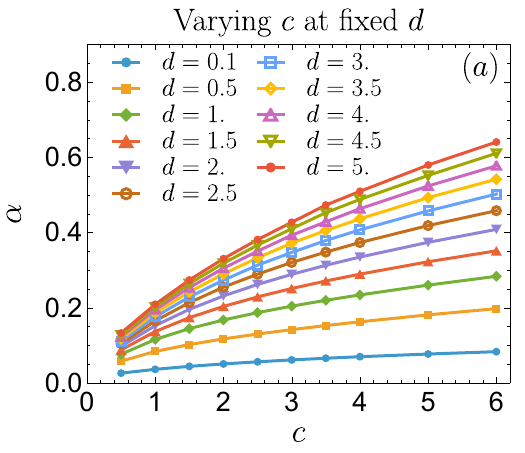}
  \end{minipage}
  \begin{minipage}{0.4\textwidth}
    \centering
    \includegraphics[width=\linewidth]{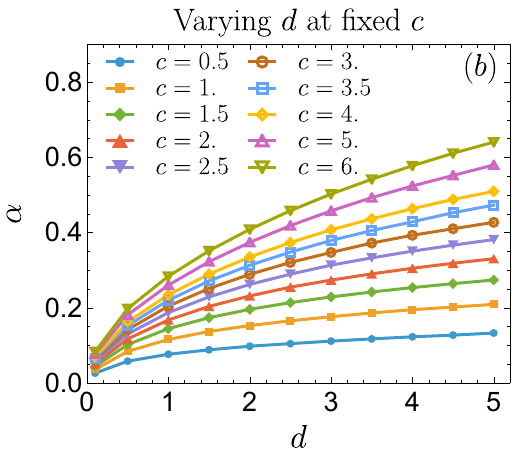}
  \end{minipage}
  \caption{Density of athermality $\alpha$ associated with the BEC-to-repulsive-LL quench. (a) $\alpha$ as a function of the postquench interaction $c>0$ at fixed particle density $d$. (b) $\alpha$ as a function of $d$ at fixed postquench $c$.}
  \label{fig:LL_main}
\end{figure}

In the repulsive regime $c>0$, thanks to the overlap formulas, Ref.~\cite{de_nardis_solution_2014} obtained an analytic expression for the saddle-point root density $\varrho_{\text{\tiny GGE}}(\lambda)$, which for completeness we report in Appendix~\ref{appendix:LL}. A remarkable feature of this quench is that, unlike $\varrho_{\beta^*,\mu^*}(\lambda)$ which has an exponential decay, $\varrho_{\text{\tiny GGE}}(\lambda)$ decays only polynomially (as $\sim \lambda^{-4}$) for large $|\lambda|$~\cite{de_nardis_solution_2014}.

For the attractive quench $c<0$, starting again from the overlap formulas, Refs.~\cite{piroli_multiparticle_2016, piroli_quantum_2016} obtained an explicit set of recursive equations from which one can easily obtain the GGE $\ell$-particle-string root densities $\varrho_{\ell}(\lambda)$ (up to large values of $\ell$). The equations, together with details on the TBA description of LL at $c<0$, are given in Appendix~\ref{appendix:LL}. A very interesting feature of the stationary state reached in such a quench is that the density of bounded particles $d_{\rm bound}=\sum_{\ell = 2}^\infty d_\ell = \sum_{\ell = 2}^\infty \ell \int_{-\infty}^\infty \varrho_{\ell}(\lambda)$ is generally higher than that of unbounded ones $d_1$, hence most of the spectral weight is on bound states. Furthermore, perhaps counter-intuitively, the lower the value of $|c|$ the higher it is the contribution of bound states of many particles.

\subsection{Results}
\begin{figure}[t!]
  \centering
  \begin{minipage}{0.4\textwidth}
    \centering
    \includegraphics[width=\linewidth]{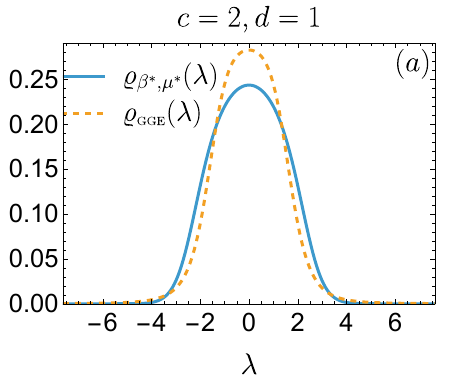}
  \end{minipage}
  \begin{minipage}{0.4\textwidth}
    \centering
    \includegraphics[width=\linewidth]{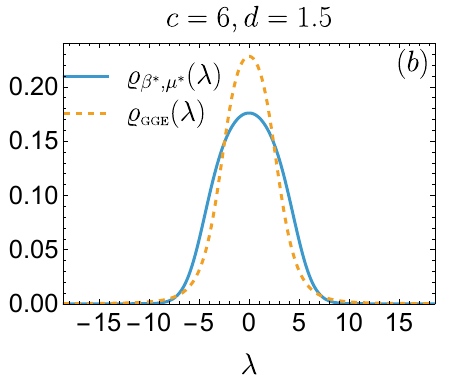}
  \end{minipage}
  \caption{Comparison between $\varrho_{\beta^*,\mu^*}(\lambda)$ (continuous line) and $\varrho_{\text{\tiny GGE}}(\lambda)$ (dashed line) in the repulsive regime for different values of $c>0$ and density of particles $d$.}
  \label{fig:LL_root_densities}
\end{figure}
 
In the context of the BEC to LL quench, we can vary the postquench Hamiltonian by tuning the interaction $c$ at fixed $\ket{\psi(0)}$, or change the initial state at fixed postquench $c$ by varying the initial density of particles $d$. 

\begin{figure}[b!]
    \centering
    \includegraphics[width=0.38\linewidth]{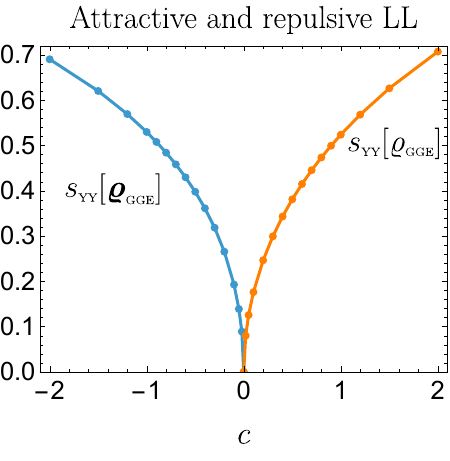}
    \caption{GGE entropy density for repulsive $c>0$ ($s_{\text{\tiny}}[\varrho_{\text{\tiny GGE}}]$) and attractive $c<0$ ($s_{\text{\tiny}}[\bs \varrho_{\text{\tiny GGE}}]$) interactions in the LL quench from the BEC state \eqref{eq:GSNLL} with density of particles $d=1$. The point $c=0$ corresponds a trivial zero due to absence of the quench.}
    \label{fig:attractiveLL}
\end{figure}

For $c>0$, we plot in \cref{fig:LL_main}  the athermality density $\alpha$ in both cases. The curves for $\alpha$ grow in a simple fashion with increasing $d$ or $c$, and do not present any remarkable behaviour (e.g.~singularities or dips) similar to the one seen for the XY and XXZ chains. In light of the results from \cref{sec:XY,sec:XXZsec}, we interpret this as a consequence of the absence of quantum phase transitions in the $T = 0$ phase diagram of LL for $c>0$. We notice, however, that while the growth of $\alpha$ trivially correlates with the growth of the initial energy density $e=d^2 c$ as a function of $d$ and $c$, the functional forms of their increase are quantitatively very different. The increasing trend of $\alpha$ as a function of $d$ and $c$ turns out to be extremely similar to that of its components $s_{\text{\tiny YY}}[\varrho_{\beta^*,\mu^*}]$ and $s_{\text{\tiny YY}}[\varrho_{\text{\tiny GGE}}]$, which are plotted in \cref{fig:LL_main_2} of Appendix~\ref{appendix:LL}. 
In \cref{fig:LL_root_densities} we show the root densities $\varrho_{\beta^*,\mu^*}(\lambda)$ and $\varrho_{\text{\tiny GGE}}(\lambda)$ for $c = 2, d = 1$ and $c=6, d = 1.5$ (see also Ref.~\cite{de_nardis_solution_2014} for similar plots). We observe that even for fairly big changes in $c$, the differences between the two root densities remain qualitatively very similar. This is the underlying reason for the lack of a rich structure in $\alpha$, cf.~\cref{fig:root_densities_1} for the XY chain.

In \cref{fig:attractiveLL} we plot the GGE entropy density as the interaction varies from the repulsive ($c>0$) to the attractive ($c<0$) regime at a fixed density $d=1$. We observe that no striking features appear in the bulk of the two regimes. The trivial zero of the Yang-Yang density at $c = 0$ arises simply because at this point no quench occurs, due to our choice of the initial state \eqref{eq:GSNLL} (GS of the $c=0$ theory). However, we notice that here the singular nature of the trivial zero is apparent already from the plot in \cref{fig:attractiveLL} of the GGE entropy (strong singularity), while at the trivial zeros of the XY chain (\cref{sec:XY}) it becomes evident only in higher-order derivatives of $s_{\text{\tiny YY}}[\varrho_{\text{\tiny GGE}}]$ (weaker singularity).

We expect non-trivial singular points in $s_{\text{\tiny YY}}[\varrho_{\text{\tiny GGE}}]$ to emerge at $c=0$ for different initial states that are not eigenstates of the free theory. While explorations of this possibility are constrained by the limited results available~\cite{he_integrable_2023} for exact overlaps in non-free initial states, it is certainly an interesting question to address in future work, because at $c=0$ a radical change in the low energy spectrum occurs which is more drastic than that associated with standard quantum critical points.

\section{Harmonic chain}
\label{sec:harmonicchain}

We finally explore how the conclusions from the previous sections change in a model whose underlying excitations have bosonic nature, unlike those of XY, LL and XXZ which obey a (generalized) Pauli principle. We stress that the emergence of a generalized Pauli principle is ubiquitous in the interacting integrable context~\cite{korepin_quantum_1993}. Therefore, to avoid it, we focus on a simple free bosonic model: the harmonic chain (HC) with PBC
\begin{equation}
    H_{\rm HC}(m) = \frac{1}{2}\sum_{j=1}^L \big[ p_j^2 + m^2 x_j^2 + (x_{j+1}-x_j)^2 \big] \ .
\end{equation}
Here $x_j$ denotes the position of an oscillator centered around site $j$ and $p_j$ its momentum, satisfying the canonical commutation relations $[x_j,p_\ell]=i \delta_{j,\ell}$. Using QFT terminology, the mass $m$ controls how tightly bound to their associated sites the oscillators are. In the limit $m\to 0$ the model becomes gapless and $m = 0$ represents a quantum critical point  whose low energy physics is described by a massless free boson CFT. 
The Hamiltonian can be easily diagonalized in Fourier space by introducing standard creation ($a^\dag_k$) and annihilation ($a_k$) operators with canonical commutation relations $[a_k,a_{k'}^\dag]=\delta_{k,k'}$
\begin{equation}
    x_j = \frac{1}{\sqrt{L}}\sum_k e^{i k j}\frac{1}{\sqrt{2 \varepsilon_m(k)}}\big(a_k + a_{-k}^\dag\big) \ , \qquad \qquad p_j = \frac{i}{\sqrt{L}}\sum_k e^{i k j}\sqrt{\frac{\varepsilon_m(k)}{2}}\big(a_{-k}^\dag - a_{k}\big)\ ,
\end{equation}
leading to the diagonal form
\begin{equation}
    H_{\rm HC}(m) = \sum_k \varepsilon_m(k)a_k^\dag a_k + E_0  \ , \qquad \qquad \varepsilon_m(k) = \sqrt{m^2+2(1-\cos k)} \ .
\end{equation}
The thermal occupation function $n_{\beta}(k) = 2 \pi \varrho_{\beta}(k)$ at inverse temperature $\beta$ is given by the Bose-Einstein distribution
\begin{equation}
\label{eq:nbetaHC}
    n_{\beta}(k) = \frac{1}{\exp[\beta \varepsilon_m(k)]-1} \ .
\end{equation}
We focus on quenches from the ground state $\ket{\psi(0)}=\ket{{\rm GS};m_0}$ of $H_{\rm HC}(m_0)$ and time-evolve with $H_{\rm HC}(m)$. The GGE density $n_{\text{\tiny GGE}}(k)$ is obtained as~\cite{calabrese_quantum_2007}
\begin{equation}
\label{eq:nspHC}
    n_{\text{\tiny GGE}}(k) = \braket{{\rm GS};m_0 |a_k^\dag a_k|{\rm GS};m_0} = \frac{1}{4}\left( \frac{\varepsilon_{m_0}(k)}{\varepsilon_m(k)}+\frac{\varepsilon_{m}(k)}{\varepsilon_{m_0}(k)}\right)-\frac{1}{2} \ ,
\end{equation}
where in the previous equation $a_k$ refers to the modes diagonalizing the final Hamiltonian $H_{\rm HC}(m)$. As before the inverse temperature $\beta^*$ is fixed by the initial energy density. A major difference compared to the previous sections is that the expression for the Yang-Yang entropy density must be modified to account for bosonic statistics
\begin{equation}
\label{eq:sBose}
    s[n] = \int_{-\pi}^\pi \frac{dk}{2\pi} \left[-n(k)\ln n(k) + (1+n(k))\ln(1+n(k))\right] \ .
\end{equation}
With the knowledge of $n_{\text{\tiny GGE}}(k)$ and $n_{\beta^*}$ we can use \cref{eq:sBose} to compute the GE and GGE entropy densities and hence the athermality $\alpha$. 

\begin{figure}[t!]
  \centering
  \begin{minipage}{0.328\textwidth}
    \centering
    \includegraphics[width=\linewidth]{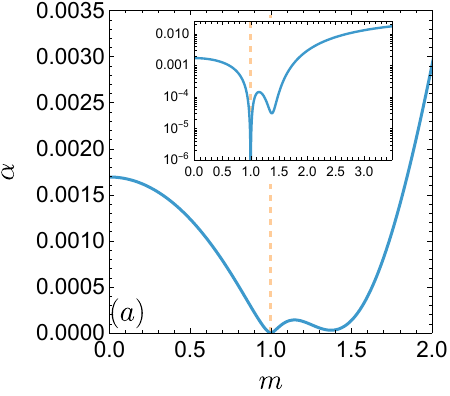}
  \end{minipage}
  \hfill 
  \begin{minipage}{0.328\textwidth}
    \centering
    \includegraphics[width=\linewidth]{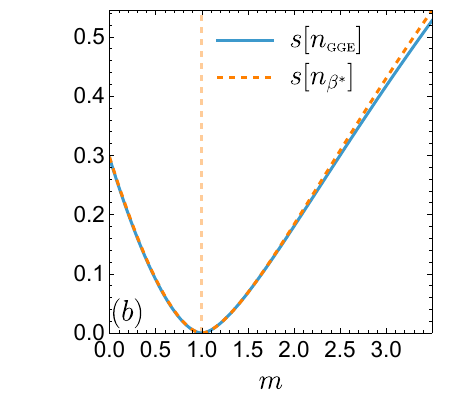}
  \end{minipage}
  \hfill 
  \begin{minipage}{0.328\textwidth}
    \centering
    \includegraphics[width=\linewidth]{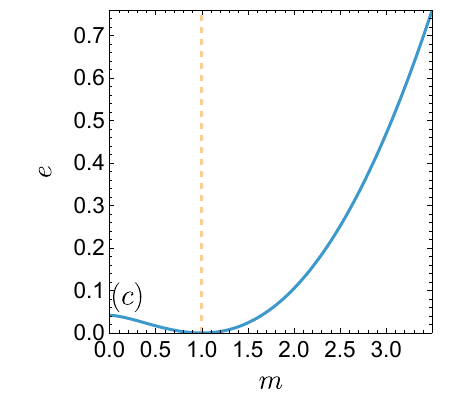}
  \end{minipage}
  \caption{Densities of athermality $\alpha$ (a), GGE and GE entropy (b) and energy $e$ (c) as a function of $m$ in the quench from the GS with $m_0 = 1$. The vertical dashed lines indicate the position of the trivial zero $m = m_0$. The inset in (a) shows the same data in log-scale.}
  \label{fig:main_HC}
\end{figure}

\begin{figure}[b!]
  \centering
  \begin{minipage}{0.328\textwidth}
    \centering
    \includegraphics[width=\linewidth]{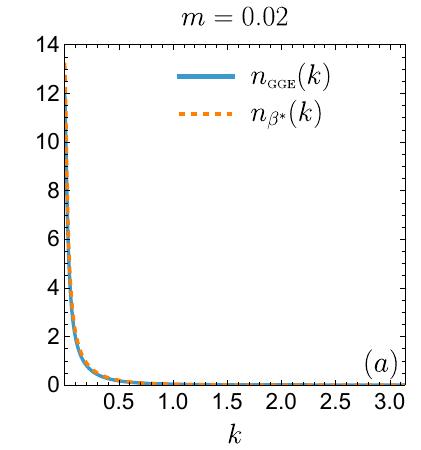}
  \end{minipage}
  \hfill 
  \begin{minipage}{0.328\textwidth}
    \centering
    \includegraphics[width=\linewidth]{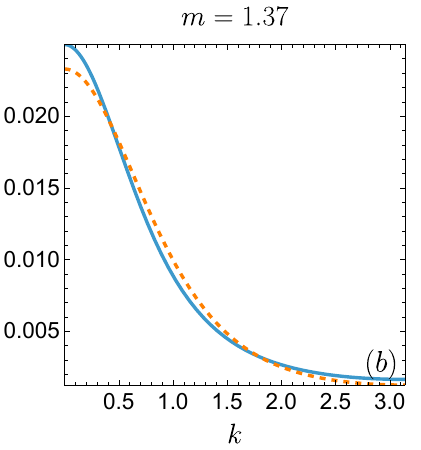}
  \end{minipage}
  \hfill 
  \begin{minipage}{0.328\textwidth}
    \centering
    \includegraphics[width=\linewidth]{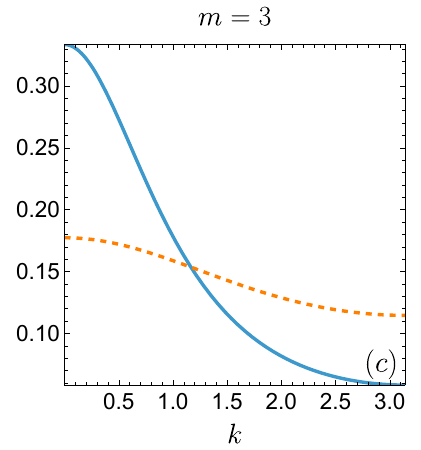}
  \end{minipage}
  \caption{GGE and thermal occupation functions $n_{\text{\tiny GGE}}(k)$ and $n_{\beta^*}(k)$ for a few values of the postquench mass $m$ in the quench from the GS with $m_0=1$. (a) $m = 0.02$ close to criticality; (b) $m = 1.37$, corresponding to the non-trivial minimum in the athermality of \cref{fig:main_HC}(a); (c) $m = 3$.}
  \label{fig:main2_HC}
\end{figure}

In \cref{fig:main_HC} we plot, as a function of the postquench mass $m$, the densities of (a) athermality $\alpha$, (b) entropies $s[n_{\text{\tiny GGE}}]$ and $s[n_{\beta^*}]$, and (c) the energy $e$, in a quench starting from the GS of $H_{\rm HC}(m_0 = 1)$. We find the behaviours in the case $m_0 =1$ to be representative of generic mass quenches in the HC. We note from \cref{fig:main_HC}(a) that $\alpha$ is at least one order of magnitude smaller than that found in previous sections. This is not a consequence of the overall magnitude of the entropies $s[n]$ being lower, but of the fact that $s[n_{\text{\tiny GGE}}]$ and $s[n_{\beta^*}]$ differ very little relative to their magnitude, as evident from \cref{fig:main_HC}(b). Importantly, this behaviour is observed despite the $\mathcal{O}(1)$ injected energy density in the system, see \cref{fig:main_HC}(c). We find that the curve for $\alpha$ generally displays the following features: (i) it has a trivial zero at $m = m_0$, as expected; (ii) it has a smooth local minimum at some value $m_{\rm min}$ to the right of the trivial zero $m = m_0$; (iii) the critical point $m=0$ is a local maximum approached in a smooth way, hence we find no dips at criticality.

It is instructive to understand how these results emerge from the occupation functions $n_{\text{\tiny GGE}}(k)$ and $n_{\beta^*}(k)$, plotted in \cref{fig:main2_HC} again for the $m_0 = 1$ quench. As we approach criticality, we find a pinning mechanism at infinity around the momentum $k=0$, as evident from \cref{fig:main2_HC}(a). This is expected for $m = 0$ because of the divergence at $k = 0$ both in \cref{eq:nspHC} and \cref{eq:nbetaHC} (the inverse temperature $\beta^*$ remains finite at $m = 0$), due to the absence of a gap $\varepsilon_0(0)=0$. Interestingly, this does not give rise to a dip at criticality in $\alpha$, despite the fact that (when observed on the full scale of values they take) the occupations functions look more similar close to criticality than for any other value of $m$. In \cref{fig:main2_HC}(b) we plot the occupation functions at a value $m\approx m_{\rm min}$. The scale of values taken by the occupation functions is greatly reduced compared to $m \approx 0$, and the curves for $n_{\text{\tiny GGE}}(k)$ and $n_{\beta^*}(k)$ happen to be very similar at this point, which is intermediate in their respective crossovers from the $m \approx 0$ to the $m\gg1$ behaviours. In \cref{fig:main2_HC}(c) we plot the occupation functions at a larger value of $m$ ($m = 3$). Here, while the curves for $n_{\text{\tiny GGE}}(k)$ and $n_{\beta^*}(k)$ appear visibly very different, we notice from \cref{fig:main_HC}(a) and (b) that they only give rise to very small values of $\alpha$. This is quite different from what we quantitatively found in the previous sections, where clear differences in the root densities always gave rise to significant differences in the entropy densities. The source of this discrepancy is the bosonic statistics intrinsic in the current definition \eqref{eq:sBose} of entropy. Recalling the definition of athermality in \cref{eq:athermality} as a relative entropy, this suggests in the spirit of quantum Stein's lemma~\cite{hiai_proper_1991, nagaoka_strong_2000, vedral_role_2002} that a bosonic GGE is more difficult to distinguish from its closest GE compared to the (effectively) fermionic context.

\section{The time evolution of athermality after a quench}
\label{sec:timevathermality}

A natural question, which turns out to admit a rather simple answer, concerns the time evolution of the athermality following a quantum quench. As already discussed, only reduced density matrices of finite subsystems relax to stationary values, which are described by a GE in nonintegrable systems and by a GGE in integrable ones. The notion of athermality in the quench context is therefore intrinsically local, and its dynamical behaviour can be analysed by focusing on a sufficiently large subsystem $A$, where boundary contributions are negligible compared to the extensive part of the entropies involved.
With this approximation in mind, 
denoting by $\rho_{\text{\tiny GE}}^{(A)}$ the reduced density matrix of the Gibbs ensemble at the effective inverse temperature $\beta^\ast$ discussed and calculated in the previous sections, and $\rho^{(A)}(t)\equiv{\rm Tr}_B[\ket{\psi(t)}\!\bra{\psi(t)}]$, we simply obtain
\begin{equation}
{\cal A}(t)
\equiv
S(\rho^{(A)}(t)\Vert \rho_{\text{\tiny GE}}^{(A)})
\simeq
S_\text{\tiny GE}^{(A)}-S^{(A)}(t)\ ,
\label{At1}
\end{equation}
where we have used
\begin{equation}
S_\text{\tiny GE}^{(A)}\equiv
-{\rm Tr} \left[\rho_{\text{\tiny GE}}^{(A)}\ln \rho_{\text{\tiny GE}}^{(A)}\right]
\simeq
-{\rm Tr}\left[\rho^{(A)}(t)\ln \rho_{\text{\tiny GE}}^{(A)}\right]
\label{sime}.
\end{equation}
This follows from the fact that $\rho_{\text{\tiny GE}}^{(A)}$ is (for large subsystem sizes) approximately a Gibbs state of the local Hamiltonian $H_A$, i.e.~$
\ln \rho_{\text{\tiny GE}}^{(A)} \simeq -\beta^\ast H_A - \ln Z_A$,
and from
$
{\rm Tr}[\rho^{(A)}(t) H_A]
=
{\rm Tr}[\rho_{\text{\tiny GE}}^{(A)} H_A] \ \forall \ t$ (we are implicitly assuming the initial state and quench to be homogeneous).
In fact, the right-hand side of \cref{At1} could have been obtained directly from \cref{eq:athermalitysimple} given the approximation for $\rho_{\text{\tiny GE}}^{(A)}$ above.

At this point, the time evolution of the athermality after a homogenous quantum quench in an integrable model, for a subsystem $A$ of length $L_A$ embedded in an infinite system, can be obtained immediately from the quasiparticle picture of entanglement growth~\cite{Calabrese_ev_2005,alba_entanglement_2017,calabrese_ln}. Indeed, using Eq.~\eqref{At1} and the quasiparticle picture for $S^{(A)}(t)$, in the scaling limit $t,L_A\to\infty$ with $t/L_A$ fixed, one straightforwardly finds
\begin{equation}
{\cal A}(t)
\simeq
S_\text{\tiny GE}^{(A)}
-\sum_\ell \int d\lambda \,
s_{\text{\tiny YY}}^{(\ell)}(\lambda)\,
\min\!\big[2|v_\ell(\lambda)|t,L_A\big] \ ,
\label{At}
\end{equation}
where the index $\ell$ labels the quasiparticle species (or string types in the Bethe-ansatz language), $s_{\text{\tiny YY}}^{(\ell)}(\lambda)$ is the corresponding GGE Yang-Yang entropy function (i.e.~the integrand in \cref{eq:YYentdensXXZ}) and $v_\ell(\lambda)$ is the quasiparticle velocity.
Note that for large times, Eq. \eqref{At} converges to the GGE athermality, as expected.

We can repeat the same reasoning that led to Eq. \eqref{sime} (generalized to all higher quasilocal charges) to argue that 
\begin{align}
&S_\text{\tiny GGE}^{(A)} \equiv
-{\rm Tr}\left[\rho_{\text{\tiny GGE}}^{(A)}\ln \rho_{\text{\tiny GGE}}^{(A)}\right]
\simeq -{\rm Tr}\left[\rho^{(A)}(t)\ln \rho_{\text{\tiny GGE}}^{(A)}\right] \ ,\\
&-{\rm Tr}\left[\rho_{\text{\tiny GGE}}^{(A)}\ln \rho_{\text{\tiny GE}}^{(A)}\right]
\simeq
-{\rm Tr}\left[\rho_{\text{\tiny GE}}^{(A)}\ln \rho_{\text{\tiny GE}}^{(A)}\right] = S_\text{\tiny GE}^{(A)} \ .
\end{align}  
Then we have 
\begin{equation}
{\cal A}(t)
\simeq 
S(\rho_{\text{\tiny GGE}}^{(A)}\Vert \rho_{\text{\tiny GE}}^{(A)})+
S(\rho^{(A)}(t)\Vert \rho_{\text{\tiny GGE}}^{(A)}) \ .
\end{equation}
We note that $S(\rho^{(A)}(t)\Vert \rho_{\text{\tiny GGE}}^{(A)})$ coincides with the one obtained in Ref.~\cite{Ares_simpler_2025}, despite arising from a conceptually different approximation and from a different physical perspective. In the present case, the result emerges naturally from the resource-theoretic notion of athermality.

Moving briefly the attention to ergodic systems that thermalise in the conventional sense, the athermality
\begin{equation}
{\cal A}(t)
=
S(\rho^{(A)}(t)\Vert \rho_\text{\tiny GE}^{(A)})
\simeq
S_\text{\tiny GE}^{(A)}-S^{(A)}(t),
\end{equation}
vanishes at long times. As a result, it provides a natural probe of the quantum Mpemba effect in closed systems~\cite{ares2023entanglement,Ares_rev_2025,Calabrese_rev_2026}. Indeed, essentially the same quantity has already appeared in recent studies of anomalous relaxation and Mpemba phenomena~\cite{Ares_simpler_2025}, although without being interpreted as a measure of athermality. 
Our analysis therefore places these previous observations within a broader unified framework, showing that the accelerated approach to equilibrium can be viewed as accelerated consumption of the resource associated with non-thermality.
This observation is consistent with the findings of Ref.~\cite{summer_2026}, which provide a resource-theoretical unification of Mpemba effects.

\section{Conclusions}

In this work, we have introduced and studied the concept of \emph{athermality} as a quantitative measure of the difference between the local stationary state of an integrable system after global quantum quenches (GGE) and the thermal state (GE) to which the system would have locally relaxed in the absence of integrability. Within the framework of quantum resource theory, we have defined athermality as the relative entropy between a given state and the \emph{closest} GE, which possess the same average energy. In the context of integrable quenches, this identification provides a particularly transparent interpretation of athermality as a measure of the memory retained by the system of the additional conservation laws associated with integrability. We stress that the athermality can be employed more generally to characterize arbitrary GGEs, which arise in several physical contexts even beyond the quench setup.

We evaluated the athermality  in a broad range of paradigmatic integrable models, including the free XY spin chain, the XXZ spin chain, the  Lieb-Liniger gas, and the harmonic chain. 

The most surprising outcome of our work concerns the role of quantum criticality. We found that the athermality often becomes anomalously small when the post-quench Hamiltonian is tuned to a quantum critical point. Furthermore, the athermality systematically develops a nonanalyticity at criticality. This behaviour is a priori surprising because GGEs describe highly excited states located deep in the many-body spectrum, far from the ground state where the quantum phase transition occurs. Nevertheless, signatures of zero-temperature criticality survive and remain clearly visible in the stationary state. We have argued that the dips at criticality of the athermality may be related to the well-known thermalisation properties of conformal boundary states~\cite{Calabrese_2006,calabrese_quantum_2007}, which become increasingly relevant in the scaling limit near criticality. However, the extreme suppression of the athermality observed in several models remains only partially understood from a physical perspective and hence deserves further investigation.

Our results also establish a direct connection between athermality and non-equilibrium dynamics. For sufficiently large subsystems, the time evolution of the athermality after an integrable quantum quench is governed by the same quasiparticle picture that describes the growth of entanglement entropy. As a consequence, athermality naturally emerges as a probe of the quantum Mpemba effect, both in non-ergodic and ergodic systems. From this perspective, accelerated relaxation corresponds to an accelerated depletion of the resource associated with non-thermality.

The present work opens a number of interesting directions for future research. On the theoretical side, it would be valuable to obtain a deeper microscopic understanding of the relation between quantum criticality and the suppression of athermality at finite energy density, and verify if the phenomenology we uncovered here survives in quenches from non-integrable initial states~\cite{bertini_quantum_2017, bertini_entanglement_2018, gibbins_quench_2024}.
It would also be interesting to investigate whether similar phenomena occur in non-integrable systems, where the notion of stationary GGE is absent but prethermal regimes may still emerge~\cite{moeckel_interaction_2008,  kollar_generalized_2011,  marcuzzi_prethermalization_2013, essler_quench_2014, 
bertini_prethermalization_2015}. In certain regimes, a modified quasiparticle description has been shown to hold~\cite{Bertini_2020_PT}. In such cases, the framework developed here is expected to extend straightforwardly. Another natural direction concerns the study of athermality in open quantum systems~\cite{Fazio_2025} and in monitored many-body dynamics~\cite{skinner2019measurement,li2019measurement,fisher2023random}, where thermalisation competes with dissipation and measurement-induced effects.
A final interesting research direction would be to investigate athermality in the many situations where a time-dependent GGE emerges as an effective description of the dynamics~\cite{Lange2017,Lange2018,bouchoule2020,t_GGE_exp,ulcakar_generalized_2025,lumia2025,travaglino2026}. In such cases, an additional layer of complexity arises because the reference thermal state itself becomes time dependent. As a consequence, the evolution of the athermality would reflect not only the dynamics of the system but also the evolution of the thermal ensemble against which it is being compared, potentially leading to a much richer phenomenology.

More generally, our results suggest that the athermality provides a powerful and physically transparent lens through which to view integrability, thermalisation, and non-equilibrium quantum dynamics. We hope that it will prove useful in future studies of generalized equilibrium states and of the rich dynamical phenomena that arise in isolated quantum many-body systems.

\section*{Acknowledgments}
We thank Filiberto Ares, Colin Rylands, and Enej Ilievski for useful discussions.  We acknowledge financial support from the Royal Society through the University Research Fellowship No.\ 201101 (B.B.), and the European Commission through the ERC-AdG grant MOSE No.\ 101199196 (R.S.\ and P.C.).

\begin{center}
    {\huge Appendices}
\end{center}

\appendix
\section{TBA and GGE entropy density}
\label{appendix:RDMabelian}

The spectrum of free and interacting integrable models is known, and the eigenstates $\ket{\bs \lambda}$ are parametrized by a set of $N$ scalars $\bs \lambda=(\lambda_1, \ldots, \lambda_N)$~\cite{korepin_quantum_1993, takahashi_thermodynamics_1999}. In the free case these represent the free single-particle momenta of the excitations on a chain of size $L$, e.g.~$\lambda_j = 2 \pi I_j/L, \ I_j \in \mathbb{Z}$. In the presence of interactions, where they are known as rapidities, the entries of $\bs \lambda$ are in general complex and satisfy a nontrivial set of boundary constraints known as Bethe equations~\cite{korepin_quantum_1993}. One of the major difference with respect to free models due to interactions is that modifying a single rapidity $\lambda_j$ produces changes in the full set $\bs \lambda$ (via the Bethe equations). The eigenstates $\ket{\bs\lambda}$ simultaneously diagonalize all the extensively many (quasi)local charges $Q^{(n)}$ obtained from the logarithmic derivatives of the model's transfer matrices~\cite{korepin_quantum_1993, takahashi_thermodynamics_1999, slavnov_algebraic_2022, ilievski_quasilocal_2016}.

In the TBA framework~\cite{yang_thermodynamics_1969, takahashi_thermodynamics_1999} one considers directly the limit of $L \to \infty$, where it is more convenient to focus on classes of eigenstates that feature an identical distribution $\varrho(\lambda)$ of rapidities, defined for asymptotically large $L$ as
\begin{equation}
\label{eq:defrholambda}
   \lfloor L \varrho(\lambda) \Delta \lambda \rfloor = \text{number of rapidities $\lambda_j$ in $[\lambda, \lambda + \Delta \lambda$]} \ .
\end{equation}
For any $\varrho(\lambda)$ a distribution of holes $\varrho^{(h)}(\lambda)$ can be defined from thermodynamic limit of the Bethe equations, and it represents the interacting analogue of the density of unoccupied momenta in free models. 
In models possessing multiple species of quasiparticles, e.g.~bound states (see Appendices \ref{appendix:XXZ} and \ref{appendix:LL}), the single root density $\varrho(\lambda)$ is replaced by a set $\bs \varrho(\lambda)=\{\varrho_\ell(\lambda)\}_{\ell=1}^{\ell_{\rm max}}$, where $\varrho_\ell(\lambda)$ denotes the root density for the $\ell$-th species~\cite{takahashi_thermodynamics_1999}. Analogously $\varrho^{(h)}(\lambda) \to \bs \varrho^{(h)}(\lambda)$. Each $\bs \varrho(\lambda)$ defines a \emph{macrostate} containing a number of distinct eigenstates (or microstates) characterized by the Yang-Yang entropy density $s_\text{\tiny YY}[\bs \varrho(\lambda)]$, defined (conceptually) as
\begin{equation}
    s_\text{\tiny YY}[\bs \varrho(\lambda)]=\lim_{L \to \infty} \frac{\ln \big[\text{number of eigenstates with distribution $\bs \varrho(\lambda)$}\big]}{L}  \ .
\end{equation}
Explicit expressions for $s_\text{\tiny YY}[\bs \varrho(\lambda)]$ in terms of $\bs \varrho(\lambda)$ and $\bs \varrho^{(h)}(\lambda)$ in all the models of interest are reported directly in the main text, see \cref{eq:YYxy,eq:YYentdensXXZ,eq:sBose}.
Importantly, eigenstates belonging to the same macrostate $\bs \varrho(\lambda)$ have identical local properties (e.g.~expectation values for any local operator $O$)~\cite{korepin_quantum_1993}.

By maximising $s_\text{\tiny YY}[\bs \varrho]$ with respect to $\bs \varrho(\lambda)$ under the constraints of fixed energy density $e=e[\bs \varrho]$ (and particle density $n=n[\bs \varrho]$ for models with a conventional U(1) charge) one obtains the thermal root densities $\bs \varrho_{\beta}(\lambda)$ that exponentially dominate GE averages at the corresponding inverse temperature $\beta$ (and chemical potential $\mu$). Very similarly, averages with respect to GGEs from \cref{eq:defGGE} are exponentially dominated by a corresponding non-thermal macrostate $\bs \varrho_{\text{\tiny GGE}}(\lambda)$, which can be determined from a similar TBA maximization procedure provided that all Lagrange multipliers $\beta_n$ from \cref{eq:defGGE} are specified~\cite{mossel_generalized_2012}. 

In the quantum quench context, the Lagrange multipliers $\beta_n$ are in principle fixed by the initial expectation values $q^{(n)}_0=\braket{\psi(0)|Q^{(n)}|\psi(0)}/L$. The latter can be computed efficiently~\cite{fagotti_stationary_2013}. However, the determination of $\beta_n$ for all $n$ starting from known values of $q_0^{(n)}$ is difficult and in general not a viable route~\cite{fagotti_stationary_2013}. The problem has been solved using three alternative strategies:
\begin{enumerate}
    \item Whenever exact expressions for the overlaps $\braket{\psi(0)|\bs \lambda}$ are known, the quench action~\cite{caux_time_2013, caux_quench_2016} formalism can be employed to obtain $\bs \varrho_{\text{\tiny GGE}}(\lambda)$ from a generalized saddle-point analysis~\cite{caux_time_2013, de_nardis_solution_2014, wouters_quenching_2014, brockmann_quench_2014, pozsgay_correlations_2014} similar to the one that determines $\bs \varrho_{\beta}(\lambda)$. The quench action perspective demonstrates that $\bs \varrho_{\text{\tiny GGE}}(\lambda)$ represents the macrostate carrying the overwhelming thermodynamic weight in the eigenstate decomposition of $|\psi(0)\rangle$. Importantly, this method entirely bypasses the charges $Q^{(n)}$ and their initial values $q_0^{(n)}$, and therefore it can be employed also in cases in which the densities $q_0^{(n)}$ of the charges are not well defined. This is for example the case for the BEC-to-LL quench~\cite{de_nardis_solution_2014} considered in \cref{sec:LL}.
    \item It was shown in Refs.~\cite{wouters_quenching_2014, brockmann_quench_2014, ilievski_complete_2015, ilievski_stringcharge_2016, piroli_exact_2016} in the paradigmatic context of the XXZ chain that it is possible to bypass the computation of $\beta_n$ and obtain the root densities $\bs \varrho_{\text{\tiny GGE}}(\lambda)$ directly from the initial-state expectation value~\cite{fagotti_stationary_2013} of a set of generating functions $\{X_j(\lambda)\}$ of all the (quasi)local charges $Q^{(n)}$. This is achieved due to the remarkable fact that in the thermodynamic limit the implicit dependence of $\braket{\psi(0)|X_j(\lambda)|\psi(0)}$ on $\bs \varrho_{\text{\tiny GGE}}(\lambda)$ can be analytically inverted~\cite{ilievski_complete_2015, ilievski_stringcharge_2016}. 
    \item Refs.~\cite{piroli_from_2017, piroli_non_2018, piroli_integrable_2019a, piroli_integrable_2019b} devised a method to obtain $\bs \varrho_{\text{\tiny GGE}}(\lambda)$ via a quantum transfer matrix~\cite{klmper_thermodynamics_1993, suzuki_spinons_1999, klumper_integrability_2004} approach that bypasses the need for the overlaps $\braket{\psi(0)|\bs \lambda}$ or the construction of the charges $Q^{(n)}$. It is applicable for quenches from integrable initial states~\cite{piroli_what_2017} due to the fact that these preserve the integrability of the quantum transfer matrix at the boundary. 
\end{enumerate}

In this work we have made use of results obtained from all these approaches.

\section{Proof of singular behaviour of GGE entropy in XY chain}
\label{appendix:XYchainProof}

As discussed in \cref{sec:nonanbh}, the non-analytic behaviour of $s_{\text{\tiny YY}}[\varrho_{\text{\tiny GGE}}]$ at quantum critical points is inherited from the singular points at criticality of $\varrho_{\text{\tiny GGE}}(k)$. We now prove this fact analytically, despite an explicit expression for $s_{\text{\tiny YY}}[\varrho_{\text{\tiny GGE}}]$ is not available. For simplicity we focus on the quantum critical point $h  = 1$ along the Ising line, where generically we set $\gamma \neq 0$, $\gamma_0 \neq 0$, $h_0 \neq 1$. Other critical points can be treated similarly. 

Our starting point are the expressions for $n_{\text{\tiny GGE}}(k)= 2 \pi \varrho_{\text{\tiny GGE}}(k)$ from \cref{eq:rhospXY}, in particular
\begin{equation}
    n_{\text{\tiny GGE}}(k) = \frac{1-\Omega_k}{2} \ , \qquad \qquad  \Omega_k = \frac{\cos^2(k) + \gamma \gamma_0 \sin^2(k)+h h_0 - (h+h_0)\cos(k)}{\varepsilon_{\gamma,h}(k)\varepsilon_{\gamma_0,h_0}(k)/J^2} \ ,
\end{equation}
where $-1 \le \Omega_k \le 1$. Singular points of $s_{\text{\tiny YY}}[\varrho_{\text{\tiny GGE}}]$, if present, come from a small integration around $k = 0$ in \cref{eq:YYxy}, therefore we focus on the Yang-Yang integral restricted to a very small interval $[-k_0,k_0]$ with $k_0\ll1$. Indeed, the integral outside of this interval can only yield analytic contributions in $h$ due to the absence of singularities in $n_{\text{\tiny GGE}}(k)$ away from $k = 0$. Introducing the small parameter $\delta = h -1$, we expand both the numerator and denominator of $\Omega_k$ around $k = 0$ and $\delta = 0$
\begin{align}
\label{eq:expansion1}
    \varepsilon_{\gamma,h}(k)/J &= \sqrt{\delta^2 + \gamma^2 k^2} \cdot \sqrt{1+\Delta}  \ , \qquad \qquad \Delta = \frac{\delta k^2 + \sigma k^4 + \mathcal{O}(k^6)}{\delta^2 + \gamma^2 k^2} \ , \\
    \varepsilon_{\gamma_0,h_0}(k)/J &= |h_0 - 1|+\mathcal{O}(k^2) \ , \\
    \text{numerator}(\Omega_k) &= \delta (h_0 - 1)+\frac{h_0 - 1 + 2 \gamma \gamma_0}{2}k^2 +\frac{\delta}{2}k^2 + \mathcal{O}(k^4) \ ,
\end{align}
where $\sigma = \frac{1}{4}-\frac{\gamma^2}{3} - \frac{\delta}{12}$ in \cref{eq:expansion1}. Given that $\Delta$ vanishes in the limit $\delta \to 0$ at $k$ finite or $k \to 0$ at $\delta$ finite, and in the scaling limit $k \sim \delta \to 0$, its appearance in the denominator can be Taylor expanded and this leads us to
\begin{equation}
    \Omega_k = \frac{c_1 \delta + c_2 k^2 + c_3 \delta k^2 + \mathcal{O}(k^4)}{\sqrt{\delta^2 +\gamma^2 k^2}} \, \big(1 + \mathcal{O}(k^2)\big) \, \big(1-\Delta/2 + \mathcal{O}(\Delta^2)\big) \ .
\end{equation}
Here the constants $c_i$ do not depend on $k$ or $\delta$. We now consider the Yang-Yang integrand function $g(n)=-n\ln n -(1-n)\ln(1-n)$ from \cref{eq:rhospXY}, which is analytic around $n = 1/2$ in the full fermionic range $n \in [0,1]$, from which we obtain
\begin{equation}
\label{eq:expansiongXY}
    g\left(\frac{1-\Omega_k}{2}\right) = \ln 2 - \sum_{m = 1}^\infty \frac{\Omega_k^{2m}}{2m(2m-1)} \ .
\end{equation}
The leading order singularity in $s_{\text{\tiny YY}}[\varrho_{\text{\tiny GGE}}]$ comes from terms in \cref{eq:expansiongXY} that give rise to the following known integrals (here $m\ge 1$)
\begin{equation}
    \int_{-k_0}^{k_0}dk \, \frac{\delta^{2m}}{(\delta^2+\gamma^2 k^2)^m} = a_m |\delta|+ \text{analytic}(\delta)  \qquad \qquad \delta \neq 0\ ,
\end{equation}
where $a_m$ are $m$-dependent constants (for example $a_1 = \pi/|\gamma|, a_2 = \pi/(2|\gamma|),a_3 = 3\pi /(8|\gamma|), \ldots$) and $\text{analytic}(\delta)$ indicates non-singular contributions coming from some simple rational functions of $\delta, \gamma$ and $k_0$, and from the smooth term $\delta\arctan[\delta/(k_0 \gamma)]$. All the other terms in \cref{eq:expansiongXY}
 give rise to singularities of higher order or to analytic contributions in $\delta$. For example, again for $m\ge1$, we get
 \begin{align}
    \int_{-k_0}^{k_0}dk \, \frac{k^{4m}}{(\delta^2+\gamma^2 k^2)^m} &= b_m |\delta|^{2m+1}+ \text{analytic}(\delta)  \qquad \qquad \delta \neq 0\ , \\
    \int_{-k_0}^{k_0}dk \, \frac{k^{2m}\delta^m}{(\delta^2+\gamma^2 k^2)^m} &= d_m \delta^{m}|\delta|+ \text{analytic}(\delta)  \qquad \qquad \ \, \, \, \delta \neq 0\ , 
 \end{align}
where $b_m$ and $d_m$ are $m$-dependent constants. This proves that the expansion of $s_{\text{\tiny YY}}(h)$ (now denoted as a function of the transverse field) around $h = 1$ is singular, and takes the form (we denote with $A_i$ coefficients that are $h$-independent)
 \begin{equation}
     s_{\text{\tiny YY}}(h) = \text{analytic}(h) + A_1 |h-1| + A_2 (h-1)|h-1|+A_3 |h-1|^3 +\ldots \ .
 \end{equation}

\begin{figure}[!h]
    \centering
    \begin{minipage}{0.3\textwidth}
        \includegraphics[width=\linewidth]{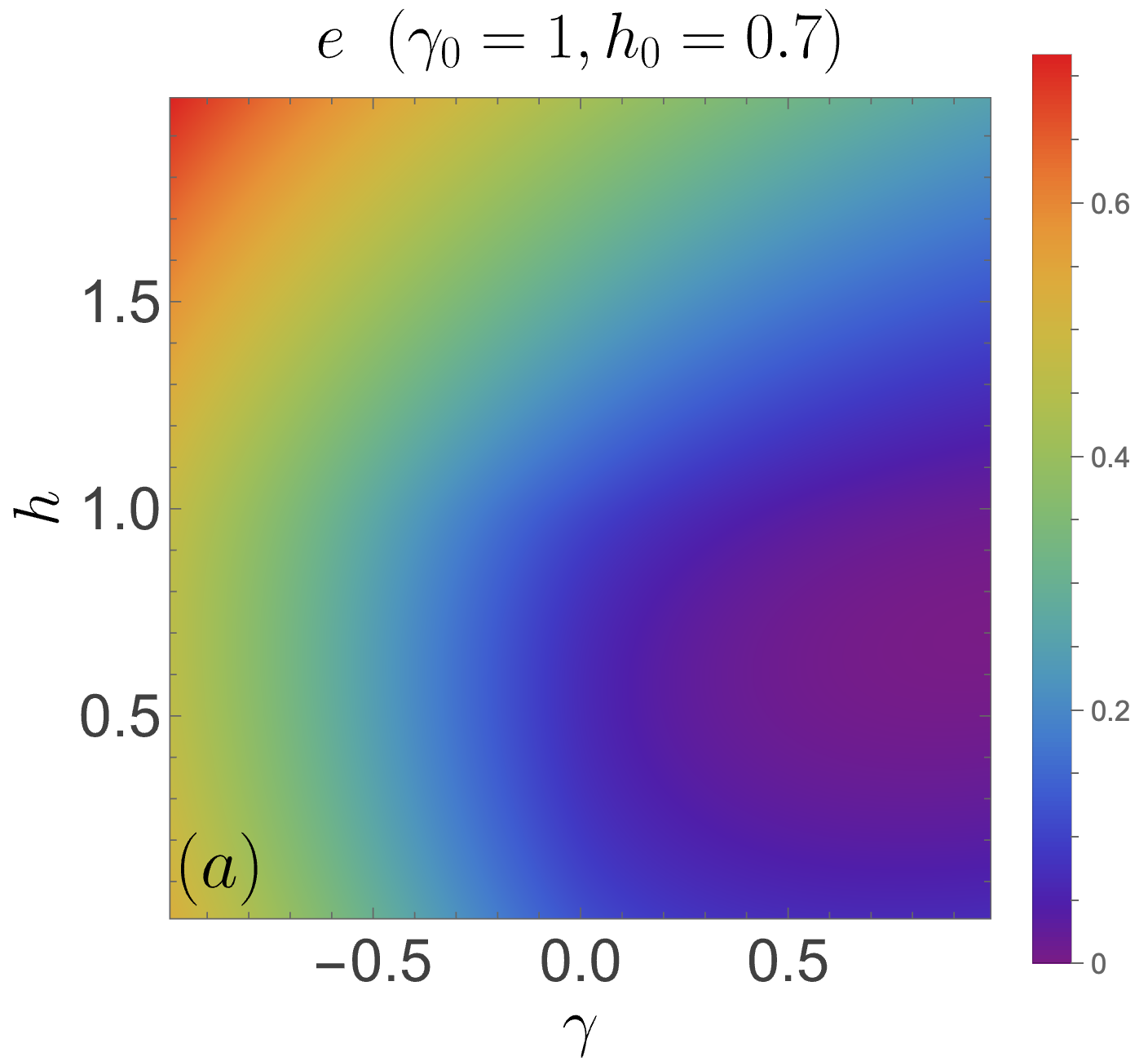}
    \end{minipage}\hfill
    \begin{minipage}{0.3\textwidth}
        \includegraphics[width=\linewidth]{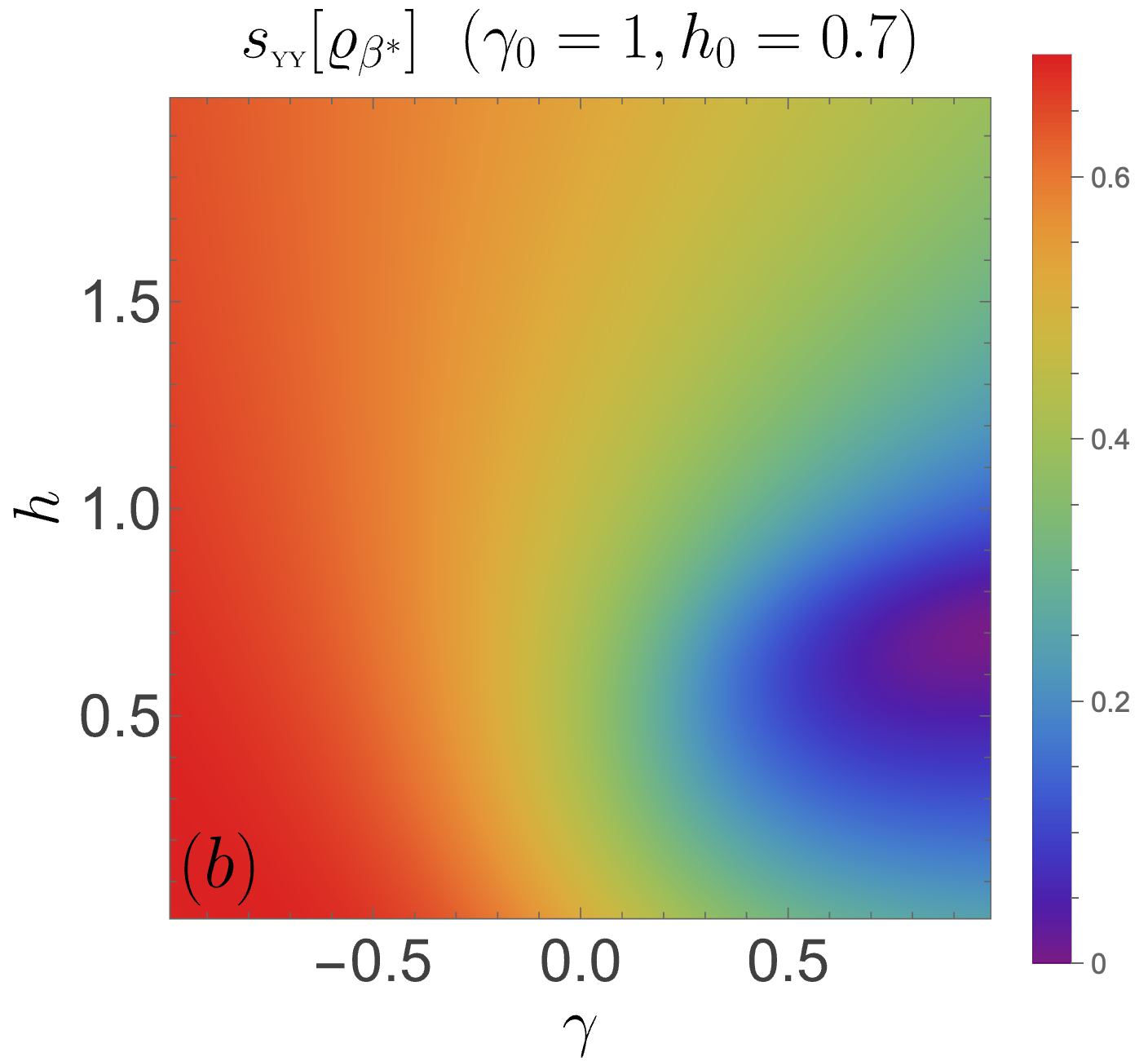}
    \end{minipage}\hfill
    \begin{minipage}{0.3\textwidth}
        \includegraphics[width=\linewidth]{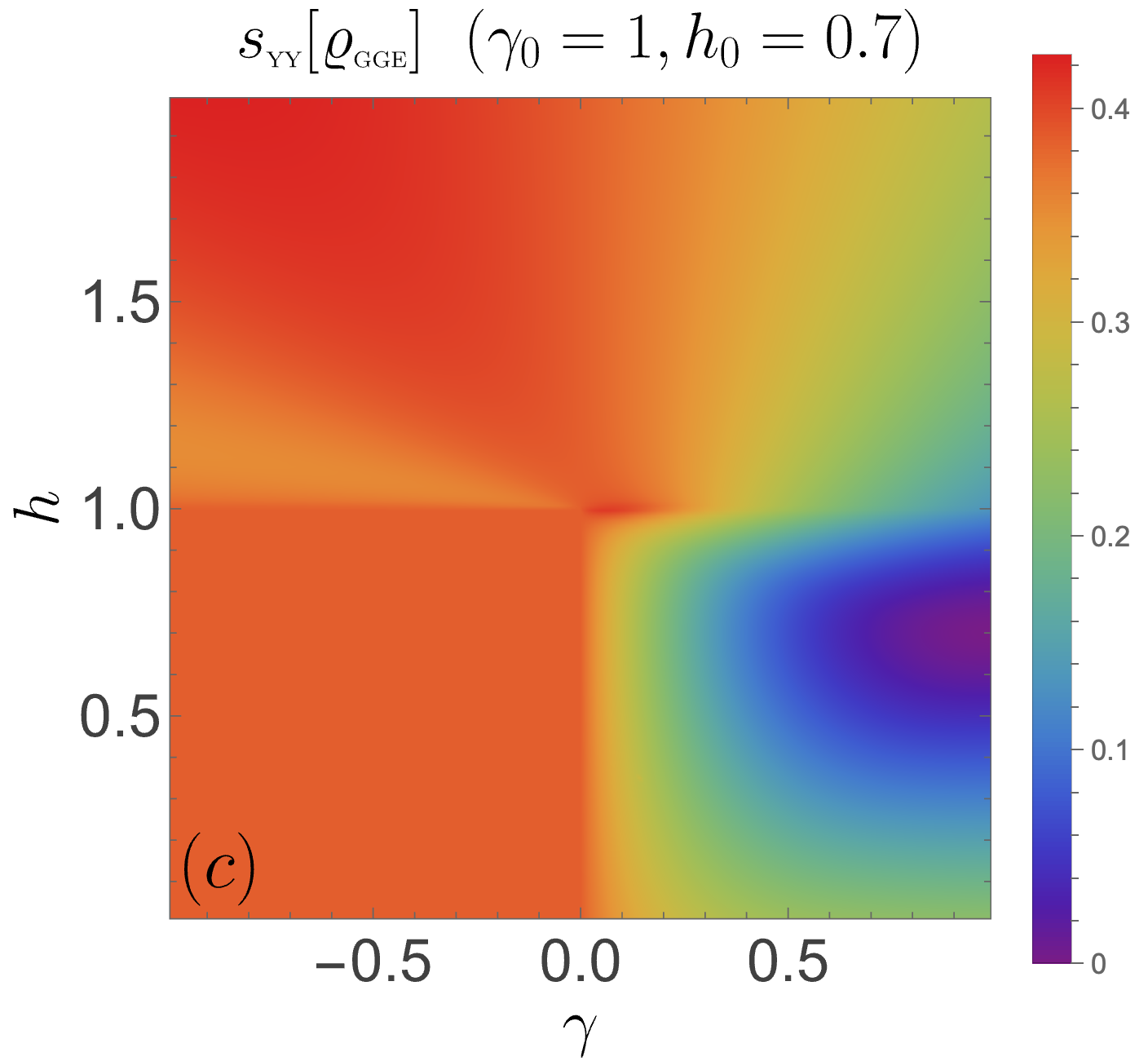}
    \end{minipage}

    \vspace{1em} 

    \begin{minipage}{0.3\textwidth}
        \includegraphics[width=\linewidth]{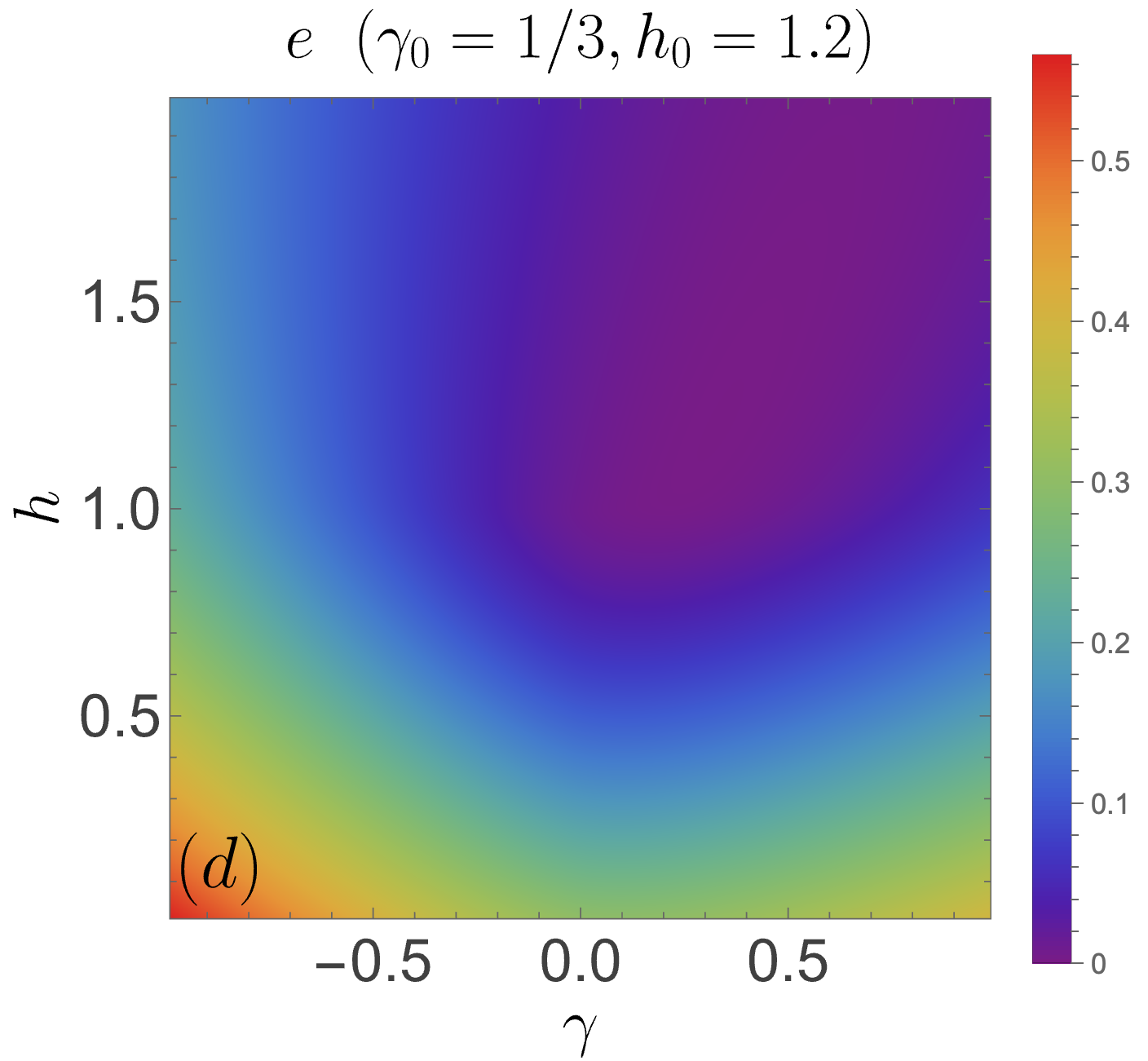}
    \end{minipage}\hfill
    \begin{minipage}{0.3\textwidth}
        \includegraphics[width=\linewidth]{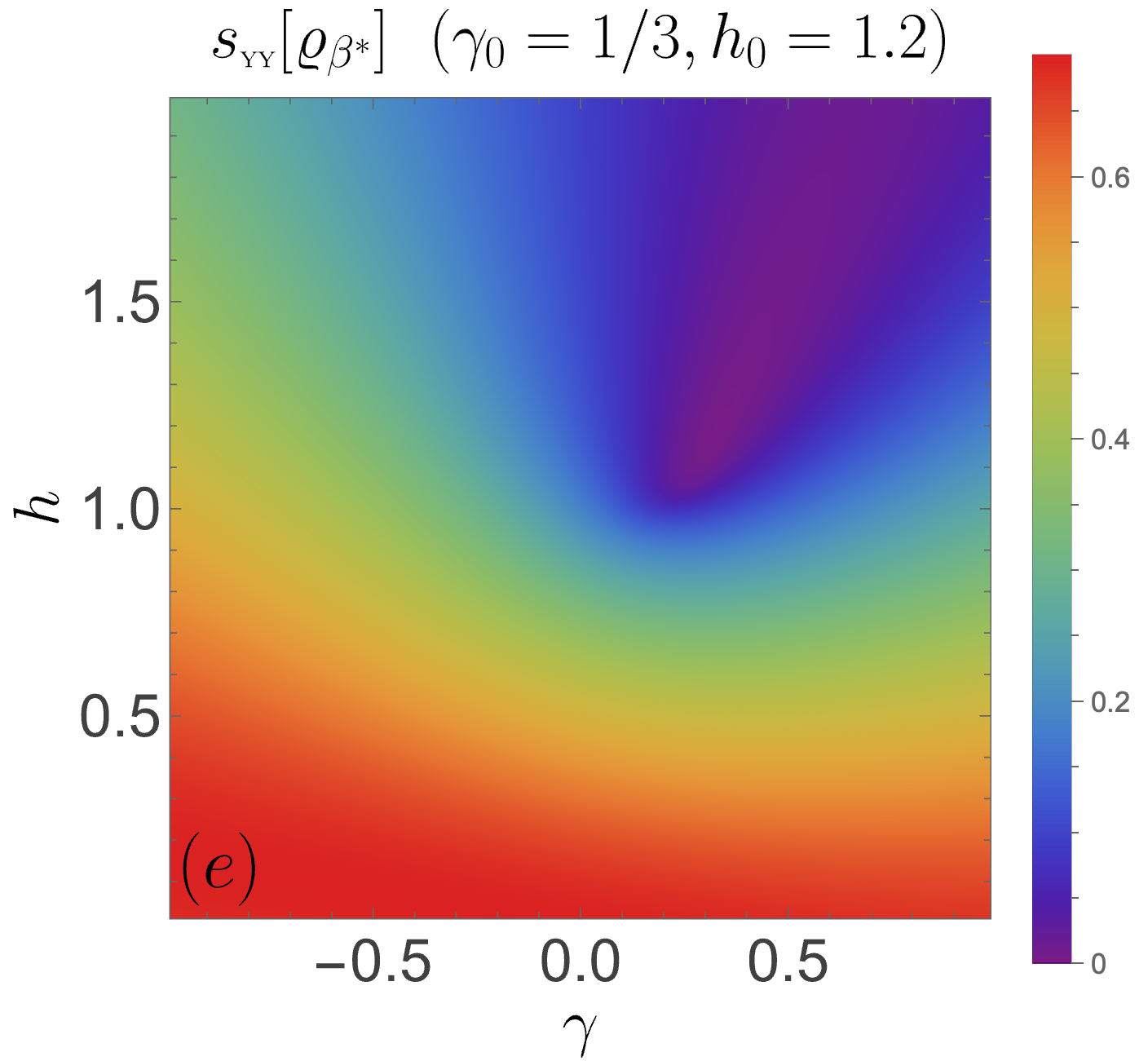}
    \end{minipage}\hfill
    \begin{minipage}{0.3\textwidth}
        \includegraphics[width=\linewidth]{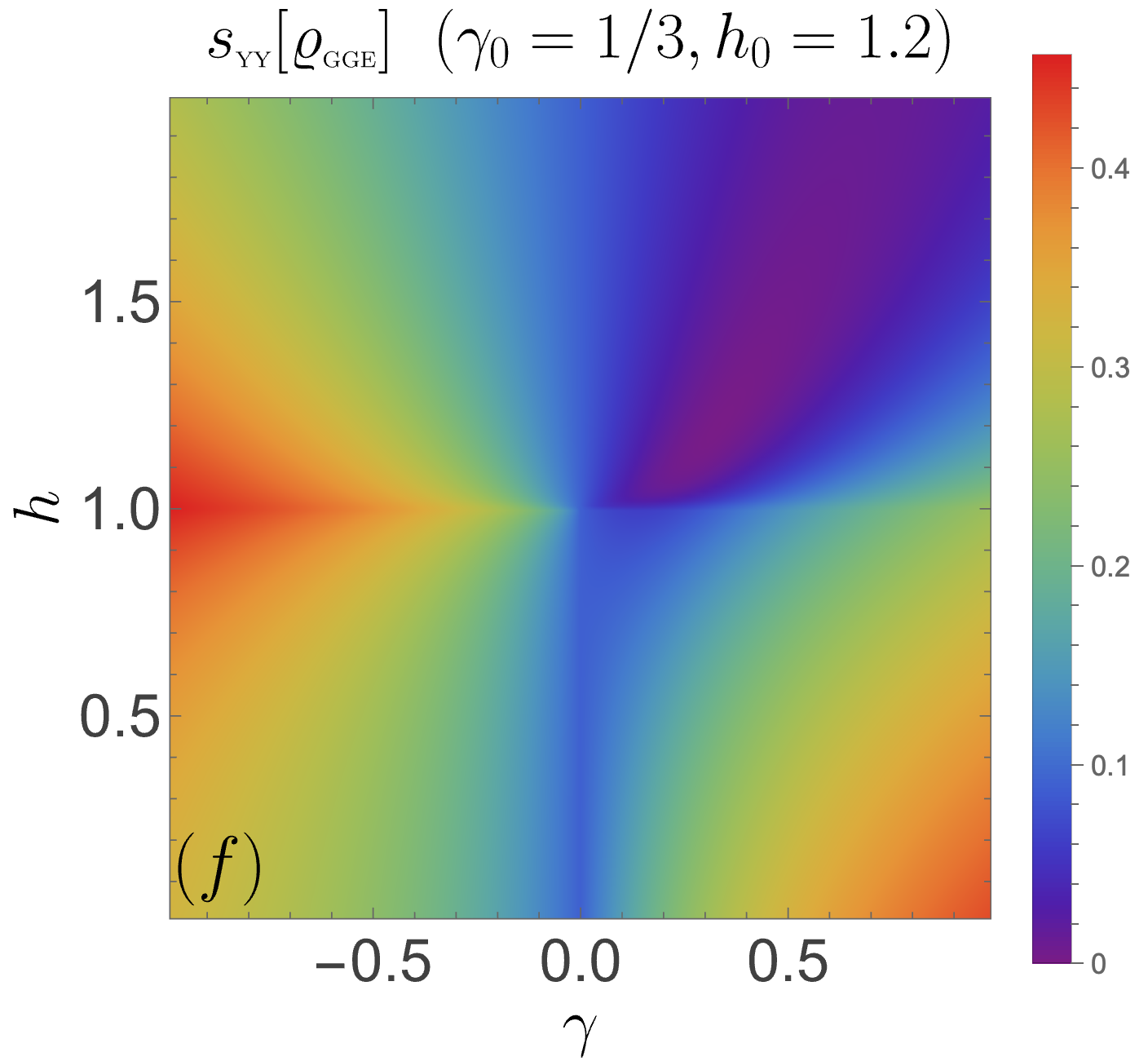}
    \end{minipage}

    \vspace{1em}

    \begin{minipage}{0.3\textwidth}
        \includegraphics[width=\linewidth]{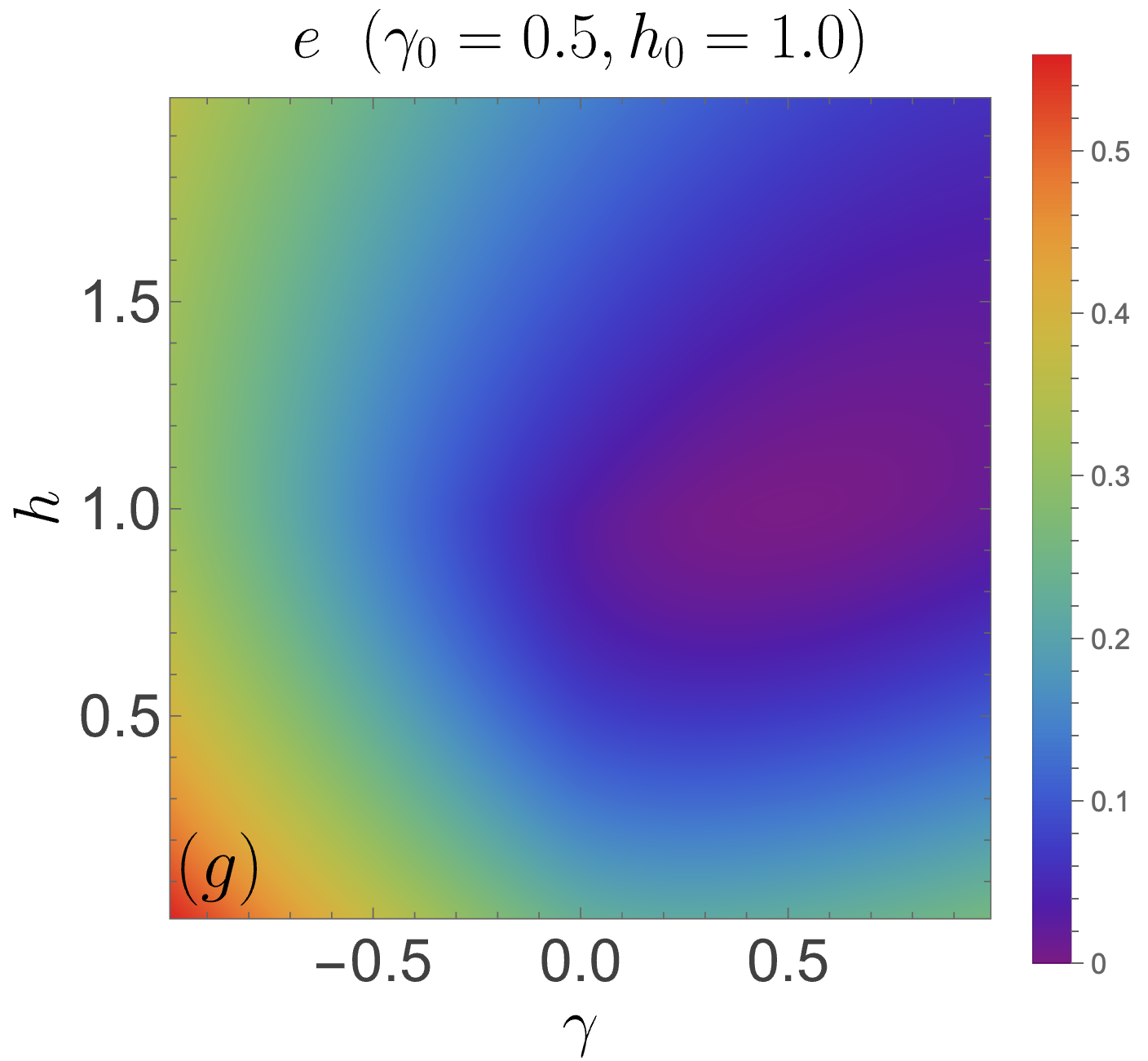}
    \end{minipage}\hfill
    \begin{minipage}{0.3\textwidth}
        \includegraphics[width=\linewidth]{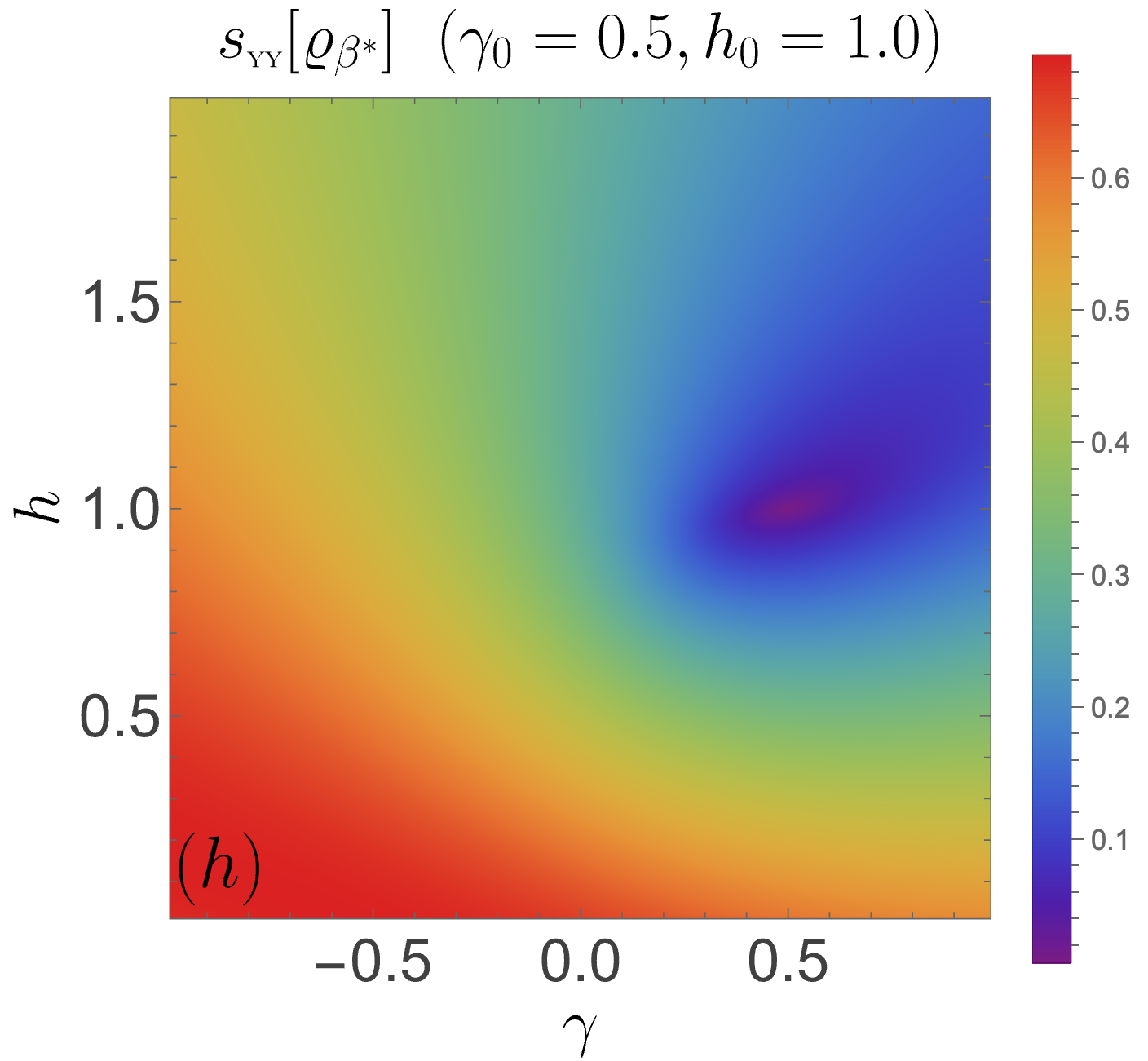}
    \end{minipage}\hfill
    \begin{minipage}{0.3\textwidth}
        \includegraphics[width=\linewidth]{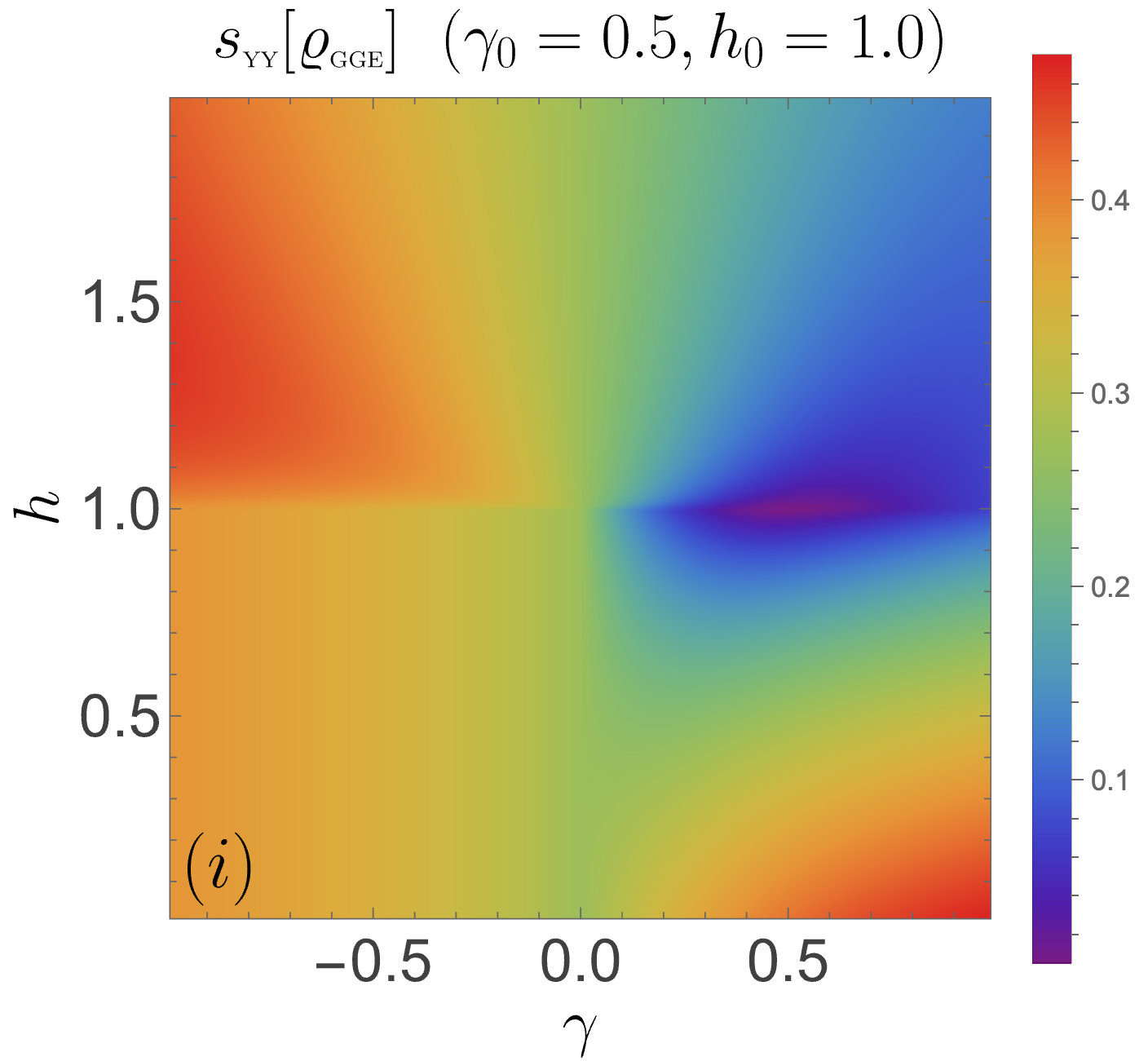}
    \end{minipage}

    \caption{Density plots for $e$, $s_{\text{\tiny YY}}[\varrho_{\beta^*}]$ and $s_{\text{\tiny YY}}[\varrho_{\text{\tiny GGE}}]$, associated with the athermality plots of \cref{fig:main_XY_DP}. (a)-(c) Subplots associated with \cref{fig:main_XY_DP}(a). (d)-(f) Subplots associated with \cref{fig:main_XY_DP}(b). (g)-(i) Subplots associated with \cref{fig:main_XY_DP}(c). In computing the energy density $e$ we have set the ground state energies $E_{\rm e}$ and $E_{\rm o}$ from \cref{eq:XYHdiag} to zero.}
    \label{fig:XY_e_sGE_sGGE}
\end{figure}

\section{Additional results for XY chain}
\label{appendix:XYchain}

\begin{figure}[t!]
    \centering
    \includegraphics[width=1\linewidth]{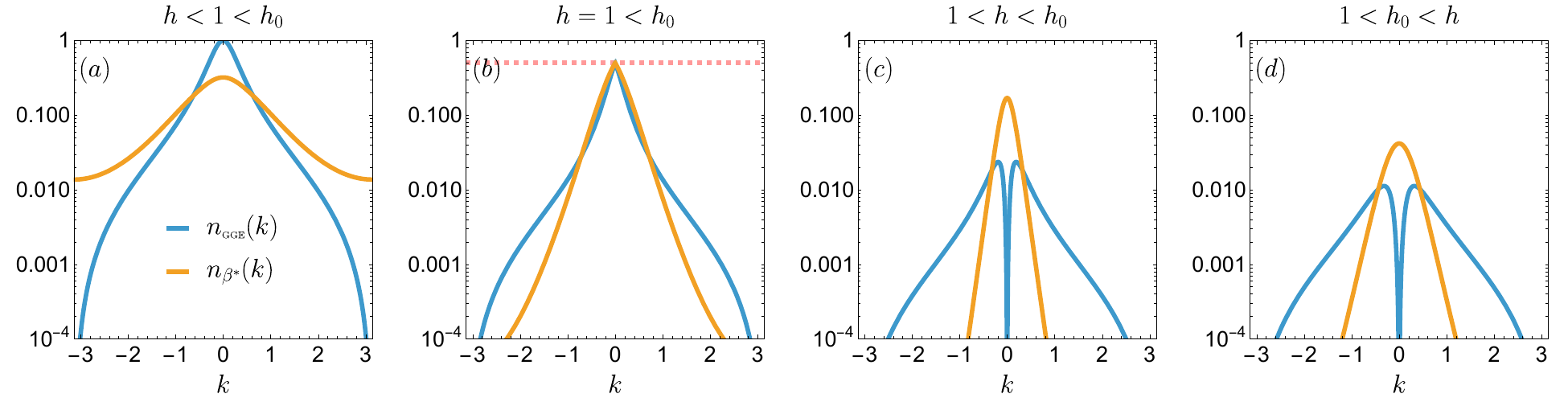}
    \caption{Log-scale comparison between the occupation functions $n_{\text{\tiny GGE}}(k)$ and $n_{\beta^*}(k)$, for different choices of $h$ in the TFIC ($\gamma = \gamma_0=1$) with $h_0=1.3$: (a) $h = 0.7$; (b) $h = 1$ (critical), with red dashed line indicates the pinning at $1/2$; (c) $h = 1.15$;  (d) $h = 1.5$.}
    \label{fig:root_densities_2}
\end{figure}

In \cref{fig:XY_e_sGE_sGGE} we show density plots for the postquench energy density $e$, Gibbs entropy density $s_{\text{\tiny YY}}[\varrho_{\beta^*}]$ and GGE entropy density $s_{\text{\tiny YY}}[\varrho_{\text{\tiny GGE}}]$ related to the athermality plots of \cref{fig:main_XY_DP}. It is evident that there is little correlation between the athermality density $\alpha$ from \cref{fig:main_XY_DP}, or $s_{\text{\tiny YY}}[\varrho_{\text{\tiny GGE}}]$ from \cref{fig:XY_e_sGE_sGGE}(c), (f) and (i), and the injected energy density $e$ of \cref{fig:XY_e_sGE_sGGE}(a), (d) and (g). This confirms that $\alpha$ and $s_{\text{\tiny YY}}[\varrho_{\text{\tiny GGE}}]$ capture interesting extra features of the quench at late times. \cref{fig:XY_e_sGE_sGGE}(c), (f) and (i) also showcase the non-analytic behaviour of $s_{\text{\tiny YY}}[\varrho_{\text{\tiny GGE}}]$ at the critical lines, analytically proved in the previous subsection. \\

\cref{fig:root_densities_2} shows log-plots for the occupations functions $n_{\text{\tiny GGE}}(k)$ and $n_{\beta^*}(k)$ in TFIC, similarly to those of \cref{fig:root_densities_1}, but here starting from an initial state in the paramagnetic phase $h_0 > 1$ (as opposed to $h_0<1$ of \cref{fig:root_densities_1}). The conclusions regarding the singularities and the pinning mechanism are identical to those in \cref{sec:nonanbh}. \\

\begin{figure}[t!]
    \centering
    \begin{minipage}{0.32\textwidth}
        \includegraphics[width=\linewidth]{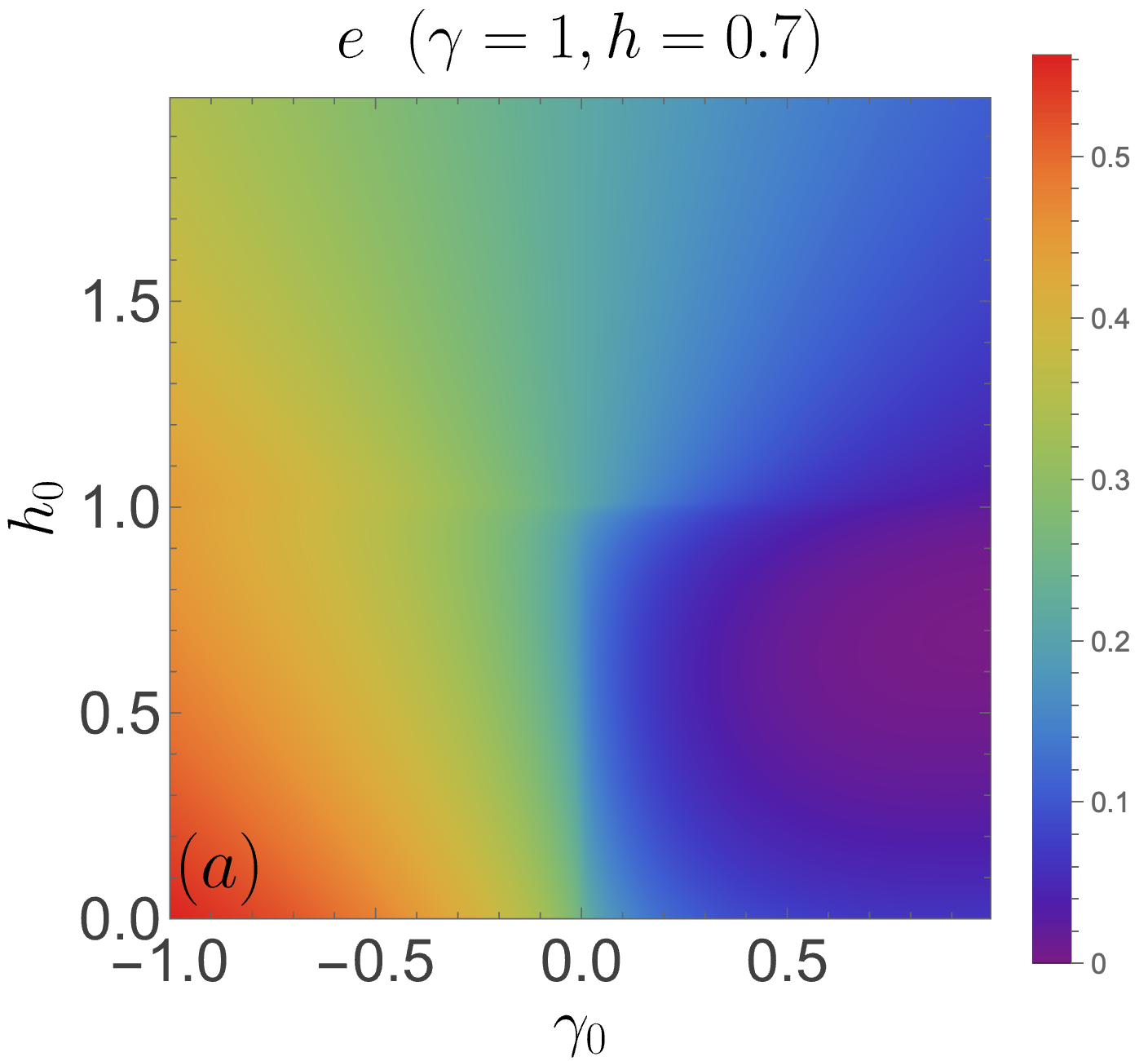}
    \end{minipage}\hfill
    \begin{minipage}{0.32\textwidth}
        \includegraphics[width=\linewidth]{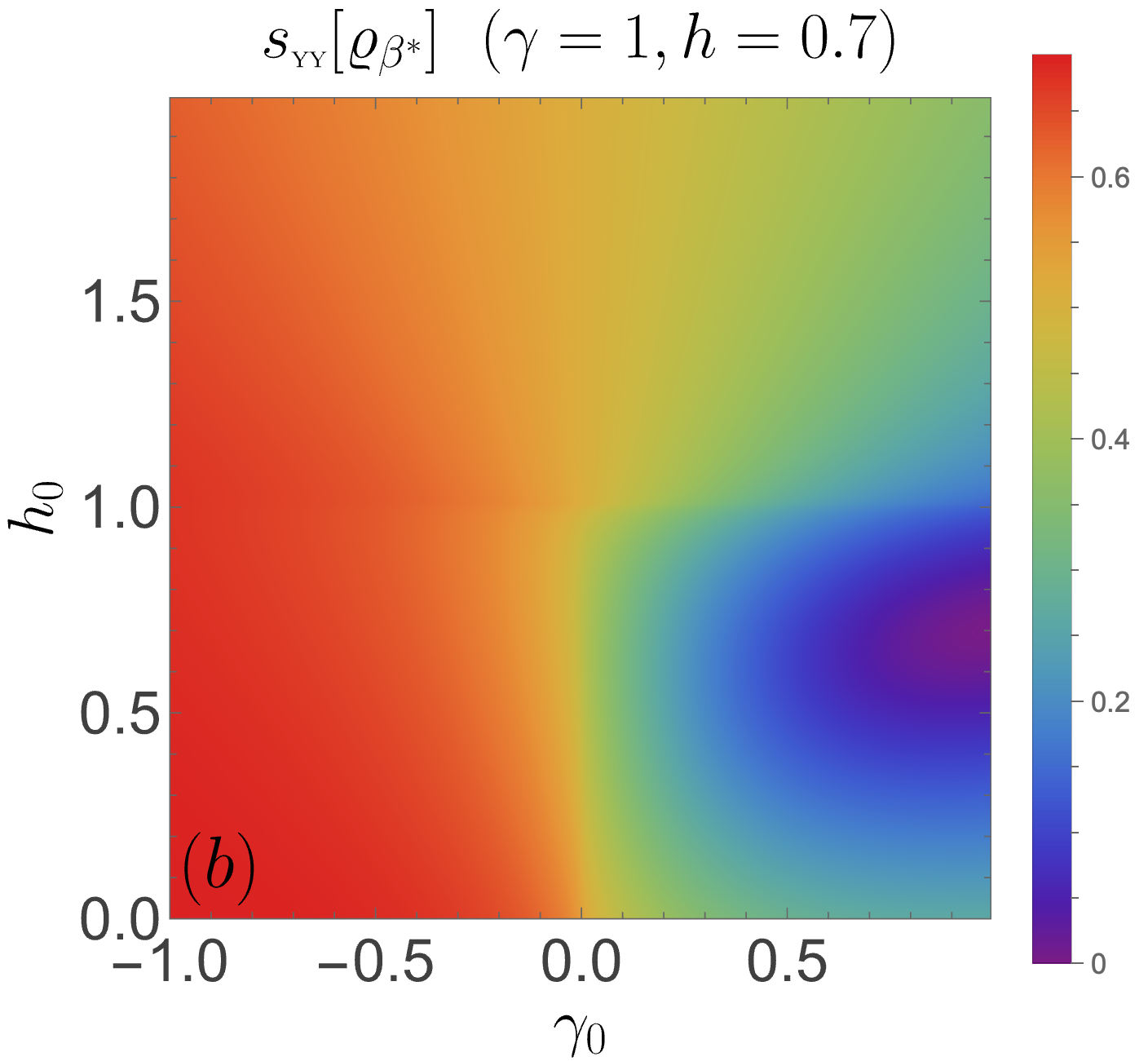}
    \end{minipage}\hfill
    \begin{minipage}{0.32\textwidth}
        \includegraphics[width=\linewidth]{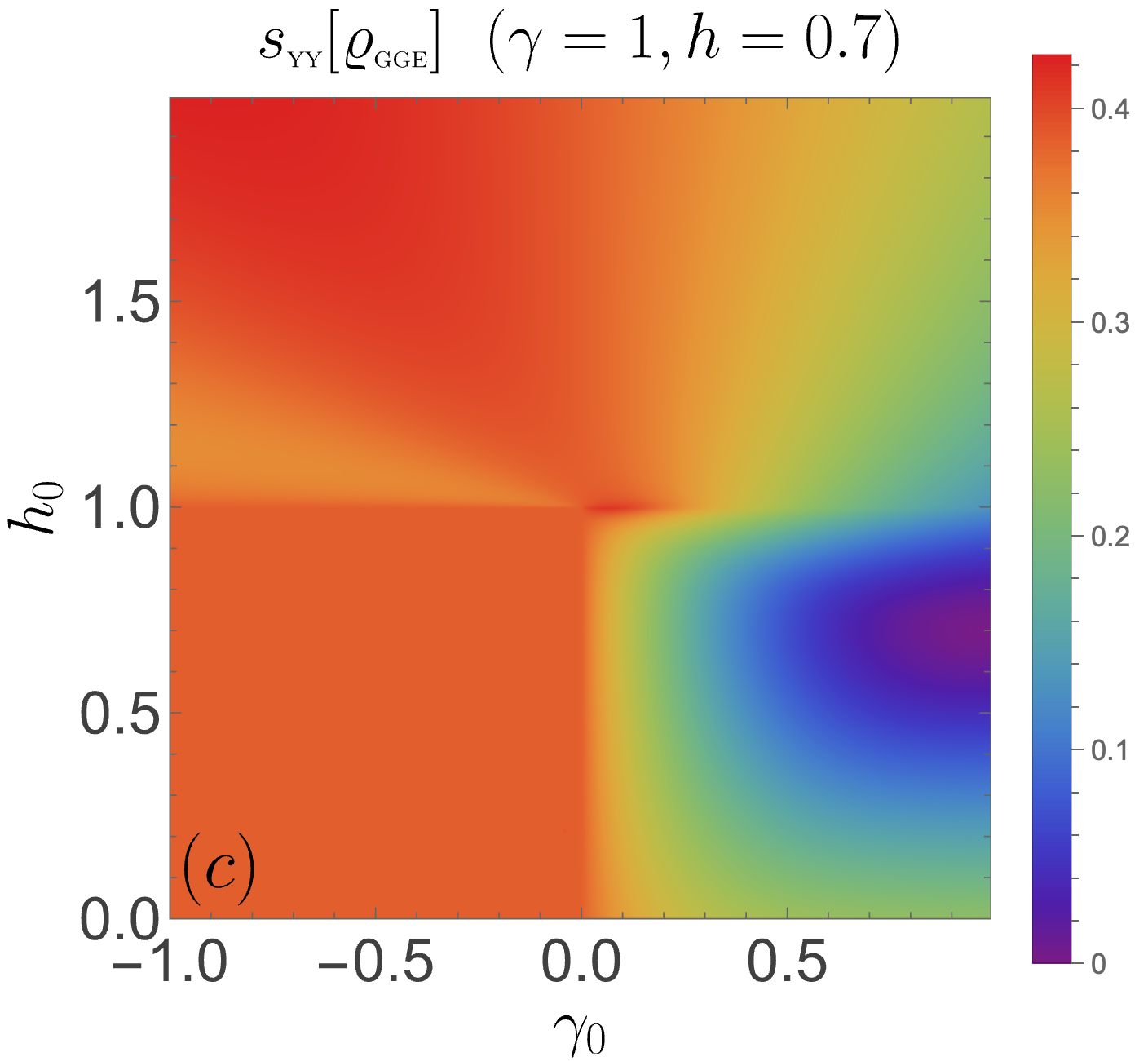}
    \end{minipage}

    \vspace{1em} 

    \begin{minipage}{0.32\textwidth}
        \includegraphics[width=\linewidth]{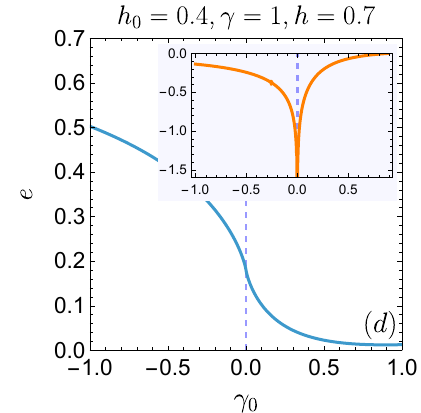}
    \end{minipage}\hfill
    \begin{minipage}{0.32\textwidth}
        \includegraphics[width=\linewidth]{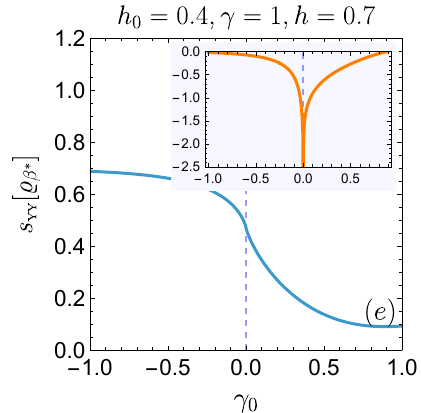}
    \end{minipage}\hfill
    \begin{minipage}{0.32\textwidth}
        \includegraphics[width=\linewidth]{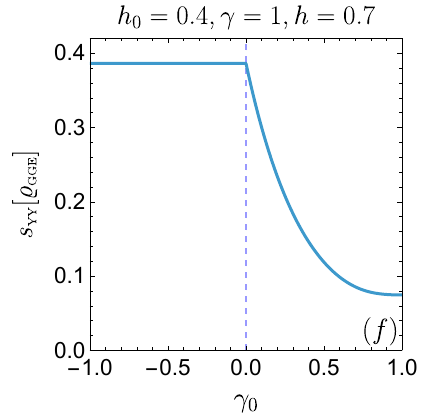}
    \end{minipage}

    \vspace{1em}

    \caption{Density plots and cuts for $e$, $s_{\text{\tiny YY}}[\varrho_{\beta^*}]$ and $s_{\text{\tiny YY}}[\varrho_{\text{\tiny GGE}}]$, associated with the athermality plots of \cref{fig:main_XY_DP_2}(a). The subplots (d), (e) and (f) correspond,  respectively, to horizontal cuts in the density subplots (a), (b) and (c). Insets in (d) and (e) show numerically computed derivatives of the curves in the main plot. In computing the energy density $e$ we set $E_{\rm e}$ and $E_{\rm o}$ from \cref{eq:XYHdiag} to zero.}
    \label{fig:XY_e_sGE_sGGE_2}
\end{figure}

\cref{fig:XY_e_sGE_sGGE_2} shows density plots and cuts for $e$, $s_{\text{\tiny YY}}[\varrho_{\beta^*}]$ and $s_{\text{\tiny YY}}[\varrho_{\text{\tiny GGE}}]$ as a function of $\gamma_0$ and $h_0$, for the same postquench parameters of the athermality plot in \cref{fig:main_XY_DP_2}(a). As discussed in \cref{sec:XYvaryingH}, now also $e$ and $s_{\text{\tiny YY}}[\varrho_{\beta^*}]$ present singularities as a function of $\gamma_0$ and $h_0$, as evident in particular from the cuts in \cref{fig:XY_e_sGE_sGGE_2}(d) and (e).

\section{TBA and additional results for XXZ chain}
\label{appendix:XXZ}

Eigenstates of the XXZ chain \eqref{eq:HXXZ} with magnetization $\sigma_{\rm tot}^z=L-2N$ are constructed via the Bethe ansatz as plane-wave superpositions of product states hosting $N$ spin flips (particles) with respect to the fully polarized state~\cite{korepin_quantum_1993, takahashi_thermodynamics_1999}. The regimes $\Delta > 1$ and $0<\Delta < 1$ differ significantly in the structure of the Bethe ansatz~\cite{takahashi_thermodynamics_1999}, and we analyze them separately.

\subsection{TBA for $\Delta > 1$}
The $N$ complex momenta $k_j$ entering the plane-wave expansion of the Bethe wavefunction are related to the rapidities $\lambda_j$ by
\begin{equation}
    e^{i k_j} = \frac{\sin\left(\lambda_j + i \phi/2  \right)}{\sin\left(\lambda_j - i \phi/2  \right)} \ , \qquad \qquad \cosh(\phi)=\Delta > 1 \, , \ \phi>0 \ .
\end{equation}
The solutions $\bs \lambda$ of the Bethe equations corresponding to normalizable wavefunctions with $N$ particles include strings patterns of $n_\ell$ particles
\begin{equation}
\label{eq:stringHXXZDeltag1}
    \lambda_j = \lambda^{(0)}+i\frac{\phi}{2}(n_\ell + 1 - 2 j) + \delta_j \qquad \qquad j = 1, \ldots, n_\ell \ ,
\end{equation}
where $n_\ell\in[1,N]$, $\lambda^{(0)}$ is the real string center belonging to the interval $(-\pi/2,\pi/2]$, and $\delta_j$ are exponentially small (with system size) corrections to the perfect string pattern (that can be neglected in the thermodynamic treatment). The case of $n_\ell = 1$ corresponds to unbound magnon excitations, while all strings $n_\ell\ge2$ give rise to bound states. 
In TBA a macrostate is specified by the particle and hole densities $\varrho_\ell(\lambda)$ and $\varrho_\ell^{(h)}(\lambda)$ associated with all $n_\ell$-particle strings (we fix the convention $n_\ell = \ell$), where $\lambda$ refers to the center of the string in \cref{eq:stringHXXZDeltag1}. The two densities are related by the following non-linear integral equations~\cite{takahashi_thermodynamics_1999}
{\allowdisplaybreaks
\begin{align}
\label{eq:rhorhohXXZDg1}
    &\varrho_\ell(\lambda) + \varrho_\ell^{(h)}(\lambda) = a_\ell(\lambda) - \bigg[\sum_{j=1}^N T_{\ell j}   \star  \varrho_j\bigg](\lambda) \ ,  \qquad \qquad  a_\ell(\lambda) = \frac{1}{\pi} \frac{\sinh(\ell\phi)}{\cosh(\ell\phi) - \cos(2\lambda)}  \ , \\
    & T_{\ell j}(\lambda) = (1 - \delta_{\ell,j}) a_{|\ell -j|}(\lambda) + 2a_{|\ell -j|+2}(\lambda) + 2a_{|\ell -j|+4}(\lambda)+ \ldots + 2a_{\ell +j-2}(\lambda) + a_{\ell +j}(\lambda) \ ,
\end{align}
}
\noindent
where we denoted with $\star$ the convolution
\begin{equation}
[g \star h](y) = \int_{-\pi/2}^{\pi/2} \mathrm{d}x \, g(y - x) h(x) \ .
\end{equation}
The thermal macrostate $\bs \varrho_{\beta}$ at finite inverse temperature $\beta$ is obtained, using \cref{eq:rhorhohXXZDg1}, from the following set of coupled integral equations in terms of $\eta_\ell(\lambda) = \varrho_\ell^{(h)}(\lambda)/\varrho_\ell(\lambda)$~\cite{takahashi_thermodynamics_1999, bertini_lowtemperature_2018,Bertini_universal_2018}
{\allowdisplaybreaks
\begin{align}
\ln \eta_\ell(\lambda) &= - J\beta \, \pi \sinh(\phi) s(\lambda)\delta_{\ell,1} + \Big[s \star \ln\big[(1 + \eta_{\ell-1})(1 + \eta_{\ell+1})\big]\Big](\lambda), \quad \ell \geqslant 1 \ , \\
\eta_0(\lambda) &\equiv 0 \ , \qquad \qquad \qquad \lim_{\ell \to \infty} \frac{\ln \eta_\ell(\lambda)}{\ell} = 0 \ , \\
s(\lambda) &= \frac{1}{2\pi} \sum_{j=-\infty}^{\infty} \frac{\mathrm{e}^{2ij\lambda}}{\cosh(j\phi)} = \frac{1}{2 \phi} \sum_{j = -\infty}^{\infty} {\rm sech} \bigg[ \frac{\pi(\lambda - j/2)}{\phi}\bigg] \ .
\label{eq:sfun}
\end{align}
}
\noindent
The previous sets of integral equations involving the periodic functions (with period $\pi$) $a_\ell(\lambda)$ and $s(\lambda)$ can be numerically solved in an efficient way by working in Fourier space (Fourier series and associated convolution theorem). In practice we use the trapezoidal rule to discretize the interval $(-\pi/2,\pi/2]$ and employ the Fast Fourier Transform (FFT). 
The GE root densities $\bs \varrho_{\beta}(\lambda)$ obtained from the integral equations above automatically feature the correct density of particles
\begin{equation}
    \frac{1}{2} = d = \sum_{\ell = 1}^\infty \ell \int_{-\pi/2}^{\pi/2} d\lambda \, \varrho_{\ell}(\lambda) \ .
\end{equation}
The value of $\beta^*$ is set by equating the GE energy density to the energy density $e$ set by the initial state $\ket{\psi(0)}$
\begin{equation}
   e = \sum_{\ell = 1}^\infty \int_{-\pi/2}^{\pi/2} d\lambda \, \varrho_{\ell}\, \epsilon_\ell(\lambda) \ , \qquad \qquad \qquad \epsilon_\ell(\lambda) = - \pi J \sinh(\phi) a_\ell(\lambda)  \ .
\end{equation}

\subsection{TBA for $0<\Delta < 1$}

In the gapless regime the $N$ complex momenta $k_j$ in the Bethe wavefunction are related to rapidities $\lambda_j$ by 
\begin{equation}
    e^{i k_j} = \frac{\sinh\left(\lambda_j + i \gamma/2  \right)}{\sinh\left(\lambda_j - i \gamma/2  \right)} \ , \qquad \qquad 0<\cos(\gamma)=\Delta<1 , \, \ \gamma \in (0, \pi/2)\ .
\end{equation}
For the choice $\gamma = \pi/(p+1)$ from \cref{eq:Deltam1} the $\ell$-th allowed string type, containing $n_\ell$ particles, is
\begin{equation}
\label{eq:stringHXXZDeltal1}
    \lambda_j = \lambda^{(0)}+i\frac{\gamma}{2}(n_\ell + 1 - 2 j) + i \frac{\pi}{4}(1-v_\ell) + \delta_j \qquad \qquad j = 1, \ldots, n_\ell, \quad  v_\ell = \bigg\{
    \begin{aligned} 
    +1 \ \ &{\rm if} \ \  n_\ell > 1 \ , \\
    \pm 1 \ \  &{\rm if} \ \  n_\ell = 1 \ .\ 
    \end{aligned}
\end{equation}
Here: $\ell = 1, \ldots, p+1$; $v_\ell$ is known as parity; we choose the convention $n_\ell = \ell$ for $\ell \le p$ and $n_{p+1} = 1$; the real string centers $\lambda^{(0)}$ live on the full real line $(-\infty,\infty)$; $\delta_j$ are again negligible exponentially suppressed corrections.

The TBA equations analogous to \cref{eq:rhorhohXXZDg1} are~\cite{takahashi_thermodynamics_1999}
{\allowdisplaybreaks
\begin{align}
\label{eq:rhorhohXXZDl1}
    \varrho_\ell(\lambda) + &\varrho_\ell^{(h)}(\lambda) = a_\ell(\lambda) - \bigg[\sum_{j=1}^{p+1} T_{\ell j}   \star  \varrho_j\bigg](\lambda) \ ,\\
    a_\ell(\lambda) &= \frac{1}{\pi} \frac{\sin(\gamma n_\ell)}{v_\ell \cosh(2\lambda) -  \cos(\gamma n_\ell)} \equiv a_{n_\ell}^{(v_\ell)}(\lambda)  \ , \\
     T_{\ell j}(\lambda) &= (1 - \delta_{n_\ell, n_j}) a_{|n_\ell -n_j|}^{(v_\ell v_j)}(\lambda) + 2a_{|n_\ell -n_j|+2}^{(v_\ell v_j)}(\lambda) + \ldots + 2a_{n_\ell +n_j-2}^{(v_\ell v_j)}(\lambda) + a_{n_\ell +n_j}^{(v_\ell v_j)}(\lambda) \ ,
\end{align}
}
\noindent
where now $\star$ denotes the convolution
\begin{equation}
[g \star h](y) = \int_{-\infty}^{\infty} \mathrm{d}x \, g(y - x) h(x) \ .
\end{equation}
Also in this case it is possible to rewrite \cref{eq:rhorhohXXZDl1} in a decoupled form~\cite{takahashi_thermodynamics_1999, bertini_nonequilibrium_2023} which is convenient for the numerical solution. 
The thermal macrostate $\bs \varrho_{\beta}(\lambda)$ is obtained, using \cref{eq:rhorhohXXZDl1}, from the set~\cite{takahashi_thermodynamics_1999}
{\allowdisplaybreaks
\begin{equation}
\begin{aligned}
    \ln[1+\eta_0(\lambda)] &\equiv \pi\beta J \sin(\gamma) \delta(\lambda) \ , \\
    \ln \eta_j &= s \star \ln(1+\eta_{\ell-1}) + s \star \ln(1+\eta_{\ell+1})  \quad \qquad \text{for } 1 \le \ell \le p-2, \\
    \ln \eta_{p-1} &= s \star \ln(1+\eta_{p-2}) + s \star \ln\big[(1+\eta_p)(1+\eta_{p+1}^{-1})\big] \ , \\
    \ln \eta_p &=  - \ln \eta_{p+1} = s \star \ln(1+\eta_{p-1}) \ ,
\end{aligned}
\end{equation}
}
\noindent
where 
\begin{equation}
    s(\lambda) = \frac{1}{2\gamma \cosh(\pi \lambda/\gamma)} \ .
\end{equation}
The GGE macrostate $\bs \varrho_{\text{\tiny GGE}}(\lambda)$ associated with the Néel initial state, i.e.~\cref{eq:tiltedNeel} for $\theta = 0$, is obtained by combining \cref{eq:rhorhohXXZDl1} with the following recursive set of equations~\cite{piroli_non_2018, bertini_nonequilibrium_2023}
\begin{equation} 
\begin{aligned}
\ln \eta_1 &= (1 + \delta_{p,2})s \star \ln(1 + \eta_2) - D(\lambda) \ , \\ 
 \ln \eta_\ell &= s \star \big[\ln(1 + \eta_{\ell-1}) + (1 + \delta_{p,j+1}) \ln(1 + \eta_{\ell+1})\big] + (-1)^\ell D(\lambda) \qquad {\rm for} \ 1 < \ell < p \ , \\ 
\ln \eta_p &= \bigg\{
\begin{aligned}
    s \star \ln(1 + \eta_{p-1}) + D(\lambda) \qquad &p \text{ even} \ , \\
    s \star \ln(1 + \eta_{p-1}) \qquad &p \text{ odd} \ , \\
\end{aligned}\\
\ln \eta_{p+1} &= -\ln \eta_p, 
\end{aligned}
\end{equation} 
where the function $D(\lambda)$ is given by 
\begin{equation} D(\lambda) = \ln \left[ \left( \coth \frac{(p+1)\lambda}{2} \right)^2 \right] \ . 
\end{equation}
\begin{figure}[b!]
    \centering
    \begin{minipage}[b]{0.235\textwidth}
        \centering
        \includegraphics[width=\textwidth]{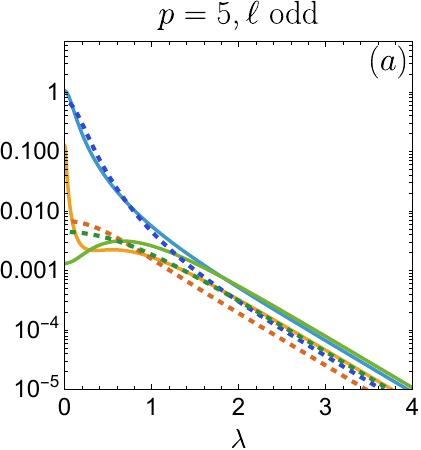}
    \end{minipage}
    \begin{minipage}[b]{0.235\textwidth}
        \centering
        \includegraphics[width=\textwidth]{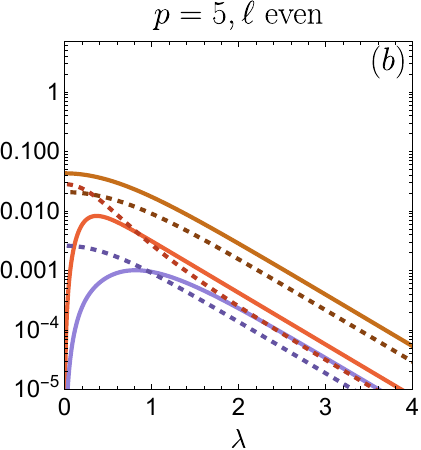}
    \end{minipage}
    \hfill
    \begin{minipage}[b]{0.235\textwidth}
        \centering
        \includegraphics[width=\textwidth]{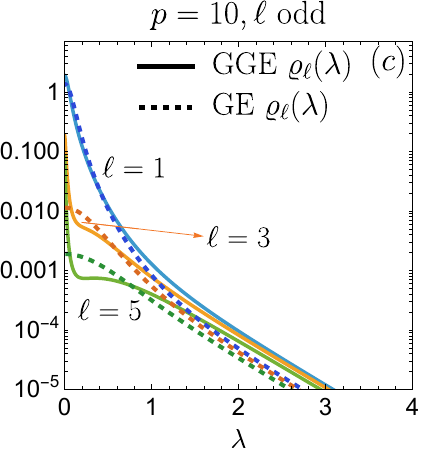}
    \end{minipage}
    \begin{minipage}[b]{0.235\textwidth}
        \centering
        \includegraphics[width=\textwidth]{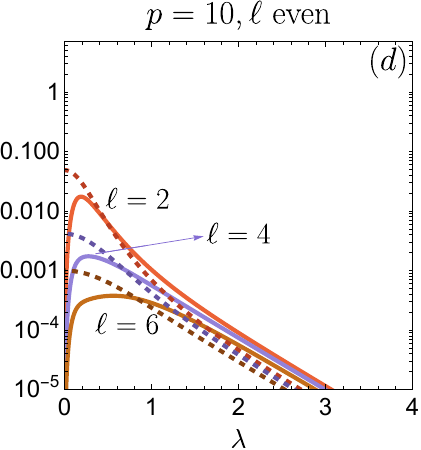}
    \end{minipage}
    
    \vspace{0.2cm} 

    \begin{minipage}[b]{0.235\textwidth}
        \centering
        \includegraphics[width=\textwidth]{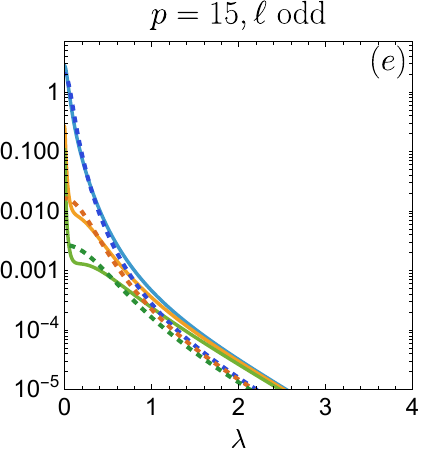}
    \end{minipage}
    \begin{minipage}[b]{0.235\textwidth}
        \centering
        \includegraphics[width=\textwidth]{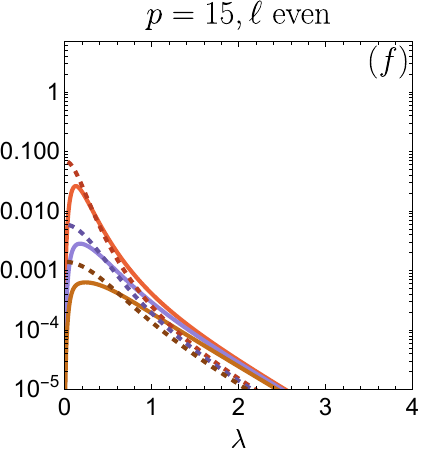}
    \end{minipage}
    \hfill
    \begin{minipage}[b]{0.235\textwidth}
        \centering
        \includegraphics[width=\textwidth]{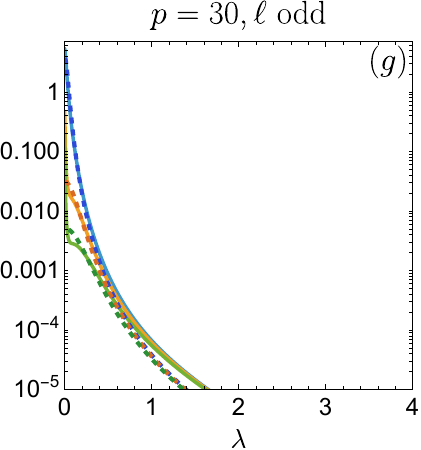}
    \end{minipage}
    \begin{minipage}[b]{0.235\textwidth}
        \centering
        \includegraphics[width=\textwidth]{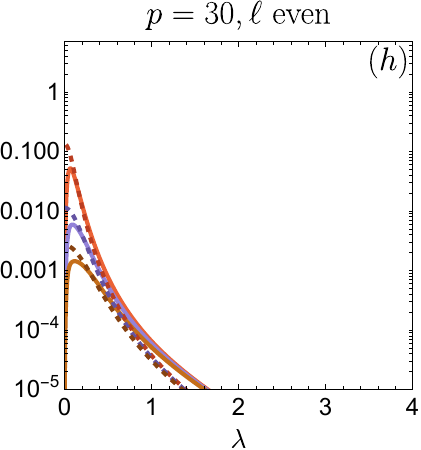}
    \end{minipage}
    
    \caption{Log-scale plots for the GE (dashed lines) and GGE (continuous lines) root densities $\varrho_{\ell}(\lambda)$  for $1\le\ell\le 6$ and for a few values of $0<\Delta<1$ in the Néel quench ($\theta = 0$). The integer $p$ refers to the parametrization \eqref{eq:Deltam1} of the anisotropy $\Delta$, with $p \to \infty$ corresponding to the critical limit $\Delta \to 1$.}
    \label{fig:XXZ_root_densities_Dl1}
\end{figure}

\begin{figure}[b!]
    \centering
    \begin{minipage}[b]{0.235\textwidth}
        \centering
        \includegraphics[width=\textwidth]{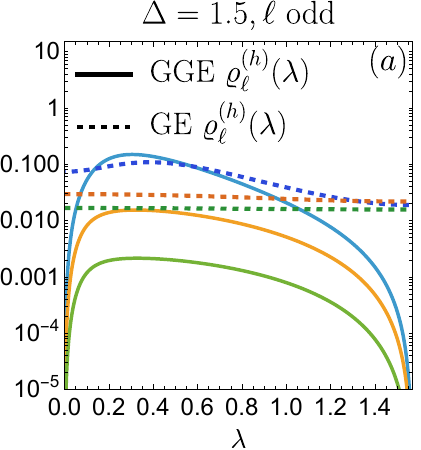}
    \end{minipage}
    \begin{minipage}[b]{0.235\textwidth}
        \centering
        \includegraphics[width=\textwidth]{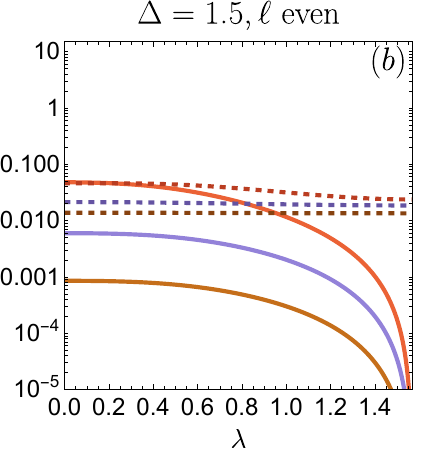}
    \end{minipage}
    \hfill
    \begin{minipage}[b]{0.235\textwidth}
        \centering
        \includegraphics[width=\textwidth]{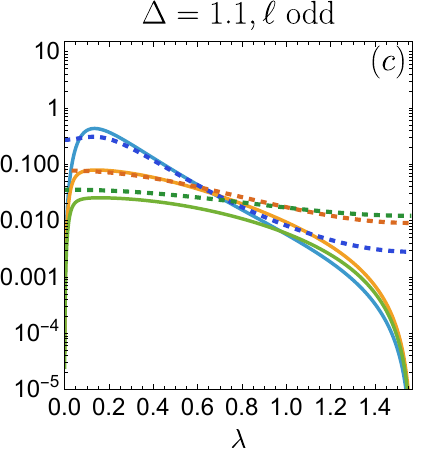}
    \end{minipage}
    \begin{minipage}[b]{0.235\textwidth}
        \centering
        \includegraphics[width=\textwidth]{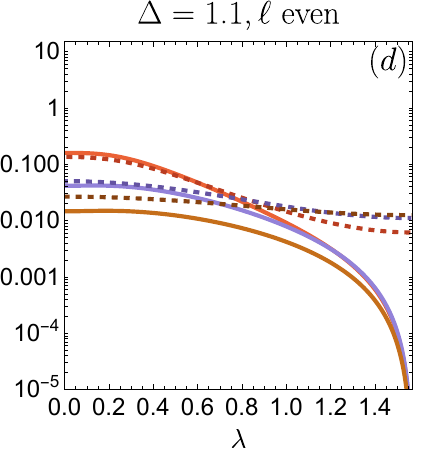}
    \end{minipage}
    
    \vspace{0.2cm} 

    \begin{minipage}[b]{0.235\textwidth}
        \centering
        \includegraphics[width=\textwidth]{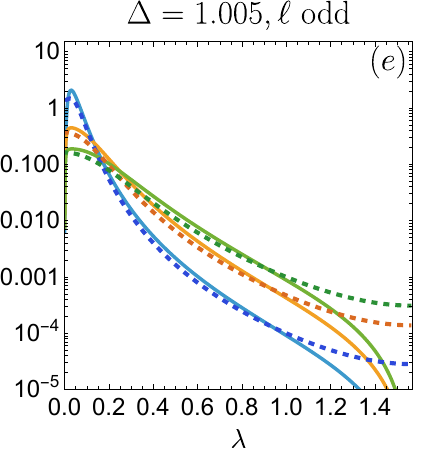}
    \end{minipage}
    \begin{minipage}[b]{0.235\textwidth}
        \centering
        \includegraphics[width=\textwidth]{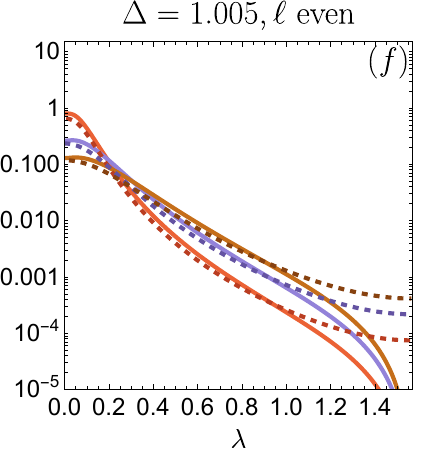}
    \end{minipage}
    \hfill
    \begin{minipage}[b]{0.235\textwidth}
        \centering
        \includegraphics[width=\textwidth]{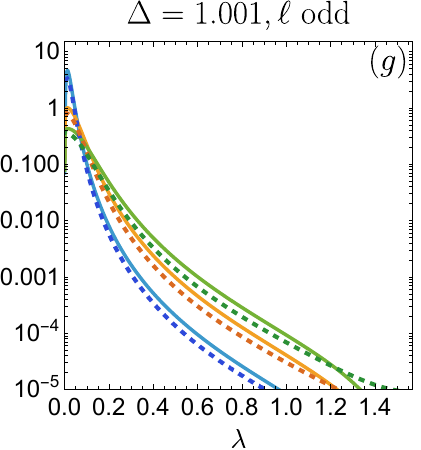}
    \end{minipage}
    \begin{minipage}[b]{0.235\textwidth}
        \centering
        \includegraphics[width=\textwidth]{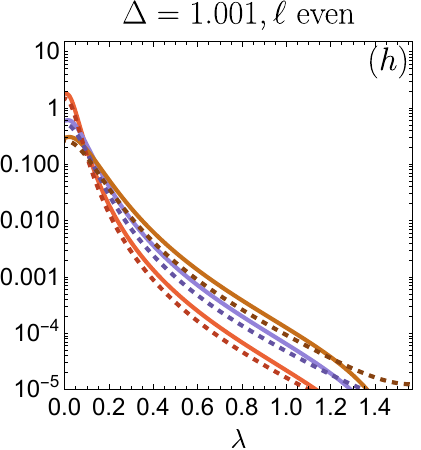}
    \end{minipage}
    
    \caption{Log-scale plots for the GE (dashed lines) and GGE (continuous lines) hole densities $\varrho_{\ell}^{(h)}(\lambda)$  for $1\le\ell\le 6$ and $\Delta>1$ in the Néel quench ($\theta = 0$). The color scheme is the same as previous figures, i.e.~$\ell = 1,3,5$ corresponding to blue, orange, green and $\ell = 2,4,6$ to red, purple, brown.}
    \label{fig:XXZ_root_holedensities_Dg1}
\end{figure}

Considerations similar to the ones of the previous section apply for the numerical solution of the sets of integral equations. Again, the GE and GGE root densities obtained from the integral equations above automatically feature the correct density of particles
\begin{equation}
    \frac{1}{2} = d = \sum_{\ell = 1}^{p+1} n_\ell \int_{-\pi/2}^{\pi/2} d\lambda \, \varrho_{\ell}(\lambda) \ ,
\end{equation}
while the value of $\beta^*$ for the GE distribution $\bs \varrho_{\beta^*}$ is set by (again denoting as $e$ the energy density in the initial state)
\begin{equation}
   e = \sum_{\ell = 1}^{p+1} \int_{-\infty}^{\infty} d\lambda \, \varrho_{\ell}(\lambda)\, \epsilon_\ell(\lambda) \ , \qquad \qquad \qquad \epsilon_\ell(\lambda) = - \pi J \sin(\gamma) a_\ell(\lambda)  \ .
\end{equation}

\subsection{Additional results}

Similarly to \cref{fig:XXZ_root_densities_Dg1} of the main text ($\Delta > 1$, $\theta = 0$), in \cref{fig:XXZ_root_densities_Dl1} we show that a pinning mechanism for $\varrho_\ell(\lambda)$ occurs also in the gapless regime $0<\Delta < 1$ at values \eqref{eq:Deltam1} parametrized by the positive integer $p$. 

In \cref{fig:XXZ_root_holedensities_Dg1} we show plots for the hole densities $\varrho_\ell^{(h)}(\lambda)$ analogous to those for $\varrho_\ell(\lambda)$ of \cref{fig:XXZ_root_densities_Dg1} in the gapped regime $\Delta > 1$ and for the Néel quench $\theta = 0$. Again, we find that a very similar pinning mechanism around $\lambda = 0$ occurs. 

\begin{figure}[h!]
  \centering
  \begin{minipage}{0.27\textwidth}
    \centering
    \includegraphics[width=\linewidth]{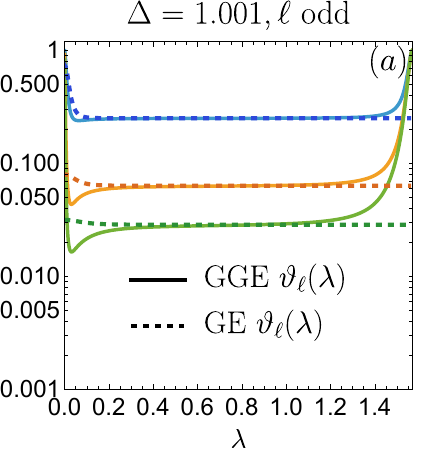}
  \end{minipage}
  \hspace{0.5cm}
  \begin{minipage}{0.27\textwidth}
    \centering
    \includegraphics[width=\linewidth]{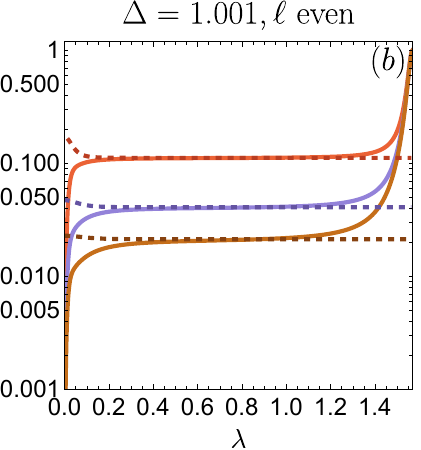}
  \end{minipage}
  \caption{Log-scale plots for the GE (dashed lines) and GGE (continuous lines) filling functions $\vartheta_{\ell}(\lambda) = \varrho_\ell(\lambda)/\varrho_\ell^{\rm (tot)}(\lambda)$  for $1\le\ell\le 6$ (same color scheme as above) and $\Delta=1.001$ in the Néel quench ($\theta = 0$).}
  \label{fig:filling_theta_XXZ}
\end{figure}

In \cref{fig:filling_theta_XXZ} we show close to criticality ($\Delta = 1.001$) the filling functions $\vartheta_\ell(\lambda)=\varrho_\ell(\lambda)/\varrho_{\ell}^{(\rm tot)}(\lambda)$ in the Néel quench $\theta = 0$. We observe that for $\vartheta_\ell(\lambda)$ the similarity between the GGE and GE curves emerges as a common plateau at intermediate values of $\lambda$, instead of a sharp pinning mechanism. In this respect, we remark that the filling functions $\vartheta_\ell(\lambda)$ do not contain all the information needed to reconstruct the entropy densities that define the athermality, as they must be supplemented with knowledge of the total density of vacancies $\varrho_{\ell}^{(\rm tot)}(\lambda)$.

\section{TBA and additional results for LL}
\label{appendix:LL}

\subsection{TBA in the repulsive regime}

For $c>0$, due to the absence of bound states, the Bethe spectrum is particularly simple~\cite{korepin_quantum_1993}. All $N$-particle eigenstates $\ket{\bs \lambda}$ at finite $L$ are uniquely identified by a set of $N$ \emph{real} rapidities $\bs \lambda=\{\lambda_1, \ldots,\lambda_N\}$, such that
\begin{equation}
\label{eq:higherchargesLL}
    Q^{(n)}\ket{\bs \lambda} = \sum_{j=1}^N \lambda_j^n \ket{\bs \lambda} \qquad n = 1, 2, \ldots \ ,
\end{equation}
where $Q^{(2)}=H_{\rm LL}$. Unlike in the XXZ chain, the rapidities $\lambda_j$ coincide with the momenta $k_j$ entering the plane-wave expansion of the Bethe wavefunctions.

To compute the density of athermality \eqref{eq:defAthDensity} we need the GE and GGE macrostates $\varrho_{\beta^*,\mu^*}(\lambda)$ and $\varrho_{\text{\tiny GGE}}(\lambda)$. 
The thermal macrostate $\varrho_{\beta^*,\mu^*}(\lambda)$ is given by the solution of the following set of nonlinear integral equations~\cite{korepin_quantum_1993}
\begin{align}
    &\frac{\varrho(\lambda)}{\varrho(\lambda)+\varrho^{(h)}(\lambda)} = \frac{1}{e^{\beta \varepsilon(\lambda)}+1} \ , \\
    &\varepsilon(\lambda) = \lambda^2 - \mu - \frac{1}{2\pi\beta} \int_{-\infty}^\infty d \sigma \, K(\lambda - \sigma) \ln\Big[e^{-\beta \varepsilon(\sigma)}+1\Big] \ ,\\
    &\varrho(\lambda)+\varrho^{(h)}(\lambda) = \frac{1}{2\pi}\Big[ 1+\int_\infty^\infty d\sigma \, K(\lambda - \sigma) \varrho(\sigma)\Big]  \ ,
\end{align}
where $K(x)=2c/(x^2 + c^2)$ and the values of $\beta^*$ and $\mu^*$ are obtained by requiring
\begin{equation}
\label{eq:LLbeta&mu}
    \int_\infty^\infty d\lambda \, \varrho_{\beta^*,\mu^*}(\lambda) = d \ , \qquad \qquad \int_\infty^\infty d\lambda \,  \lambda^2\varrho_{\beta^*,\mu^*}(\lambda) = e \ ,
\end{equation}
with $d$ and $e$ denoting, respectively, the density of particles and energy associated with the initial state in \cref{eq:GSNLL}. 

The GGE saddle-point root density $\varrho_{\text{\tiny GGE}}(\lambda)$ for the BEC quench has been obtained analytically in Ref.~\cite{de_nardis_solution_2014}, and reads
\begin{equation}
\begin{aligned}
\label{eq:spcint}
    \varrho_{\text{\tiny GGE}}(\lambda) &= \frac{1}{2\pi}\frac{a(\lambda/c)}{1+a(\lambda/c)}\left[ 1+\sqrt{\tau}\frac{I_{2-2i\lambda/c}(4\sqrt{\tau})}{I_{1-2i\lambda/c}(4\sqrt{\tau})}+\sqrt{\tau}\frac{I_{2+2i\lambda/c}(4\sqrt{\tau})}{I_{1+2i\lambda/c}(4\sqrt{\tau})}\right] \ , \\
    a(z) &= \frac{2\pi \tau}{z \sinh(2\pi z)}I_{1-2iz}(4\sqrt{\tau})I_{1+2iz}(4\sqrt{\tau}) \ ,
\end{aligned}
\end{equation}
where $I_p(x)$ denotes the modified Bessel function of the first kind of order $p$, and $\tau=e^{\mu/2}$ enforces the densities set by the BEC state $\ket{\psi(0)}$ according to $d=\tau c$ and $e=\tau^2 c^3 = d^2 c$. The GGE hole density $\varrho_{\text{\tiny GGE}}^{(h)}(\lambda)$, needed to compute $s_{\text{\tiny YY}}[\varrho_{\text{\tiny GGE}}]$, is simply given by $\varrho_{\text{\tiny GGE}}^{(h)}(\lambda)=\varrho_{\text{\tiny GGE}}(\lambda)/a(\lambda/c)$. 

\subsection{TBA in the attractive regime}

In the attractive regime $c<0$ the rapidities $\bs \lambda$ solving the Bethe equations and associated with normalizable wavefunctions are generally complex, reflecting the existence of bound states~\cite{takahashi_thermodynamics_1999}. An $\ell$-particle string takes the form
\begin{equation}
    \lambda_j = \lambda^{(0)}+i\frac{|c|}{2}(\ell + 1 - 2 j) + \delta_j \qquad \qquad j = 1, \ldots, \ell \ ,
\end{equation}
where $\lambda^{(0)}\in(-\infty,\infty)$ denotes the real string center and $\delta_j$ are exponentially small corrections that can be neglected for $L$ large. 
The GGE root densities $\bs \varrho_{\text{\tiny GGE}}(\lambda)$ for the $c<0$ quench from the BEC state \eqref{eq:GSNLL} have been derived in Ref.~\cite{piroli_multiparticle_2016, piroli_quantum_2016}. Defining $\eta_\ell(\lambda)=\varrho_{\ell}^{(h)}(\lambda)/\varrho_{\ell}(\lambda)$ and $\tilde \lambda=\lambda/|c|$, they are obtained from the following recursive set of equations
\begin{equation} 
\begin{aligned}
\eta_1(\lambda) &= \frac{\tilde \lambda^2 \left[ 1 + 4\tau + 12\tau^2 + (5 + 16\tau)\tilde \lambda^2 + 4\tilde \lambda^4 \right]}{4\tau^2(1 + \tilde \lambda^2)} \ , \\
\eta_\ell(\lambda) &= \frac{\eta_{\ell-1}\left(\lambda + \frac{i|c|}{2}\right) \eta_{\ell-1}\left(\lambda - \frac{i|c|}{2}\right)}{1 + \eta_{\ell-2}(\lambda)} - 1 \quad \qquad \ell \ge 2 \ , \\
\varrho_{\ell}(\lambda) &= \frac{\tau}{4\pi} \frac{\partial_\tau \eta_\ell^{-1}(\lambda)}{1 + \eta_\ell^{-1}(\lambda)} \ , 
\end{aligned} 
\end{equation}
where $\eta_0(\lambda) \equiv 0$ and $\tau = d/|c|$. The set above automatically enforces
\begin{align}
    \sum_{\ell=1}^\infty d_\ell &= d  \ , \quad \qquad \quad \qquad d_\ell = \ell \int_{-\infty}^\infty d \lambda \varrho_{\ell}(\lambda) \ , \\
    \sum_{\ell=1}^\infty e_\ell &= e = -d^2 |c| \ , \qquad e_\ell = \int_{-\infty}^\infty d \lambda \varrho_{\ell}(\lambda) \epsilon_\ell(\lambda) \ , \qquad  \ \epsilon_\ell(\lambda) = \ell \lambda^2 - \frac{c^2 \ell (\ell^2 -1)}{12} \ .
\end{align}
The previous equations for the particle and energy density can be used to check that, in the algorithmic determination of $\varrho_{\ell}(\lambda)$ up to a given cutoff $\ell_{\rm max}$, convergence in $\ell_{\rm max}$ is reached.

\begin{figure}[t!]
    \centering
    \begin{minipage}{0.4\textwidth}
        \centering
        \includegraphics[width=\linewidth]{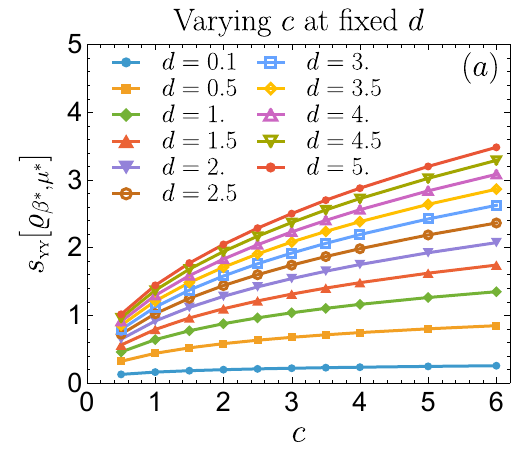}
    \end{minipage}\hspace{3pt}
    \begin{minipage}{0.4\textwidth}
        \centering
        \includegraphics[width=\linewidth]{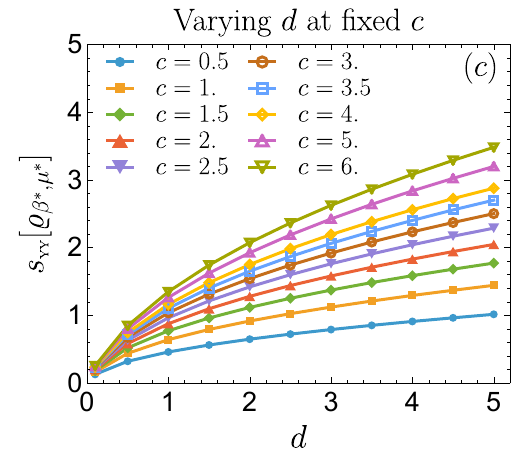}
    \end{minipage}

    \vspace{5pt} 

    \begin{minipage}{0.4\textwidth}
        \centering
        \includegraphics[width=\linewidth]{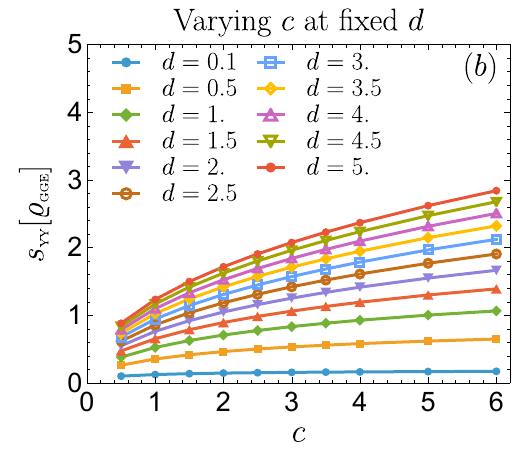}
    \end{minipage}\hspace{3pt}
    \begin{minipage}{0.4\textwidth}
        \centering
        \includegraphics[width=\linewidth]{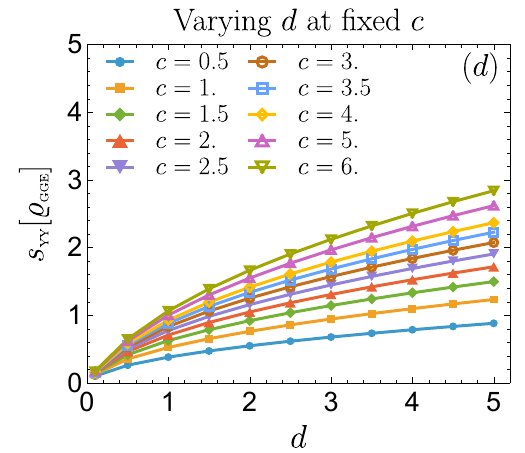}
    \end{minipage}

    \caption{Thermal and GGE entropy densities for the BEC to repulsive ($c>0$) LL quenches associated with \cref{fig:LL_main}. (a) and (b) show, respectively, $s_{\text{\tiny YY}}[\varrho_{\beta^*,\mu^*}]$ and $s_{\text{\tiny YY}}[\varrho_{\text{\tiny GGE}}]$ as functions of the postquench interaction $c$ at fixed particle density $d$. (c) and (d) show, respectively, $s_{\text{\tiny YY}}[\varrho_{\beta^*,\mu^*}]$ and $s_{\text{\tiny YY}}[\varrho_{\text{\tiny GGE}}]$ as functions of $d$ at fixed postquench $c$.}
    \label{fig:LL_main_2}
\end{figure}
\subsection{Further plots}

In \cref{fig:LL_main_2} we show the thermal and GGE entropy densities $s_{\text{\tiny YY}}[\varrho_{\beta^*,\mu^*}]$ and $s_{\text{\tiny YY}}[\varrho_{\text{\tiny GGE}}]$ from which the athermality $\alpha$ of \cref{fig:LL_main} (repulsive $c>0$ regime) has been obtained. 

\printbibliography 

\end{document}